\documentclass[a4paper,11pt]{article}
\pdfoutput=1
\usepackage{jheppub}
\usepackage{amsmath}
\usepackage{amssymb}
\usepackage{color}
\usepackage{graphicx}
\usepackage{hyperref}
\usepackage{xspace,slashed}
\usepackage[utf8]{inputenc}
\usepackage{subcaption}
\usepackage{multirow}
\usepackage{algorithm}
\usepackage{algorithmic}
\usepackage{stackengine}
\usepackage{enumitem}

\usepackage{booktabs}

\makeatletter
\g@addto@macro\bfseries{\boldmath}
\makeatother

\newcommand{\as}{\alpha_s}
\newcommand{\tp} {{\tilde p}}
\newcommand{\mathd}{\mathrm{d}}
\newcommand{\tmop}[1]{\ensuremath{\operatorname{#1}}}
\newcommand\nf{n_{f}}
\newcommand\Tf{T_{F}}
\newcommand\Ca{C_A}
\newcommand\Cf{C_F}
\newcommand\zqq{\zeta_{q{\bar q}}}
\newcommand\zqg{\zeta_{q g}}
\newcommand{\dtl}[1]{\hat{#1}} \newcommand{\dtp}{\dtl{p}}
\newcommand{\approxczero}{\underset{c\to 0}{\approx}}

\definecolor{azure}{rgb}{0.0, 0.5, 1.0}

\newcommand{\OXaff}{Rudolf Peierls Centre for Theoretical Physics,
  Clarendon Laboratory, Parks Road, Oxford OX1 3PU, UK}
\newcommand{\WADaff}{Wadham College, Oxford OX1 3PN, UK}
\newcommand{\TTPaff}{Institute for Theoretical Particle Physics,
  KIT, 76128 Karlsruhe, Germany}
\newcommand{\MILaff}{Universit\`a di Milano-Bicocca and INFN,
  Sezione di Milano-Bicocca, Piazza della Scienza 3, 20126 Milano, Italy}

\preprint{
  \begin {flushright}
    OUTP-21-21P, TTP21-026, P3H-21-056
  \end{flushright}
}

\title{
On linear power corrections in certain collider observables
}
\author[a,b]{Fabrizio Caola,}
\author[a]{Silvia Ferrario Ravasio,}
\author[c]{Giovanni Limatola,}
\author[d]{Kirill Melnikov,}
\author[c]{Paolo Nason}
\affiliation[a]{\OXaff}
\affiliation[b]{\WADaff}
\affiliation[c]{\MILaff}
\affiliation[d]{\TTPaff}

\abstract{ We study linear power corrections ${\cal O}(\Lambda_{\rm
    QCD}/Q)$ to certain collider observables.  We present arguments
  that prove that such corrections cannot appear in observables that
  are inclusive with respect to QCD radiation, such as total cross
  sections as well as rapidity and transverse momentum distributions
  of color-neutral particles.
  Although our calculations are carried
  out in a simplified framework, our arguments and conclusions are
  applicable, with some reservations, to  processes both at lepton and
  hadron colliders.
  We also show how an improved understanding
  of the origin of linear power corrections allows us to 
  simplify their calculation. As an application, we compute the
  leading non-perturbative corrections to the 
  $C$-parameter and the thrust in $e^+e^-$ annihilation in a generic
  three-jet configuration. 
 }

\begin{document}

\maketitle 
\section{Introduction}
\label{sec:intro}
An important part of the LHC physics program consists in the
exploration of phenomena that occur at distances that are between a
hundred and a thousand times smaller than the size of the
proton. Thanks to the celebrated properties of Quantum Chromodynamics
(QCD) such as asymptotic freedom and factorization, physics at such
distances can be described using perturbation theory, where elusive
quarks and gluons play the role of fundamental physical degrees of
freedom.

Corrections to this perturbative picture are expected to be small,
suppressed by ratios of the non-perturbative QCD parameter
$\Lambda_{\rm QCD} \sim 0.3~{\rm GeV}$ and the typical energy scale
$Q$ of the process (or observable) under consideration. This hard scale
$Q$ typically ranges from a few tens to a few hundred GeV. It
follows that these non-perturbative effects may change perturbative
predictions by a relative amount proportional to $(\Lambda_{\rm
  QCD}/Q)^n \sim (0.01)^n - (0.001)^n$ which, depending on the value
of $Q$ and of the exponent $n$, varies  from a percent for $n=1$ and $Q
\sim 30~{\rm GeV}$ to a permille and even smaller values for larger
values of $Q$ and $n$.

Perturbative predictions for cross sections and distributions, on the
other hand, are controlled by powers of the strong coupling constant
$\alpha_s(Q) \sim 0.1$ for $Q \sim 30 - 100~{\rm GeV}$.  Currently,
perturbative computations are often performed to second or even third
order in the expansion in $\alpha_s$, leading to theoretical
predictions for hard processes which are typically accurate to within
one to ten percent~\cite{Heinrich:2020ybq}.  Further development of
methods for perturbative calculations in QCD may improve the precision
of such theoretical predictions, perhaps by about an order of
magnitude. If this happens even for a few selected processes and
observables, perturbative predictions at this precision will have to
be supplemented with non-perturbative corrections {\it provided that
  $\mathcal O((\Lambda_{\rm QCD}/Q)^n)$ contributions with $n=1$
  exist for a particular process or observable}.  On the other hand,
non-perturbative corrections with $n> 1$ are too small to be of any
relevance for most hard processes at the LHC.

For color-singlet decay rates, deep-inelastic scattering structure
functions and inclusive decays of heavy quarks~\cite{Bigi:1993fe}, it is
well-known that operator product expansion techniques allow one to
conclude that fully inclusive observables do not receive linear power
corrections.  However, it is currently not known how to generalize
these results to more differential observables and to the case of
hadron-hadron collisions. Indeed, at present there is no full theory of
non-perturbative corrections to short-distance processes at lepton and
hadron colliders. It is then not possible to predict the exponent $n$
for a generic process or observable, let alone compute the
contribution of $O((\Lambda_{\rm QCD}/Q)^n)$ terms precisely. However,
it is well-understood that one source of non-perturbative corrections
is present within perturbation theory itself. Indeed, the appearance
of the Landau pole in the strong coupling constant leads to an
intrinsic ambiguity when integrating over soft momenta.  Since such an
ambiguity will have to cancel with contributions that arise from
physics beyond perturbation theory, it can be used as an estimate of,
at least, some non-perturbative contributions.

It is well-known that the ambiguity related to the appearance of the
Landau pole can be studied within the approximation of a large (and
negative) number of massless fermion species (see
Ref.~\cite{Beneke:1998ui} for a review).  This approach is
particularly simple if no gluons appear in a given process at leading
order. Indeed, in such cases the appearance of linear power
corrections can be investigated by computing $\mathcal O(\alpha_s)$
corrections to the process (and observable) under consideration that
originate from virtual exchanges and real emissions of \emph{massive
  gluons}, in the limit of a small gluon mass $\lambda$, see
Ref.~\cite{Beneke:1998ui}.\footnote{Since by assumption the underlying
  process does not contain gluons, this does not lead to any issue
  with gauge invariance.} The presence of terms that are \emph{linear}
in $\lambda$ implies that a particular observable receives leading
power corrections that are of the type $\mathcal O(\Lambda_{\rm
  QCD}/Q)$, while their absence can be interpreted as an indication
that non-perturbative corrections are further suppressed.

In the context of high-energy collider physics, early studies of
linear power corrections were mostly focused on studying shape
variables in electron-positron
collisions~\cite{Manohar:1994kq,Webber:1994cp,
  Akhoury:1995sp,Dokshitzer:1995zt, Nason:1995np,Dokshitzer:1995qm,
  Dasgupta:1996ki,Beneke:1997sr,Dokshitzer:1997ew,
  Dokshitzer:1997iz,Dokshitzer:1998pt}\footnote{For recent work in
  this context, see e.g.~\cite{Agarwal:2020uxi}.}, on the heavy-quark mass
definition~\cite{Beneke:1994sw,Bigi:1994em}, on the Drell-Yan
process~\cite{Beneke:1995pq,Dasgupta:1999zm,Korchemsky:1996iq} and on
jets~\cite{Korchemsky:1994is,Dasgupta:2007wa}.
Recently, first attempts were made to extend such studies to more
complicated processes that could be considered as proxies for realistic
processes at hadron colliders.  In particular, appearances of linear
power corrections in top production and decay processes and in the
transverse momentum distribution of the $Z$ bosons produced in
photon-hadron collisions were studied in
Refs.~\cite{FerrarioRavasio:2018ubr,FerrarioRavasio:2020guj},
respectively.  In both cases, calculations were performed numerically
for finite values of the gluon mass $\lambda$.  The presence or absence of
linear power corrections was established by a numerical extrapolation
to vanishing values of $\lambda$.

In the case of the $Z$ transverse momentum distribution studied in
Ref.~\cite{FerrarioRavasio:2020guj}, no evidence of linear power
corrections was found.  Although this result is fully sufficient for
phenomenological purposes, it is interesting to understand if the
presence or absence of such linear corrections in hard processes can
be deduced on more general grounds. This is what we set out to do in
this paper. Unfortunately, we are not yet able to perform an analysis
of fully realistic processes since  we have to restrict ourselves to cases
where there are no gluons at leading order.  Apart from this rather 
substantial restriction, we keep our discussion general. Whenever we
are interested in a process that \emph{does} contain gluons at leading
order (e.g. the $Z$ transverse momentum distribution, or $e^+e^-$
event shapes in the three-jet region) we follow the approach of
Ref.~\cite{FerrarioRavasio:2020guj} and use photons as proxies for hard
gluons. 
Our main findings can be stated as follows: 
\begin{itemize}
\item no linear powers of $\lambda$ arise from virtual corrections in
  generic hard processes with massless partons;
\item observables that are inclusive with respect to momenta of
  colored final state particles do not receive ${\cal O}(\lambda)$ and,
  therefore, ${\cal O}(\Lambda_{\rm QCD}/Q)$ power corrections.
  Observables of this type include e.g. total cross sections as well
  as kinematic distributions of colorless particles.
\end{itemize}
From these findings it immediately follows that no linear power
corrections appear in the inclusive Drell-Yan cross
section~\cite{Beneke:1995pq} and in the rapidity distribution of the
Drell-Yan pair~\cite{Dasgupta:1999zm}, at least away from the
kinematic boundaries.  Similarly, they also imply that the $Z$
transverse momentum distribution computed in the simplified model of
photon-hadron collisions studied in
Ref.~\cite{FerrarioRavasio:2020guj} also does not receive linear power
corrections, even if rapidity cuts on the $Z$ boson are applied.

Our results have interesting implications for event-shape studies in
$e^+e^-$ annihilation.
Recently, this topic has received renewed
attention in relation to the extractions of the strong coupling
constant $\alpha_s$ from event shapes.  Indeed, it was argued in
Ref.~\cite{Luisoni:2020efy} that a better control on non-perturbative
corrections is crucial for a reliable determination~of~$\alpha_s$.\footnote{
  We stress that the power corrections we are discussing here are not
  related with the so called next-to-leading-power soft corrections
  (see \emph{e.g.} refs.~\cite{vanBeekveld:2019prq,Ebert:2018gsn}
  and references therein).
  In the latter,
  the next-to-leading power refers to the power of a resummation variable, rather than
  to the power of a non-perturbative correction.}
In particular, in Ref.~\cite{Luisoni:2020efy} the standard approach to
computing power corrections, that consists in extrapolating them from
the two-jet to the three-jet region, was criticized.  By considering
shape variables like the $C$-parameter that exhibit two Sudakov
regions (one near the two-jet limit and the other at the three-jet
symmetric point), the authors of Ref.~\cite{Luisoni:2020efy} argued
that the coefficient of the linear power correction near the two-jet
region cannot be reliably extrapolated to the three-jet one.  With our
formalism, we can compute the coefficient of the power corrections in
the three-jet region for several shape variables, irrespective of the
presence of Sudakov regions.\footnote{We stress however that at the
  present stage we are only able to obtain robust results for
  processes of the form $e^+ e^-\to q \bar q + \gamma$, i.e. using
  photons as proxies for gluons. We will speculate on the full QCD
  generalization in Section~\ref{sec:shapevars}.}%

This paper is organized as follows.  In Section~\ref{sec:phiex} we study
non perturbative corrections in a toy model, namely the production of
two scalar color-charged particles in the decay of a massive vector
boson. Within this toy model, we argue that the decay rate in this case is
free of linear ${\cal O}(\Lambda_{\rm QCD}/Q)$ corrections. Rather
than presenting new results, the purpose of this section is to
illustrate basic features of our approach and to provide arguments
that can be generalized to more complex cases. Such generalization is
discussed in Section~\ref{sec:gencase}, which is devoted to more complex
processes with additional hard particles in both the initial and final
states. There, we generalize arguments given in Section~\ref{sec:phiex}
and argue that also in more complex cases linear ${\cal
  O}(\Lambda_{\rm QCD}/Q)$ terms are not present for observables that
are inclusive with respect to QCD radiation.

In Sections~\ref{sec:shapevars-ex} and~\ref{sec:shapevars} we consider
the implications of our result for the calculation of shape variables
in $e^+e^-$ annihilation. In particular, in
Section~\ref{sec:shapevars-ex} we discuss a specific observable, namely the
$C$-parameter, and show how our formalism can be applied to compute
non-perturbative corrections to it in an approximation where the
splitting of a massive gluon into a $q \bar q$ pair is neglected. In
Section~\ref{sec:shapevars} we present a general framework for dealing
with a broader class of shape variables. Using this framework, we
compute linear power corrections to both the thrust and the
$C$-parameter distributions and compare these results with a numerical
calculation at finite $\lambda$, extrapolated to $\lambda\to 0$. We
find consistent results, confirming our analytical findings.  We
conclude in Section~\ref{sec:conclusions}.

This paper also contains several appendices.  In
Appendix~\ref{app:int} we describe the computation of integrals
relevant for our study.  In Appendix~\ref{app:cali}, we detail the
analytic calculation of the various integrals that we use in our
analysis of the $C$-parameter in Section~\ref{sec:shapevars-ex}.  In
Appendix~\ref{app:largenfcalc}, we report technical details of the
calculation of shape variables in the large-$\nf$ limit that we
discuss in Section~\ref{sec:shapevars}. Finally, in
Appendix~\ref{app:twoJetLim} we study non-perturbative corrections
to the $C$-parameter in the two-jet limit.

\section{A toy  model: vector-boson decay to scalars}
\label{sec:phiex}
In this section, we consider the decay of a spin-one boson into two
colored charged scalars $\phi$.
Our goal is to understand linear power corrections in this model and
present arguments that, on the one hand, can be easily verified in
this simple case and, on the other hand, are sufficiently general to
be applicable in more complex situations.  Because of this, we refrain
as much as possible from using the exact form of the various matrix
elements relevant for this calculation, and instead focus on their
general structure.

We investigate power corrections to the process
\begin{equation}
  V(q) \to \phi(p_1) + \bar\phi(p_2).
  \label{eq2.1}
\end{equation}
Following the discussion in the introduction, we do this by
computing ${\cal O}(\alpha_s)$ corrections to this process in a
QCD-like theory where the gluon has a small mass $\lambda$, and by
checking whether or not such corrections contain terms that are linear
in $\lambda$.
To keep our analysis as simple as possible, in this section we only
consider the total decay rate.\footnote{It is well-known~\cite{Beneke:1998ui}
  that the total decay rate does not receive linear power corrections.
  However, as we have stressed in the introduction, we study this process
  as a first step towards establishing more general results.}
  We will discuss more complicated
processes and observables in Section~\ref{sec:gencase}. As stated
earlier, we only consider the case of massless scalars,
$p_1^2=p_2^2=0$.

We begin with the analysis of virtual corrections. There are two
contributions that need to be studied -- the wave-function
renormalization constant for the external $\phi$ particles and the
one-loop matrix element.  We start with the former. We work in
dimensional regularization, define the space-time dimension as
$d=4-2\epsilon$, and use the Feynman gauge for simplicity. The scalar's
self-energy reads
\begin{equation}
  \Sigma(p^2) = (\sigma_1\, p^2 + \sigma_2\,\lambda^2) B(p,\lambda) +
  ~{\rm terms~that~do~not~depend~on~}p^2, 
\end{equation}
where 
\begin{equation}
  B(p,\lambda) = \int\frac{d^d
    k}{(2\pi)^d}\frac{1}{(k^2-\lambda^2)(k+p)^2},
  \label{eq:genb}
\end{equation}
and $\sigma_{1,2}$ are two constants whose specific form is
irrelevant for our discussion. Using Feynman parameters,
$B(p,\lambda)$ can be written as
\begin{equation}
  B(p,\lambda) = i\frac{\Gamma(\epsilon)}{(4\pi)^{d/2}}
  \int\limits_0^1 \mathd x\,
  x^{-\epsilon}\left[\lambda^2-p^2(1-x)\right]^{-\epsilon}.
\end{equation}
This representation makes it apparent that
\begin{equation}
  B(p,\lambda)|_{p^2 = 0} \sim \lambda^{-2\epsilon},~~~~~~
  \lambda^2 \left.\frac{\partial
    B(p,\lambda)}{\partial p^2}\right|_{p^2 = 0} \sim \lambda^{-2\epsilon},
  \label{eq:zscaling}
\end{equation}
which in turn implies that the wave-function renormalization constant
$Z_\phi$ does not contain terms that are linear in
$\lambda$.

We then move to the one-loop matrix element.  There are three diagrams
that contribute to it.  Using the Passarino-Veltman
reduction~\cite{Passarino:1978jh}, one can express this matrix element
through four scalar integrals.  Omitting color indices for simplicity,
we schematically write
\begin{equation}
\begin{split} 
  & {\cal M}_{1-\rm loop}(p_1,p_2,\lambda) =
  c\left(p_1,p_2,\lambda^2\right) C(p_1,p_2,\lambda)
  + b_1\left(p_1,p_2,\lambda^2\right) B(p_1,\lambda) \\
  &\quad\quad +
  b_2\left(p_1,p_2, \lambda^2\right) B(p_2,\lambda)
  + b_3\left(p_1,p_2,\lambda^2\right)
  B(p_1+p_2,0) + a\left(p_1,p_2,\lambda^2\right) A(\lambda),
\end{split}
\label{eq:genPV}
\end{equation}
where $B(p,\lambda)$ is given in Eq.~\eqref{eq:genb} and the
other loop integrals are defined as\footnote{We note that since
  $C(p_1,p_2,\lambda)$ is finite we can evaluate it
  directly in $d=4$.}
\begin{equation}
\begin{split} 
C(p_1,p_2,\lambda) = \int \frac{{\rm d}^4 k }{(2\pi)^4}
\frac{1}{(k^2-\lambda^2) (k+p_1)^2 (k-p_2)^2 },~~~ A(\lambda) = \int \frac{{\rm
    d}^d k }{(2\pi)^d} \frac{1}{(k^2-\lambda^2)}.
\end{split} 
\end{equation}
As indicated in Eq.~\eqref{eq:genPV}, the coefficient functions $c,
b_{1,2,3}$ and $a$ are rational functions of $\lambda^2$; this is a
direct consequence of how the Passarino-Veltman reduction proceeds.
It remains to consider the scalar integrals. Eq.~\eqref{eq:zscaling}
implies that $B(p_{1,2},\lambda)\sim\lambda^{-2\epsilon}$. Also,
dimensional analysis dictates that $A \sim
\lambda^{2-2\epsilon}$. These integrals cannot then generate odd
powers of $\lambda$ upon expansion in both $\epsilon$ and $\lambda$.
The case of the scalar triangle $C$ is less trivial. However, it is
easy to see that it admits the following representation
\begin{equation}
C(p_1,p_2,\lambda) = \frac{i}{(4\pi)^2} \int \limits_{0}^{1}
\frac{{\rm d} x }{q^2 x + \lambda^2} \left [ \ln \frac{q^2 x}{\lambda^2} - i
  \pi \right ].
\end{equation}
Although it is straightforward to complete the integration over $x$
and express $C$ in terms of polylogarithmic functions, this is not
necessary as the above representation makes it obvious that the
function $C$ does not contain odd terms in the small-$\lambda$
expansion.

We then conclude that the behavior of the renormalized one-loop
virtual amplitude in the limit of small $\lambda$ is described by the
following formula\footnote{ For brevity, we refer to all possible
  ${\cal O}(\lambda^n \ln^m \lambda)$ terms as ${\cal O}(\lambda^n)$
  terms.}
\begin{equation}
  {\cal
  M}^{\rm ren}_{1-\rm loop}(p_1,p_2,\lambda) = \omega_1(p_1,p_2) \ln^2
\frac{q^2}{\lambda^2} + \omega_2(p_1,p_2) \ln \frac{q^2}{\lambda^2} +
\omega_3(p_1,p_2) + {\cal O}(\lambda^2).
\label{eq:toyvirt}
\end{equation}
The logarithmic divergences in $\lambda^2$ are the usual
soft-collinear singularities, that are canceled by analogous
contributions in the real emission terms. Apart from that,
Eq.~\eqref{eq:toyvirt} implies that the dependence on $\lambda$ in
virtual corrections starts at $\mathcal O(\lambda^2)$. We conclude
that for the process $V \to \phi \bar \phi$
virtual corrections do not induce any linear
sensitivity to infrared physics.

As the next step, we need to analyze the real-emission contribution
\begin{equation}
V(q) \to \phi(p_1) + \bar\phi(p_2) + g(k),
\end{equation}
with $k^2 = \lambda^2$. The amplitude of this process can be written
as
\begin{equation}
\begin{split} 
{\cal M}(p_1,p_2,k) & = g_s T^a_{12}\left \{ \frac{(2 p_1 + k)_\mu
  \epsilon^\mu}{(p_1 + k)^2} \mathcal M_0(p_1+k,p_2) \right.  \\
& \left.  -
\frac{(2 p_2 + k)_\mu \epsilon^\mu}{(p_2 + k)^2} \mathcal M_0(p_1,p_2+k) +
\epsilon_\mu \mathcal M_3^\mu(p_1,p_2,k) \right \}.
\label{eq2.8}
\end{split} 
\end{equation}
The two first terms on the right hand side describe emissions by
external particles while the third one describes
``structure-dependent'' radiation. In Eq.~\eqref{eq2.8}, $g_s$ is the
strong coupling, $T^a_{12}$ is the generator of the $SU(3)$ color
algebra in the fundamental representation and $\epsilon$ is
the polarization vector of the gluon. Also, $\mathcal M_0(p_1,p_2)$ is the
color-stripped matrix element with no extra emissions, while the
structure of $\mathcal M_3$ will be discussed in the
following.\footnote{A straightforward calculation shows that $\mathcal
  M_0(p_1,p_2) = e (p_2-p_1)_\mu\epsilon^\mu_V$ where $\epsilon_V$ is
  the polarization vector of the decaying vector boson and $e$ is its
  coupling to the scalars. Also, $\mathcal M_3^\mu = 2
  \epsilon_V^\mu$.}

It is obvious that if the emitted gluon is resolved,
the amplitude squared and the phase space can be
expanded in powers of $\lambda^2$. The situation is, however, more
delicate in the soft and collinear regions where \emph{a)} the amplitude
develops singularities in the $\lambda\to 0$ limit and \emph{b)} one
becomes sensitive to restrictions on the phase space induced by the
gluon mass that can be linear in $\lambda$. These regions are a potential
source of linear power corrections and we now study them in detail. 
   
We begin by discussing the emission of a soft gluon. Simple power
counting arguments show that this region could give rise to linear
power corrections. Indeed, consider a situation where the gluon energy
$\omega$ is comparable to $\lambda$, $\omega \sim \lambda$. The phase
space is proportional to $\omega\, {\rm d} \omega\, \beta\,
\theta(\omega - \lambda)$, with $\beta = \sqrt{1-\lambda^2/\omega^2}$.
Since for small $\lambda$ the real emission amplitude is expandable in
powers of $\omega$ starting with $\mathcal M \sim 1/\omega$, linear
terms in $\lambda$ could potentially be generated. To see whether this
is the case, we need to study both the matrix element and the phase
space in more detail. The power counting argument implies that linear
power corrections can only originate from next-to-leading terms in the
small-$\omega$ expansion of both the matrix element squared and the
phase space. We now discuss how to compute them.

We consider first the matrix element, and construct an expansion of
the real-emission amplitude in powers of $k$~\cite{Low:1958sn,landau}.
We write
\begin{equation}
\begin{split} 
{\cal M}(p_1,p_2,k) & = g_s T^a_{12} \left \{ \left [ \frac{(2 p_1 +
    k)^\mu }{(p_1 + k)^2} - \frac{(2 p_2 + k)^\mu
    }{(p_2 + k)^2} \right ] \mathcal M_0(p_1,p_2) \right.
\\ &
\left.  + \frac{p_{1}^\mu}{(p_1 k)} \frac{\partial
  \mathcal M_0}{\partial p_1^\alpha} k^\alpha - \frac{p_{2}^\mu
  }{(p_2 k)} \frac{\partial \mathcal M_0}{\partial p_2^\alpha}
k^\alpha + \mathcal M_3^\mu(p_1,p_2,0) +{\cal O}(k) \right \}
\epsilon_\mu.
\label{eq2.9}
\end{split} 
\end{equation}
We can determine $\mathcal M_3^\mu$ if we require that the above
expression satisfies the Ward identity which means that upon replacing
$\epsilon^\mu$ with $k^\mu$ in Eq.~(\ref{eq2.9}) we should get zero.
We find that this is achieved if the following condition is
satisfied
\begin{equation}
  \frac{\partial \mathcal M_0}{\partial p_1^\alpha} k^\alpha -
  \frac{\partial \mathcal M_0}{\partial p_2^\alpha} k^\alpha +
  \mathcal M_3^\alpha k_\alpha = 0.
\end{equation}
It follows that
\begin{equation}
  \mathcal M_3^\alpha = -\frac{\partial \mathcal M_0}{\partial p_1^\alpha}
  + \frac{\partial \mathcal M_0}{\partial p_2^\alpha}.
\end{equation}
The amplitude expanded to first subleading order in $k$ then reads
\begin{equation}
  \begin{split}
    {\cal M}(p_1,p_2,k) & = g_s T^{a}_{12} \left \{ \left [ \frac{(2
        p_1 + k)_\mu }{(p_1 + k)^2} - \frac{(2 p_2 +
        k)_\mu }{(p_2 + k)^2} \right ] \mathcal M_0(p_1,p_2)
    \right.  \\
    & \left.  + \frac{p_{1,\mu}}{(p_1 k)}
    \frac{\partial \mathcal M_0}{\partial p_1^\alpha} k^\alpha -
    \frac{p_{2,\mu}}{(p_2 k)} \frac{\partial \mathcal M_0}{\partial
      p_2^\alpha} k^\alpha -  \left ( \frac{\partial
      \mathcal M_0}{\partial p_1^\mu} -
    \frac{\partial \mathcal M_0}{\partial p_2^\mu}
    \right ) +\mathcal O(k)\right \}\epsilon^\mu.
  \end{split}
  \label{eq2.12}
\end{equation}

We now need to square this amplitude and integrate it over the phase
space of the three final state particles, working through
next-to-leading approximation in the soft limit. Although this can be done by  choosing
a particular parametrization of the three-particle phase space, we 
do this in a way
that can be generalized to more complex cases.
Consider the three-particle phase space
\begin{equation}
  \begin{split}
    &{\rm dLips}(q;p_1,p_2,k) = \\ &
    \quad\quad\frac{{\rm d}^4 p_1}{(2\pi)^3}
    \delta_+\big(p_1^2\big)\, \frac{{\rm d}^4 p_2}{(2\pi)^3}
    \delta_+\big(p_2^2\big)\, \frac{{\rm d}^4 k}{(2\pi)^3} \delta_+\big(k^2
    - \lambda^2\big)\, (2 \pi)^4 \delta^{(4)}(q - p_1 - p_2 - k).
  \end{split}
\end{equation}
To expose its dependence on $\lambda$, we follow
Ref.~\cite{DelDuca:2019ctm} and introduce a Lorentz transformation
$\Lambda$ that boosts the vector $q-k$ to the vector $\kappa q$ where
$\kappa = \sqrt{(q-k)^2/q^2}$. Specifically,
\begin{equation}
  \Lambda_{\mu\nu}(q-k)^{\nu} = \sqrt{\frac{(q-k)^2}{q^2}}q_\mu.
\end{equation}
We also find it convenient to define 
\begin{equation}
  \tilde l^\mu = \kappa^{-1} \Lambda^{\mu\nu} l_\nu,
\end{equation}
for a generic momentum $l$. It is then straightforward to obtain
\begin{equation}
  \tilde p_1^\mu + \tilde p_2^\mu = \kappa^{-1} \Lambda^{\mu\nu}
  (p_{1,\nu} + p_{2,\nu}) = \kappa^{-1} \Lambda^{\mu\nu} (
  q_\nu-k_\nu) = q^\mu.
\end{equation}
Since $\Lambda$ is a Lorentz transformation, it follows that
\begin{equation}
  \mathd^4 p_{1,2}\, \delta_+\left(p_{1,2}^2\right) =
  \kappa^4 \mathd^4\tilde p_{1,2}\,
  \delta_+\left(\kappa^2 \tilde p_{1,2}^2\right) =
  \kappa^2 \mathd^4 \tilde p_{1,2}\,
  \delta_+\left(\tilde p_{1,2}^2\right).
\end{equation}
This, together with
\begin{equation}
  \delta^{(4)}\left( q-p_1-p_2-k\right) =
  \delta^{(4)}\left( \kappa\Lambda^{-1}(q-\tilde p_1-\tilde p_2)\right) =
  \kappa^{-4} \delta^{(4)}(q-\tilde p_1-\tilde p_2),
\end{equation}
allows us to re-write the phase space as
\begin{equation}
  \begin{split}
    {\rm dLips}(q;p_1,p_2,k) = {\rm dLips}(q;\tp_1,\tp_2) \times
    \frac{{\rm d}^4 k}{(2\pi)^3} \delta_+(k^2 - \lambda^2)
    \theta\big[(q-k)^2\big],
  \end{split}
  \label{eq:psfact}
\end{equation}
where 
\begin{equation}
  {\rm dLips}(q;\tp_1,\tp_2) =
  \frac{\mathd^4 \tilde p_1}{(2\pi)^3}\delta_+\big(\tilde p_1^2\big)\,
  \frac{\mathd^4 \tilde p_2}{(2\pi)^3}\delta_+\big(\tilde p_2^2\big)\,
  (2\pi)^4 \delta^{(4)}(q-\tilde p_1 - \tilde p_2).
\end{equation}
We note that in Eq.~\eqref{eq:psfact} the dependencies on the
gluon momentum and its mass are separated from the rest of the phase
space.

We can now explicitly check whether or not soft gluons lead to
contributions proportional to $\lambda$ in the total rate. To this
end, we need to consider
\begin{equation}
  \sigma_R = N^{-1}
  \int {\rm dLips}(q;p_1,p_2,k) |{\cal M}(p_1,p_2,k)|^2,
\end{equation}
where $N$ is an irrelevant normalization factor,
expand the integrand through next-to-leading order in small $k$
and integrate over $k$.  To facilitate this, we perform the Lorentz
transformation discussed above. We write
\begin{equation}
  \sigma_R = N^{-1}\int {\rm dLips}(q;\tilde p_1, \tilde p_2)\;
  \frac{{\rm d}^4 k}{(2\pi)^3} \delta_+(k^2 - \lambda^2)
  \theta\big[(q-k)^2\big] |{\cal M}( \kappa \Lambda^{-1} \tp_1, \kappa
  \Lambda^{-1} \tp_2,k)|^2.
\end{equation}
For soft gluons, the Lorentz transformation is small.
To first order in $k$, it is straightforward
to obtain
\begin{equation}
  \Lambda_{\mu\nu}^{-1} = g_{\mu\nu} - \frac{k_\mu q_\nu-q_\mu
    k_\nu}{q^2} + \mathcal O(k^2),
\end{equation}
leading to
\begin{equation}
  p_i^\mu   = \tilde p_i^\mu  
  -\frac{1}{2}k^\mu - \frac{(k q)}{q^2}\tilde p_i^\mu +
  \frac{(k\tilde p_i)}{q^2}q^\mu + \mathcal O(k^2),~~~~ i=1,2.
  \label{eq:ptilde1}
\end{equation}
We use $q = \tilde p_1+\tilde p_2$ and rewrite Eq.~\eqref{eq:ptilde1} as
\begin{equation}
\begin{split} 
   p_1^\mu = \tilde p_1^\mu -\frac{1}{2} k^\mu - {B^{\mu
       \nu}k_\nu}, \;\;\;\; p_2^\mu = \tilde p_2^\mu -\frac{1}{2}
   k^\mu + {B^{\mu \nu} k_\nu},
\end{split} 
\end{equation}
where the antisymmetric tensor $B^{\mu \nu}$ reads 
\begin{equation}
B^{\mu \nu} = \frac{\tp_1^\mu \tp_2^\nu - \tp_2^\mu \tp_1^\nu
}{2(\tp_1\tp_2)}.
\end{equation}

We now consider the matrix element $\mathcal M(p_1,p_2,k)$. From the
discussion above it follows that it is sufficient to consider the
approximation Eq.~\eqref{eq2.12}, that we now write in terms of the
momenta $\tilde p_{i}$. To this end, we note that
\begin{equation}
(p_{1,2}+k)^2 = k^2 + 2 p_{1,2}^\mu k_\mu = 2 \tp_{1,2}^\mu k_\mu \mp 
  2k_\mu B^{\mu \nu} k_\nu = 2(\tp_{1,2}k) ,
\end{equation}
and
\begin{equation}
  2 p_1^\mu + k^\mu = 2 \tp_1^\mu - 2 B^{\mu \nu} k_\nu,~~~~~
  2 p_2^\mu + k^\mu = 2 \tp_2^\mu + 2 B^{\mu \nu} k_\nu.
\end{equation}
We also introduce the soft current
\begin{equation}
J^\mu=\frac{ \tp_1^\mu  }{(\tp_1 k)}
  -  \frac{ \tp_2^\mu }{(\tp_2 k)},
\end{equation}
and use Eq.~\eqref{eq2.12} to write the expansion of the matrix
elements in powers of $k$ as
\begin{equation}
  \begin{split}
    &{\cal M}( \kappa \Lambda^{-1} \tp_1, \kappa \Lambda^{-1} \tp_2 ,k)
     = g_s T^a_{12} \left \{ \left[J^\mu  -
    B^{\mu \nu} k_\nu \left ( \frac{ 1 }{ (\tp_1 k)} +\frac{1}{(\tp_2 k)}
      \right )\right] \mathcal M_0( \tp_1,\tp_2) \right. \\
    &\quad\quad\quad\quad\quad\quad\quad
    - \frac{J^\mu}{2} \left ( \frac{\partial \mathcal M_0}{\partial
      \tp_1^\alpha} + \frac{\partial \mathcal M_0}{\partial
      \tp_2^\alpha} \right ) k^\alpha
    - J^\mu \left (
    \frac{\partial \mathcal M_0}{\partial \tp_1^\alpha} - \frac{\partial
      \mathcal M_0}{\partial \tp_2^\alpha} \right ) B^{\alpha \beta} k_\beta
    \\
    &\quad\quad\quad\quad\quad\quad\quad
    \left.  + \frac{\tp_{1}^{\mu}}{(\tp_1 k)}
    \frac{\partial \mathcal M_0}{\partial \tp_1^\alpha} k^\alpha -
    \frac{\tp_{2}^{\mu}}{(\tp_2  k)} \frac{\partial
      \mathcal M_0}{\partial \tp_2^\alpha} k^\alpha - \left (
    \frac{\partial \mathcal M_0}{\partial \tp_{1,\mu}} - \frac{\partial
      \mathcal M_0}{\partial \tp_{2,\mu}} \right )
    +\mathcal O(k)\right \}\epsilon_\mu.
  \end{split}
  \label{eq:me2exp}
\end{equation}
Computing the square of this amplitude is straightforward. One can
replace the sum over the gluon polarizations with $-g^{\mu \nu}$ since
the Ward identity is satisfied. Contracting the soft current $J_\mu$
with the various structures that appear in Eq.~\eqref{eq:me2exp}, we
obtain
\begin{equation}
  J^\mu \frac{\tp_{1,\mu}}{(\tp_1 k)} = - J^\mu
  \frac{\tp_{2,\mu}}{(\tp_2 k)} = \frac{1}{2} J^\mu J_\mu =
  -\frac{ (\tp_1 \tp_2)}{(\tp_1 k)(\tp_2 k)}, \;\;\;
  J_\mu B^{\mu \nu} k_\nu = -1.
\end{equation}
Using these results, we find
that all derivatives of the leading order amplitude drop out from
the matrix element squared and we
obtain
\begin{equation}
  \begin{split}
  \sum_{\rm col,pol}| {\cal M}( \kappa \Lambda^{-1} \tp_1, \kappa
  \Lambda^{-1} \tp_2,k ) |^2 =
  \sum_{\rm col}  | {\cal M}_0(\tp_1, \tp_2 ) |^2\times
  \\
  g_s^2 C_F \times
  \frac{2
    (\tp_1 \tp_2)}{(\tp_1 k) (\tp_2 k) }
  \left [ 1 - \frac{(\tp_1 k)
    }{(\tp_1 \tp_2)}- \frac{(\tp_2 k) }{(\tp_1 \tp_2)}
    + {\cal O}(k^2)
  \right ].
  \end{split}
  \label{eq:polsum}
\end{equation}
In Eq.~\eqref{eq:polsum}, we need to sum over the gluon polarizations but
it is not necessary to sum over the polarizations of the decaying particle. 

We are now ready to ascertain whether soft gluon emission leads to
linear power corrections. To do this, we study the ratio
\begin{equation}
\frac{\sigma_R}{\sigma_{LO}}
= g_s^2 C_F {\cal R}(q^2),
\end{equation}
 where
 \begin{equation}
{\cal R}(q^2) = \int \frac{{\rm d}^4 k}{(2\pi)^3} \delta_+(k^2 -
\lambda^2) \theta\big[(q-k)^2\big] \frac{2 (\tp_1 \tp_2)}{(\tp_1 k) (\tp_2 k) }
\left [ 1 - \frac{(\tp_1 k) }{(\tp_1 \tp_2)}- \frac{(\tp_2 k) }{(\tp_1 \tp_2)}
  + {\cal O}(k^2) \right ].
\label{eq:calR}
 \end{equation}
In principle, we need to integrate this formula over soft gluon
momenta but it is actually easier to integrate it over all
possible values of $k$. Since the region of large gluon momenta only
gives rise to quadratic terms in $\lambda$, it is safe
to extend the integration region. The computation of the
integrals in Eq.~\eqref{eq:calR} is described in
Appendix~\ref{app:int} where it is proven that they do not contain 
terms which are linear in $\lambda$. We conclude that 
\begin{equation}
\begin{split} 
{\cal R}(q^2)  =   
     \omega^{(2)}_R(q) \ln^2 \frac{q^2}{\lambda^2}
   + \omega^{(1)}_R(q) \ln \frac{q^2}{\lambda^2} + \omega^{(0)}_R(q)  + {\cal O}(\lambda^2).
\end{split}
\label{eq3.28}
\end{equation}
Eq.~(\ref{eq3.28})   implies that the emission of soft gluons does not generate
$\mathcal O(\lambda)$ contributions to the total decay rate. 

Since we work at next-to-leading order in the soft expansion,
one may wonder whether soft \emph{scalars}
give rise to linear terms in $\lambda$. 
To study this situation, consider the matrix element in
Eq.~(\ref{eq2.8}).  We are interesting in its behavior 
in the limit $p_1\to 0$. Momentum conservation then implies
that both $p_2$ and $k$ are large. Since
$\epsilon_\mu k^\mu = 0$ and since ${\cal M}_3$ is non-singular in the $p_1
\to 0$ limit, we conclude that the matrix element is not
singular for $p_1 \to 0$.  This, together with the fact that the phase
space is proportional to $E_1 {\rm d} E_1$, allows us to conclude that
also this kinematic region cannot produce terms that are linear in
$\lambda$.

The last potentially problematic region is the hard-collinear one.
For definiteness we will consider the kinematic configuration where the gluon
momentum is collinear to the outgoing momentum $p_1$.  To study this
region, we employ the Sudakov decomposition to parametrize the gluon
momentum $k$ and write
\begin{equation}
  k = z p_1 + \beta p_2 + k_\perp.
\end{equation}
Then,
introducing $s_{12} = 2 (p_1 p_2)$, we write
\begin{equation} {\rm d}^4 k \;
\delta_+(k^2 - \lambda^2) = \frac{s_{12}}{2} \; {\rm d} z \; {\rm d}
\beta \; {\rm d}^2 \vec k_\perp \; \delta_+(s_{12} z \beta - \vec k_\perp^2 -
\lambda^2) \to \frac{ {\rm d}z \; {\rm d}^2 \vec k_\perp }{2z} \Bigg
|_{\beta \to \frac{ \lambda^2 + \vec k_\perp^2}{s_{12} z}},
\end{equation}
where in the last step we have integrated over $\beta$ to remove the
$\delta$-function. The $z\to 0,1$ limits correspond to the soft-gluon and
soft-scalar cases that we have already discussed.
In the hard-collinear region, $k_\perp$ should be integrated from zero
to some value which is large compared to the gluon mass and $z$ should
be integrated between some minimal value that is much larger than $\lambda/q$
to $z \sim 1$.  Inspection of
propagators shows that they are quadratic in
$\lambda$. Indeed,
\begin{equation}
  \frac{1}{(p_1+k)^2} = \frac{z}{\lambda^2(1+z) + \vec k_\perp^2}, \;\;\;
  \frac{1}{(p_2+k)^2} = \frac{1}{s_{12} z + \lambda^2}.
\end{equation}
We conclude that the contribution of the
hard-collinear region leads to an expansion in powers of $\lambda^2/q^2$ and
cannot produce terms that are linear in $\lambda$ \emph{unless} linear
terms in $k_\perp$ appear.

The presence of such linear terms in $k_\perp$ is, in principle,
possible in processes where more particles are involved. For example,
suppose that there is another particle with momentum $p_3$ in
the process. Then considering the collinear region $k || p_1$ and
writing the Sudakov decomposition for $p_3 = z_3 p_1 + \beta_3 p_2 +
p_{3\perp}$, we find
\begin{equation}
  \frac{1}{(p_3+k)^2 } \sim \frac{1}{ s_{12} z \beta_3 + ... + k_{\perp,\mu}
    \,p_{3 \perp}^\mu}.
\end{equation}
Since in the hard-collinear region $ s_{12} z \beta_3 \sim 1$ and
$k_\perp \sim \lambda$, we can expand the above propagator in powers
of $k_\perp$. If the integration over the directions of $k_\perp$ is
not restricted, all odd powers of $k_\perp$ disappear.\footnote{We
  note that this cancellation is not guaranteed if observables have a
  non-trivial dependence on $\vec k_\perp/|\vec k_\perp|$ in the
  collinear limit. We will come back on this issue when discussing
  event-shape variables in Section~\ref{sec:shapevars}.}
  Even powers of
$k_\perp$, on the other hand, correspond to even powers of $\lambda$.
Hence also in this case the hard-collinear region does not give rise
to linear power corrections. 
Finally, we note that, since collinear radiation is local in momentum
space, this conclusion is general and applies to any process,
regardless of its complexity.

To conclude, in this section we have studied the process $V\to
\phi\bar \phi(+g)$ and provided general arguments showing that no
contributions that are linear in the gluon mass can appear. While this
result is neither surprising nor new (see e.g.~\cite{Beneke:1998ui}
for a generic discussion of decay rates), we avoided using explicit
formulas for the matrix elements and for the phase space in the hope
that the above arguments can be then easily generalized to more
complex processes and observables.  Indeed, the next section is
devoted to the discussion of such generalization, for a broader class
of processes which occur at both leptonic and hadronic colliders.

\section{General case}
\label{sec:gencase}
We now turn to the discussion of more general cases and 
study processes at lepton and hadron colliders, with the usual
caveat that we do not consider processes that involve gluons at
leading order.\footnote{As a consequence, the ``hadron collider'' may
  become a photon-proton collider as it is done in
  Ref.~\cite{FerrarioRavasio:2020guj}.} Although our discussion is
general, for simplicity we focus on cases where only two massless
color-charged partons are present in the Born amplitude (i.e. where
there is only one emitting QCD dipole), while keeping the number of
colorless particles arbitrary. One of our motivations for studying
this case is the analysis of the $Z$ transverse momentum distribution
in photon-proton collisions in Ref.~\cite{FerrarioRavasio:2020guj}
that we would like to understand analytically. Also, we wish to
develop a general formalism that allows one to deal with non-perturbative
corrections to a large class of event-shape variables. 

Compared to the discussion of Section~\ref{sec:phiex}, if we consider
``hadron'' colliders we should also study the renormalization of
parton distribution functions (PDFs). However, at least as long as the
collinear factorization framework holds, PDFs renormalization is
process-independent, and can be then studied in deep-inelastic
scattering.  But there an operator product expansion allows one to
conclude on very general grounds that power corrections start at
$\mathcal O ((\Lambda_{\rm QCD}/Q)^2)$, and no linear terms are
present.\footnote{An explicit calculation of the collinear
  counterterms that shows that this is indeed the case can be found
  e.g. in Ref.~\cite{Beneke:1995pq}.} We then only need to discuss
virtual and real corrections, in analogy with what we did in
Section~\ref{sec:phiex}. We devote the next two subsections to this.

\subsection{Virtual corrections}
To study $\mathcal O(\lambda)$ contributions arising from virtual
corrections, we need to consider both the wave-function
renormalization constant and the one-loop amplitude. However, a simple
calculation shows that even in the case of quarks the former can be
expanded in powers of $\lambda^2$ and $\ln \lambda$, so that it cannot
give rise to linear power corrections.  Hence, we only need to
investigate one-loop amplitudes.\footnote{The situation is more
  delicate for heavy quarks. Indeed, in this case it is well-known
  that the mass renormalization counterterm can receive linear power
  corrections. see e.g.~\cite{Beneke:1998ui} for a review.}

We have discussed virtual corrections in the toy model of the previous
section where we argued that Passarino-Veltman reduction combined with
explicit formulas for scalar integrals that contribute to the one-loop
matrix element for the $V\to\bar\phi\phi$ process makes it obvious that
virtual corrections possess an expansion in powers of $\lambda^2$ and
logarithms of $\lambda$.
To generalize this discussion, we note that the Passarino-Veltman
reduction argument remains valid also for more complex processes, but the
scalar integrals that one obtains are more complicated.

The analysis
in the previous section was based on an explicit computation of the
three-point function $C(p_1,p_2,\lambda)$. To understand what
happens in the case of more complex integrals, it is useful to go back to
that computation and ask whether an expansion of the
three-point function in powers of $\lambda$ can be constructed
directly in the momentum representation.
The answer to this question is known \cite{Smirnov:1997gx}. To obtain
such an expansion, one writes the following identity\footnote{We consider
  $C(p_1,-p_2,\lambda)$ rather than $C(p_1,p_2,\lambda)$ for convenience.
  Indeed, $C(p_1,-p_2,\lambda)$ is symmetric under $1\leftrightarrow 2$
  exchange.}
\begin{equation}
  C(p_1,-p_2,\lambda) = \int \frac{{\rm d}^d k}{(2\pi)^4 } \left [
  T^\gamma_{\lambda} +T^{(1)}_{k^2}+T^{(2)}_{k^2} \right ]
  \frac{1}{(k^2-\lambda^2) (p_1+k)^2 \; (p_2+k)^2},
\end{equation}
where the three operators produce particular Taylor expansions of various
propagators.\footnote{Although this integral is finite in four
  dimensions, individual expansion terms may exhibit divergences that
  we regularize dimensionally.  In fact, it is known in this case that
  dimensional regularization is not sufficient to regularize
  $T^{(1)}_{k^2}$ and $T^{(2)}_{k^2}$ separately.  This subtlety is
  not important to us since it only affects terms that contain
  logarithms of $\lambda$.}  Specifically,
\begin{equation}
\begin{split} 
  & T^\gamma_{\lambda} \frac{1}{k^2-\lambda^2} = \frac{1}{k^2} \sum
  \limits_{j=0}^{\infty} \left ( \frac{\lambda^2}{k^2} \right
  )^j,\;\;\; T^{(i)}_{k^2} \frac{1}{ (p_i+k)^2} = \frac{1}{2(p_i k)} \sum
  \limits_{j=0}^{\infty} \left ( \frac{-k^2}{2 (p_i k)} \right )^j, \;
  i=1,2.
\end{split} 
\end{equation}
It is obvious that the operator $T^\gamma_{\lambda}$ produces terms
that only contain even powers of $\lambda$.  To see that this is also
the case for $T^{(1,2)}_{k^2}$, consider the $j$-th term in the
expansion generated  by $T^{(1)}_{k^2}$
\begin{equation}
\int \frac{{\rm d}^d k}{(2\pi)^4 } 
\frac{(-k^2)^j}{(k^2-\lambda^2) (2 p_1 k) ^{j+1} \; (p_2+k)^2}.
\end{equation}
Using the Sudakov decomposition
\begin{equation}
k = \alpha p_1 + \beta p_2 + k_\perp, 
\end{equation}
it is easy to see that upon rescaling $k_\perp \to \lambda k_\perp$
and $\alpha \to \lambda^2 \alpha$, the $j$-th term in the sum scales as $(
\lambda^2)^j$, modulo logarithmic corrections.  We conclude that the
triangle $C(p_1,-p_2,\lambda)$  can be expanded in powers of $\lambda^2$ and no linear
corrections can be generated, in agreement with the explicit result
of Section~\ref{sec:phiex}.

We note that the reason why the three operators
$T_{\lambda^2}^\gamma$, $T^{(i)}_{k^2}$, $i=1,2$, are needed to expand
the three-point function in powers of $\lambda$ is as
follows. Starting from the Sudakov decomposition, it is possible to
recognize~\cite{Smirnov:1997gx} that only three kinematic
configurations may contribute to the expansion of the three-point
function $C$ in power of $\lambda$. They are
\begin{equation}
\begin{split} 
  & \alpha \sim \beta \sim \frac{k_\perp}{s_{12}} \gg \lambda; \;\;\;
  \alpha \sim \frac{\lambda^2}{s_{12}},\; k_\perp \sim \lambda,\;
  \beta \sim 1;\;\;\;\; \beta \sim \frac{\lambda^2}{s_{12}}, \;
  k_\perp \sim \lambda, \; \alpha \sim 1,
  \label{eq3.5}
\end{split} 
\end{equation}
where $s_{12} = 2 (p_1 p_2)$. These three regimes correspond to the
$T^\gamma_{\lambda^2}, T_{k^2}^{(1)}$ and $T_{k^2}^{(2)}$ operators,
respectively (see Ref.~\cite{Smirnov:1997gx} and references therein
for more details).

We continue with the discussion of more complex scalar integrals.  A
typical case that arises e.g. in the computation of corrections to the
$Z$-boson transverse momentum distribution is the four-point function
that, in addition to the three propagators that appear in
$C(p_1,-p_2,\lambda)$, contains a further propagator that does not go
on the mass-shell in any of the singular limits (i.e. when $k$ becomes
soft or collinear to external particles). We write
\begin{equation}
D(p_1,p_2,p_3,\lambda) = \int \frac{{\rm d}^d k}{(2\pi)^d }
\frac{1}{(k^2-\lambda^2)(p_1+k)^2 \; (p_2+k)^2 \; (q+k)^2},
\end{equation}
where $p_1^2=p_2^2=p_3^2=0$, $q^2 = (p_2+p_3)^2\ne 0$.
The expansion of $D$ in powers of $\lambda$ proceeds in the same way
as for the three-point function. We find
\begin{equation}
D(p_1,p_2,p_3,\lambda) = \int \frac{{\rm d}^d k}{(2\pi)^d } \left [
T^\gamma_{\lambda} +T^{(1)}_{k^2}+T^{(2)}_{k^2} \right ]
\frac{1}{(k^2-\lambda^2)(p_1+k)^2 \; (p_2+k)^2 \; (q+k)^2}.
\end{equation}
The way the three operators act on the off-shell propagator follows
from the scalings of various Sudakov parameters in relevant regions,
cf. Eq.~(\ref{eq3.5}).  Then,
the operator $T^\gamma_{\lambda}$ does nothing to the last propagator
whereas the operators $T^{(1,2)}_{k^2}$ produce its expansion in
powers of $\alpha, k^2$ and $k_{\perp,\mu}\, q^\mu_\perp$ or in powers
of $\beta, k^2$ and $k_{\perp,\mu}\,q^\mu_\perp$, respectively.
It follows that the operator $T^{\gamma}_\lambda$ generates an expansion in
powers of $\lambda^2$.  The action of the operator $T^{(1)}_{k^2}$ leads to
integrals of the following type
\begin{equation}
\int \frac{{\rm d}^d k}{(2\pi)^d } \frac{\{ k^2, (p_2 k), k_\perp^2
  \}^j}{(k^2-\lambda^2) (2 p_1 k)^{l} \; (p_2+k)^2 \; \big[q^2 + 2(q p_2)(k
    p_1)/(p_1 p_2)\big]^m },
\label{eq:dexp}
\end{equation}
where we made use
of the fact that upon averaging over directions of $k_\perp^\mu$ all
odd powers of $q_{\perp,\mu}\,k^\mu_\perp$ disappear. Also,
numerator terms such as $(k_{\perp,\mu}\, q^\mu_\perp)^{2n}$ can be
rewritten upon azimuthal integration in terms of $k_\perp^2$ and $q_\perp^2$. 
Upon rescaling $\alpha \to \lambda^2 \alpha, k_\perp \to \lambda
k_\perp$, we observe that the integral in Eq.~\eqref{eq:dexp} is proportional to $
(\lambda^2)^j$.  The analysis of how the operator $T^{(2)}_{k^2}$ acts
on the integrand leads to the same conclusion.  It follows that also the
box integral $D(p_1,p_2,p_3,\lambda)$ can be expanded in powers of
$\lambda^2$.

Using the Passarino-Veltman procedure, higher-point integrals can be
reduced to boxes, triangles, bubbles and tadpoles. The latter two can
be straightforwardly expanded in powers of $\lambda^2$ and
$\ln\lambda$, as shown in the previous section. Box and triangle
integrals that do not develop infrared and collinear singularities in
the $\lambda\to 0$ limit can be  expanded in powers of
$\lambda^2$ in a straightforward way. In fact, for such integrals, the first correction to the
$\lambda \to 0$ limit scales as ${\cal O}(\lambda^2)$. Integrals that
do develop infrared and collinear singularities, on the other hand,
can be related to the box and triangle cases discussed
above. We therefore conclude that virtual corrections for generic
processes with massless particles do not generate
linear power corrections.

\subsection{Real radiation}
We now discuss real corrections. Specifically, we consider a generic
process $I \to F$, where $I$ and $F$ are short-hand notations
for the collection of initial and final state particles, respectively,
and study the real-emission corrections $I\to F + g$ where $g$ is
a gluon with mass $\lambda$.  We imagine that there are two and only
two massless partons with QCD charges each of which can be either in
the initial or in the final state.  We do not consider cases when one
of these partons is a gluon. On the other hand, we allow for an
arbitrary number of (massless or massive) QCD-neutral particles. 

From the discussion in Section~\ref{sec:phiex}, it follows that to expose
the infrared sensitivity we only need to consider singular kinematic
configurations. Furthermore, in Section~\ref{sec:phiex} we argued that
collinear emissions do not lead to linear power
corrections.\footnote{ This statement is valid as long as the
  observable under consideration satisfies certain properties upon
  azimuthal integration, see Section~\ref{sec:shapevars} for a discussion
  in the context of event shapes in $e^+e^-$ annihilations.} Since
collinear emissions are local in momentum space, the argument of
Section~\ref{sec:phiex} holds for generic processes. We then only need to
consider soft radiation.

When discussing the toy model of Section~\ref{sec:phiex}, we have argued
that soft scalars do not lead to linear power corrections.  One may
wonder if the same conclusion also holds for soft quarks.  To study this,
consider a quark with momentum $p^\mu = E (1, \vec n)$,
where the energy $E$ is small, $E \to 0$.  The phase space volume
element of a massless
quark is proportional to $E {\rm d} E$ and the most singular
contribution to any matrix element squared is proportional to $E/(2(pk)
+ \lambda^2)^2$ where one power of the energy in the numerator comes from
the density matrix of a soft quark.  Hence, the contribution of the
small-energy region arises from energies that are proportional to
$\lambda^2$ and is then given by
\begin{equation}
\int \limits_{0}^{E_{\rm max}} E {\rm d} E\; \frac{E}{(2(pk)
  +\lambda^2)^2} \sim c + \mathcal O (\lambda^2),
\end{equation}
where $c$ is independent of $\lambda$. 
We conclude that soft quarks cannot produce contributions that are
linear in $\lambda$.

As a result, we reach the conclusion that we only need to
investigate the emission of soft gluons. From the discussion in
Section~\ref{sec:phiex} it follows that it is sufficient to expand
both the matrix element and the phase space in the soft limit
retaining the first subleading (i.e. next-to-eikonal) terms.  In the
remainder of this section, we discuss how to do this for processes
which are more general than the one discussed in
Section~\ref{sec:phiex}. We need to consider three distinct cases: the
two QCD partons are in the final state (``final-final'' dipole), one
parton is in the initial state and the other one is in the final state
(``initial-final'' dipole) and, finally, both partons are in the
initial state (``initial-initial'' dipole).  In what follows, we study
these three cases separately. For definiteness, we will always denote
the momenta of the partons that form the dipole as $p_1$ and $p_2$,
irrespective of whether they are in the initial or in the final state.

\subsubsection{Final-final dipole}
\label{sec:ffdip}
We consider a process where colorless particles with a total momentum
$p_I$ produce final state particles with momenta
$p_{1},p_{2},p_3,...,p_N$
\begin{equation}
  p_I \to p_1 + p_2 +p_3 ... + p_N.
\end{equation}
Particles with momenta $p_{1,2}$ have QCD charges;\footnote{We remind the
  reader that $p_1$ and $p_2$ cannot be gluons.}
all other particles are colorless. We only consider cases
where the QCD-charged partons are massless, $p_1^2 = p_2^2 = 0$.
For ease of notation, in this section we will assume that also the non-QCD
final-state particles are massless.
To explore power corrections in this situation, we consider the emission
of a massive gluon with momentum $k$ and mass $\lambda$. Momentum
conservation then reads
\begin{equation}
  p_I \to p_1 + p_2 +p_3 ... + p_N + k.
\end{equation}
We note that since, by construction, there is only one gluon
participating in the process, the non-Abelian nature of QCD is
immaterial and Ward identities are trivially satisfied.

As we discussed in Section~\ref{sec:phiex}, to study soft-gluon emission
it is convenient to construct  mappings of hard final state particles 
\begin{equation}
  p_i \longrightarrow \tilde p_i = \tilde p_i(\{p_j\},k),\;\;\; i=1...N, 
\end{equation}
that preserve both the on-shell conditions $\tilde p_i^2 = p_i^2,\;
i=1...N$, and the momentum conservation constraint
\begin{equation}
p_I =
\sum_{i=1}^N \tilde p_i = \sum_{i=1}^N \; p_i + k.
\end{equation}

As we argued in Section~\ref{sec:phiex}, since we are interested in
linear power corrections, we only require these mappings to first
order in the gluon momentum $k$.  We now discuss how to construct
them. Although we only need to find one particular mapping that
satisfies the above requirements, we keep our discussion 
general as different mappings may offer different advantages and
disadvantages when used in practical applications.
One assumption is that
the mappings behave as
\begin{equation}
  p^\mu_i = \tilde p^\mu_i + K_i^{\mu\nu}k_\nu + \mathcal O(k_0^2)
  \label{eq:gensmooth}
\end{equation}
for small gluon momentum, 
where the tensors $K_{i}^{\mu \nu}$ are constructed using the momenta
$\tilde p_i$ and the metric tensor.
The momentum-conservation constraint implies
\begin{align}
\sum \tilde p^\mu_i = \sum p^\mu_i+k^\mu  \;\; \Rightarrow   \;\; \sum K^{\mu \nu}_i k_\nu = -k^\mu.
\label{eq3.7}
\end{align}
We also require our mapping to satisfy the following
form of the on-shell condition
\begin{align}
  p_i^2 = (1+ \lambda_i) {\tilde p}_i^2,
\end{align}
where $\lambda_i$ are analytic functions of momenta.  Using
Eq.~(\ref{eq:gensmooth}), we find
\begin{equation}
  2 \tilde p_{i,\mu} K_i^{\mu \nu} k_\nu = \lambda_i {\tilde p_i}^2.
\end{equation}
We note that $\lambda_i \sim \mathcal O(k)$. 

We are now in position to express the phase-space element for
final-state particles in terms of the momenta $\tp_i$.  We write
\begin{equation}
  \begin{split}
    {\rm d}{\rm Lips}(p_I;p_1,..,p_N,k) = \left[\prod\limits_{i=1}^N
      \frac{{\rm d}^4 p_i}{(2\pi)^{3}} \delta_+(p_i^2) \right] \;
    [\mathd k] \; (2\pi)^4\delta^{(4)} \left(p_I-\sum\limits_{i=1}^N
    p_i -k\right) \\ = \left[\prod\limits_{i=1}^N \frac{{\rm d}^4
        \tilde p_i}{(2\pi)^{3}} \delta_+\big({\tilde p}_i^2
      (1+\lambda_i) \big) \right] \; [\mathd k] \;
    \frac{\partial(p_1,...,p_N)}{\partial(\tilde p_1,...,\tilde p_N )}
    \; (2\pi)^4\delta^{(4)} \left(p_I-\sum\limits_{i=1}^N \tilde p_i
    \right),
     \label{eq:genps_ff}
     \end{split}
\end{equation}
with $[\mathd k] = \mathd^4 k/(2\pi)^3\; \delta_+(k^2-\lambda^2)$.
We now make use of the fact that we only need this expression to first
order in $k$.  Then, using a relation between the determinant and the
trace of a matrix that is nearly the identity matrix and the fact that
$\lambda_i \sim k$, we find
\begin{equation}
   {\rm d}{\rm Lips}(p_I;p_1,..,p_N,k) \approx {\rm d}{\rm
     Lips}(p_I;\tilde p_1,..,\tilde p_N) \; \frac{{\rm d}^4 k}{(2\pi)^3} \;
   \delta_+(k^2 - \lambda^2) \; J,
   \label{eq3.11}
\end{equation}
where
\begin{equation}
J =  1-\sum\limits_{i=1}^{N} \lambda_i +\sum\limits_{i=1}^{N}
 \frac{\partial K_i^{\mu\nu}}{\partial
   \tilde p_i^{\mu}} k_\nu.
 \label{eq3.12}
\end{equation}

To proceed further, we need to specify the mapping explicitly.  To
this end, we focus on the so-called dipole-local mappings,
i.e. mappings where the momenta of the particles that do not belong to
the radiating dipole are not transformed.
By assumption, the dipole in our case
is formed by the final state particles with momenta
$p_{1,2}$. Therefore, we choose $K_i = 0$ for $i=3,4..,N$.
Furthermore, we want to construct $K_{1,2}^{\mu \nu}$ using only
$\tilde p_{1,2}$ and the metric tensor.  Then, writing the most general form of $K_{i=1,2}^{\mu \nu}$
\begin{equation}
K_{i}^{\mu \nu} =  \left ( a_i {\tilde p}_1^\mu
{\tilde p}_1^\nu + b_i {\tilde p}_2^\mu {\tilde p}_2^\nu + c_i {\tilde
  p}_1^\mu {\tilde p}_2^\nu +d_i {\tilde p}_2^\mu {\tilde p}_1^\nu
\right )+
e_i g^{\mu \nu},
\end{equation}
and using Eq.~(\ref{eq3.7}) together with the fact that the coefficients of
the tensor do not depend on $k$, we find the following constraints 
\begin{equation}
  a_1+a_2= 0,\; b_1 + b_2 = 0,\;
c_1 + c_2 = 0,\; d_1 + d_2 = 0,\; e_1 + e_2 = -1.
\label{eq3.13}
\end{equation}
The requirement that ${\tilde p}_{i,\mu} K_i^{\mu \nu} k_\nu \propto
{\tilde p}_i^2$ leads to the following set of equations
\begin{equation}
\begin{split} 
  & a_1\, {\tilde p}_1^2 + d_1\, (\tp_1 \tp_2) + e_1\propto \tp_1^2,
  \;\;\ b_1\, (\tp_1 \tp_2) + c_1\, \tp_1^2 \propto \tp_1^2, \\
  & 
  b_2\, {\tilde p}_2^2 + c_2\, (\tp_1 \tp_2) + e_2\propto \tp_2^2,
  \;\;\ a_2\, (\tp_1 \tp_2) + d_2\, \tp_2^2 \propto \tp_2^2.
\label{eq3.14}
\end{split}
\end{equation}
In total, Eqs.~(\ref{eq3.13},\ref{eq3.14}) provide 9 equations for
the ten unknowns $a_i,b_i,c_i,d_i,e_i$. We decide to express the solution
in terms of $e_1$.
It reads
\begin{equation}
  a_{1,2} = b_{1,2} = 0,~~~ c_2 = -c_1 = \frac{1+e_1}{(\tp_1\tp_2)},
  ~~~
  d_2 = -d_1 = \frac{e_1}{(\tp_1\tp_2)},~~~ e_2 = -1 - e_1.
\end{equation}
Denoting $e_1 = -\alpha$, we  finally obtain
\begin{equation}
\begin{split} 
& K_1^{\mu \nu} = -\alpha g^{\mu \nu} - \frac{(1-\alpha) \tp_1^\mu
    \tp_2^\nu -\alpha \tp_2^\mu \tp_1^\nu}{(\tp_1 \tp_2) }, \\ &
  K_2^{\mu \nu} = -(1-\alpha) g^{\mu \nu} + \frac{(1-\alpha) \tp_1^\mu
    \tp_2^\nu -\alpha \tp_2^\mu \tp_1^\nu}{(\tp_1 \tp_2) }.
  \end{split} 
\label{eq3.24}
\end{equation}

It is now straightforward to finalize the computation of the
phase-space transformation.  We find
\begin{equation}
  \begin{gathered}
  \lambda_1 = -2(1-\alpha)\,\frac{(\tilde p_2 k)}{(\tilde p_1 \tilde p_2)},
  ~~~~
  \lambda_2 = -2 \alpha\,\frac{(\tilde p_1 k)}{(\tilde p_1 \tilde p_2)},
  \\
  \frac{\partial K_1^{\mu\nu}}{\partial\tilde p_1^{\mu}} = -
  (3-4\alpha)\frac{\tilde p_2^\nu}{(\tilde p_1 \tilde p_2)},
  ~~~~
  \frac{\partial K_2^{\mu\nu}}{\partial\tilde p_2^{\mu}} = 
  (1-4\alpha)\frac{\tilde p_1^\nu}{(p_1\tilde p_2)}.
  \end{gathered}
\end{equation}
With the help of these
equations, the Jacobian of the transformation $J$ in Eq.~(\ref{eq3.12})
is found to be
\begin{equation}
J = 1 + (1-2\alpha) \frac{(\tp_1 k) - (\tp_2 k)}{(\tp_1 \tp_2)}.
\label{eq3.19}
\end{equation}
The integration over $k$ is restricted by the same condition that we
discussed in  Section~\ref{sec:phiex}. In particular, writing
$q=\tilde p_1+\tilde p_2$ and using momentum conservation $p_1+p_2 =
q-k$, we find that the condition $(q-k)^2>0$ puts an upper bound
on the possible values of $k$. 

Before continuing, we now briefly comment on the form of the
transformation Eq.~\eqref{eq3.24}. First, if we compare it with the
analogous transformation in Section~\ref{sec:phiex}, we immediately
see that the mapping used there corresponds to the symmetric case
$\alpha=1/2$. This case is particularly simple because the phase-space
Jacobian does not receive linear corrections.  For the sake of
generality, however, in this section we will keep $\alpha$
arbitrary. Second, it is interesting to note that the mapping
Eq.~\eqref{eq3.24} automatically satisfies nice infrared
conditions. The soft limit is not particularly interesting, since for
$k\to 0$ one has $p_i = \tp_i$ by construction.  The collinear limit
is however less trivial. In this case, if we formally replace $k$ with
$\eta\, p_1$, we find $p_1 = (1-\eta)\, \tp_1$ and $p_2 = \tp_2$,
which is exactly what we expect from a collinear-safe mapping. An
analogous result holds for the $k\to \eta\, p_2$ case.\footnote{
  Although for our purposes it is sufficient to consider a mapping of the form
  Eq.~\eqref{eq3.24}, we note that in principle we could have also employed
  less smooth mappings. For example, assume that
  $K^{\mu\nu}_{1,2}$ can be generically written as
  \begin{equation}
    K^{\mu\nu}_{i}=K^{\mu\nu}_{i,\parallel}+K^{\mu\nu}_{i,\perp},
    \nonumber
  \end{equation}
  where $K_{i,\parallel}$ and $K_{i,\perp}$ satisfy the conditions
  \begin{equation}
    K^{\mu\nu}_{i,\parallel}\left(g_{\mu \alpha} -
    \frac{\tp_{1,\mu}\tp_{2,\alpha}+\tp_{1,\alpha}\tp_{2,\mu}}
         {(\tp_1\tp_2)}\right)=0,
         \qquad K^{\mu\nu}_{i,\perp} \tp_{1\nu}=K^{\mu\nu}_{i,\perp}
         \tp_{2\nu}=0.
         \nonumber
  \end{equation}
  Although $K^{\mu\nu}_{i,\perp}$ could depend in a non-trivial way on
  $(\tp_{1,2}k)$, it is easy to see that this term leads to an odd
  linear dependence on the transverse momentum component of $k$, which
  vanishes after azimuthal integration. Because of this, it is then
  possible to show that all the arguments presented in this section
  would apply to this case as well, with no significant modification.}

Having studied the phase-space transformation, we need to discuss the
matrix element and its integration.  Since  we only have one QCD
dipole, the matrix element squared summed over gluon and quark
polarizations can be written in the following way
\begin{equation}
|{\cal M}|^2(\{p_i\},k) =
\frac{A(\{p_i\},k)}{(p_1+k)^2 (p_2+k)^2}
+
\frac{B_1(\{p_i\},k)}{[(p_1+k)^2]^2}
+
\frac{B_2(\{p_i\},k)}{ [(p_2+k)^2]^2}.
\label{eq3.26}
\end{equation}
The functions $A,B_{1,2}$ are polynomials in $k$. The limiting
behavior of the function $A$ follows from the standard soft
approximation. Hence, we can write
\begin{equation}
  A(\{p_i\},k) = a_0(\{p_i\})  + a_1^\mu(\{p_i\}) k_\mu  + {\cal O}(k^2),
\end{equation}
where $a^\mu_1(\{p_i\})$ is a four-vector that, in principle, depends
on all vectors $p_i$.
To understand the contributions proportional to $B_{1,2}$,
we note that they can 
only appear from squares of diagrams where a gluon is emitted and
absorbed by the same line.  Focusing on the function $B_1$ for the
sake of definiteness, we can write
\begin{equation}
  \frac{B_1(\{p_i\},k)}{[(p_1+k)^2]^2} \propto -g^{\mu \nu}
  \frac{1}{[(p_1+k)^2]^2} {\rm Tr} \left [ (\hat p_1 + \hat k) \gamma_\mu
    \hat p_1 \gamma_\nu \left (\hat p_1 + \hat k \right).....  \right ],
\end{equation}
where we have used $\sum \epsilon_\mu \epsilon^*_\nu = -g_{\mu \nu}$ to
sum over gluon polarizations.\footnote{ The sum over gluon
  polarizations for massive gluons contains a term $k_\mu
  k_\nu/k^2$. However, this term can be dropped because of the Ward
  identities that are valid in the (abelian) problem.}  A simple
computation then gives
\begin{equation}
  B_1(\{p_i\},k) = \lambda^2\big[b_{10}(\{p_i\}) + b_{11}^\mu(\{p_i\})
    k_\mu\big]-
  (p_1+k)^2\, b_{11}^{\mu}(\{p_i\})k_\mu.
  \label{eq:bterm}
\end{equation}
A similar calculation shows that $B_2$ admits an analogous
decomposition. The term proportional to $(p_1+k)^2$ in
Eq.~\eqref{eq:bterm} removes the double pole in Eq.~\eqref{eq3.26};
therefore, it can be treated together with the term
$A/(p_1+k)^2/(p_2+k)^2$ in Eq.~\eqref{eq3.26}.  Finally, power
counting arguments show that contributions of the form
$\lambda^2/[(p_i+k)^2]^2$ are not required since, due to the
$\lambda^2$ suppression, the small $k$-region in these integrals only
produces $\mathcal O(\lambda^2)$ contributions.

What remains to do is to integrate the amplitude squared, expanded
through first order in $k$, over the gluon phase space, after the $p
\to \tilde p$ transformation is performed. To remap the matrix element
squared, we use Eq.~(\ref{eq3.26}) but we discard double poles, for the
reasons we
just explained. Writing the propagators as 
\begin{equation}
(p_{1,2}+k)^2 = 2 (\tp_i k) \pm  (1-2 \alpha) \lambda^2 \mp 2(1-2 \alpha)
  \frac{(\tp_1 k)(\tp_2 k)}{(\tp_1 \tp_2)}.
\end{equation}
and expanding them to next-to-leading order in $k \sim \lambda$, we obtain. 
\begin{equation}
\frac{1}{(p_{1,2}+k)^2} = \frac{1}{(2\tp_i k)} \left ( 1 \mp (1-2\alpha)
\frac{\lambda^2}{2 (\tp_i k)} \pm (1-2\alpha) \frac{(\tp_1 k)(\tp_2
  k)}{ (\tp_i k)(\tp_1 \tp_2) } \right ).
\end{equation}

We now consider theoretical predictions for an observable that is
inclusive with respect to  QCD radiation. It follows that
\begin{equation}
{\cal O}(p_1+p_2+k;p_3,..,p_N) = {\cal O}(\tp_1+\tp_2;\tp_3,..,\tp_N).
\end{equation}
For such observables, we can write 
\begin{equation}
  \begin{split}
    \frac{1}{\mathcal O}
    \frac{ {\rm d} \sigma}{{\rm d}{\rm
        Lips}(p_I;\tp_1,...,\tp_N)} & = N^{-1} g_s^2 C_F\;
    |{\cal M}_0(\{\tp_i\})|^2\;
    \int \; \frac{\mathd^4 k}{(2\pi)^3}\;\delta_+(k^2-\lambda^2)
    \theta\big[(q-k)^2\big]
    \\
    &\times
    \frac{2(\tp_1\tp_2)}{(\tp_1 k) (\tp_2 k)} \left [ 1
      + v^\mu k_\mu - (1-2\alpha) \left ( \frac{\lambda^2}{2 (\tp_1k)} -
      \frac{\lambda^2}{2 (\tp_2k)} \right ) \right ],
    \label{eq3.4}
  \end{split}
\end{equation}
where $N$ is an irrelevant normalization factor and 
the vector $v^\mu$ depends on momenta $\tp_i$; its
exact form is not important for our purposes.  We note that the upper
bound on $k$-integration follows from the constraint $\theta[(q-k)^2]$.
In principle, since Eq.~(\ref{eq3.4}) refers to the expansion around
the soft $k \sim \lambda$ region, the integration could have been
restricted accordingly. However, since our goal is to understand
whether ${\cal O}(\lambda)$ terms appear in the differential cross
section, we can extend the integration to all values of $k$ since
the region where $k$ is hard does not generate linear ${\cal O}(\lambda)$
terms.  A discussion of the integrals that appear in Eq.~(\ref{eq3.4}) is
given in Appendix~\ref{app:int} where we show that they can be
written as a power series in $\lambda^2$.

We conclude that arbitrary differential cross sections that are {\it
  inclusive} w.r.t. the QCD radiation are free of linear power
corrections. On the other hand, if one computes an observable that is
sensitive to gluon momenta, linear sensitivity can appear.\footnote{
  The same holds for less-inclusive definitions of cross sections,
  like e.g. the so-called longitudinal cross section in $e^+e^-$
  annihilation, which indeed receives linear power
  corrections~\cite{Beneke:1997sr}.}  We discuss this case in details in
Sections~\ref{sec:shapevars-ex} and~\ref{sec:shapevars}.

\subsubsection{Initial-final dipole}
In this section, we generalize the discussion of Section~\ref{sec:ffdip}
to the case where one of the radiating partons is in the initial state
and the other one is in the final state. This is relevant for example
for the production of a vector boson with non-vanishing transverse
momentum in hadronic collisions.  At the Born level, we write
\begin{equation}
  p_1 + p_3 \to p_2 + p_F,
\end{equation}
where we have assigned momenta in such a way that partons with momenta $p_{1,2}$
form the dipole and $p_F$ stands for the momenta of
the colorless particles.
Since our formalism in its current form cannot deal with gluons in the
Born process, we follow the same approach as in
Ref.~\cite{FerrarioRavasio:2020guj} and consider quark-photon
collisions
\begin{equation}
  q(p_1) + \gamma(p_3) \to q(p_2) + X(p_F).
\end{equation}

We are interested in constructing a local dipole mapping that involves
the partons $p_1$ and $p_2$ and that can be used to understand linear power
corrections in this process.
At variance with the discussion of the previous section, when
constructing the mapping for the initial-state parton we require that
the direction of its momentum does not change. Then, writing the
transformation for $p_1$ as
\begin{equation}
  p_1^\mu = \tp_1^\mu + (\kappa_1 k) \tp_1^\mu,
  \label{eq3.30}
\end{equation}
and using momentum conservation
\begin{equation}
  p_1 - p_2 - k = \tp_1 - \tp_2,
\end{equation}
we derive 
\begin{equation}
  p_2^\mu = \tp_2^\mu + (\kappa_1 k) \tp_1^\mu - k^\mu.
\end{equation}
Similar to the case of final-final dipoles, we require that the
on-shell conditions are not affected by the mapping.  This is
obviously the case for Eq.~(\ref{eq3.30}) which implies
\begin{equation}
p_1^2 = \big(1 + 2 (\kappa_1 k) + \mathcal O(k^2)\big)\, \tp_1^2,
\end{equation}
for any $\kappa_1$.  The equation for $p_2^2$ is more informative. We find
\begin{equation}
  p_2^2  = \tp_2^2 + 2(\tp_1\tp_2 )(\kappa_1 k) - 2 (\tp_2 k)
  +\mathcal O(k^2).
\end{equation}
Hence, to satisfy the condition  $p_2^2 \propto \tp_2^2$, we  require
\begin{equation}
2\left[(\tp_1\tp_2 )\kappa_1^\mu -\tp_2^\mu\right] k_\mu = 0.
\end{equation}
Since $\kappa_1$ is $k$-independent, it follows that 
\begin{equation}
\kappa_1^\mu  = \frac{\tp_2^\mu}{(\tp_1 \tp_2)}. 
\end{equation}
In summary, for an initial-final dipole we find the following momenta
mappings
\begin{equation}
  \begin{split}
     p_1^\mu = \left(1+\frac{(\tilde p_2 k)}{(\tilde p_1 \tilde
       p_2)}\right) \tilde p_1^\mu, \;\;\;\; p_2^\mu = \tilde p_2^\mu +
     \frac{(\tilde p_2 k)}{(\tilde p_1 \tilde p_2)}
     \tilde p_1^\mu -k^\mu.
  \end{split}
  \label{eq:mapIF}
\end{equation}
We note that also in this case this mapping is well-behaved in the
soft and collinear limits. Indeed, by construction, in the soft $k\to 0$  limit
one has $p_i\to \tp_i$. If we formally replace $k$ with $\eta\,
p_1$, we obtain $p_1 = (1+\eta)\,\tp_1$, $p_2=\tp_2$. Similarly, for
$k\to \eta\, p_2$ we obtain $p_1=\tp_1$, $p_2 = (1-\eta)\,\tp_2$.
  
Next, we consider the phase-space transformation. The calculation
proceeds exactly as in the case of the final-final dipole in the
previous section except that in the current case, we only integrate
over the momentum $p_2$. Since $p_2^2 = \tp_2^2$, the
parameter $\lambda_2$ from the previous section should be set to
zero. Also, using the result for the Jacobian
\begin{equation}
\frac{\partial(p_2)}{\partial( \tp_2 )} = 1 + \frac{(\tp_1 k)}{(\tp_1
  \tp_2)},
\end{equation}
and the momentum conservation, we obtain 
\begin{equation}
   {\rm dLips}(p_1,p_3; p_2,p_F,k) =
   {\rm dLips}(\tilde p_1, \tilde p_3; \tilde p_2, \tilde p_F) \;
   \frac{{\rm d}^4 k }{(2\pi)^3} \; \delta_+(k^2 - \lambda^2) \left (
   1 + \frac{(\tp_1 k)}{(\tp_1 \tp_2)} \right ),
   \label{eq:if_ps}
\end{equation}
where $p_3=\tp_3$. 
   Similarly to what we saw in the final-final case, the integral over
   the gluon momentum is constrained by the requirement $(q-k)^2 < 0$,
   where $q = \tp_1 - \tp_2$, which is assumed in Eq.~\eqref{eq:if_ps}.

To compute hadronic cross sections, we need to convolute the partonic
phase space and the matrix element squared with parton distribution
functions. We write
\begin{equation}
  \begin{split}
  {\rm d} \sigma_R &= \int {\rm d} x_1 {\rm d} x_3\;
  f_q(x_1)\;f_\gamma(x_3)\\
  &\times
  {\rm dLips}(x_1 P_1 , x_3 P_3 ; p_2,p_F,k) 
  \frac{|{\cal M}(x_1 P_1,x_3 P_3,p_2,...,k)|^2}{2 s_{\rm hadr} x_1 x_3},
  \end{split}
\end{equation}
where $P_{1,3}$ are the momenta of the incoming hadrons,
$s_{\rm hadr} = 2 (P_1 P_3)$ is the hadronic center-of-mass  energy squared and
$f_{q,\gamma}$ are the quark and photon parton distribution functions.
We now interpret Eq.~\eqref{eq:mapIF} as a 
transformation rule for $x_1$. Indeed, writing $p_1 = x_1 P_1$ and
$\tilde p_1 = \tilde x_1 P_1$ we find through linear order in $k$
\begin{equation}
  x_1 = \tilde x_1 + \frac{(\tilde p_2 k)}{(\tilde p_2 P_1)} = \tilde x_1
  + \xi(k,\tilde p_2).
\end{equation}
We then use the phase space transformation and obtain 
\begin{equation}
  \begin{split} 
    {\rm d} \sigma_R = & \int {\rm d} \tilde x_1 {\rm d} x_3 f_q\left
    ( \tilde x_1 + \xi(k,\tilde p_2) \right) \, f_\gamma(x_3)\; {\rm
      dLips}(\tilde x_1 P_1,x_3 P_3; \tilde p_2,p_F) \; \\ & \times
    \frac{{\rm d}^4 k }{(2\pi)^3} \; \delta_+(k^2 - \lambda^2) \left (
    1 + \frac{(P_1 k)}{(P_1 \tp_2)} \right ) \frac{\big|{\cal
        M}\big((\tilde x_1 + \xi) P_1, x_3 P_3,\tilde p_2,
      ...,k\big)\big|^2}{2 s_{\rm hadr} \big( \tilde x_1 + \xi(k,\tilde
      p_2)\big) x_3}.
  \end{split}
\end{equation}
Under the assumption that $x_1$ is a regular point, the above equation can
be expanded in $\xi$. Since $\xi$ appears in the argument of
the quark distribution function $f_q$ we write
\begin{equation}
f_q\left(\tilde x_1 + \xi \right) =
f_q(\tilde x_1) +
f'_q(\tilde x_1) \xi  + \mathcal O(k^2).
\end{equation}
Similarly, the amplitude can be expanded up to next-to-eikonal level
in a way that is analogous to what we have discussed in
Section~\ref{sec:ffdip}. The only difference is the expansion of the two
singular propagators that now read
\begin{equation}
  \begin{split}
    &\frac{1}{(p_1-k)^2} =   -\frac{1}{2(\tilde p_1 k)}
    \left (1 - \frac{(\tilde p_2 k)}{(\tilde p_1 \tilde p_2)}
    +\frac{\lambda^2}{2(p_1 k)}\right)
    + \mathcal O(\lambda),\\
    &
    \frac{1}{(p_2+k)^2} = \frac{1}{2(\tilde p_2 k)}
    \left (1-\frac{(\tilde p_1 k)}{(\tilde p_1 \tilde p_2)}
    + \frac{\lambda^2}{2(\tilde p_2 k)}\right) + \mathcal O(\lambda).
  \end{split}
\end{equation}
Combining these results, we find that we again need to consider
integrals that are identical to the ones for the final-final case.
As we have already said, all such integrals are discussed in
Appendix~\ref{app:int} where it is shown that they can be expanded in
powers of $\lambda^2$. We conclude that also in this case there are no
linear power corrections to kinematics distributions of final-state
QCD-neutral particles.
Among other things, this implies that the transverse momentum of a
vector boson does not receive linear power corrections even if
rapidity cuts are imposed, at least in our simplified
``hadron-photon'' setup.

\subsubsection{Initial-initial dipole}
In this section, we consider the case where both radiating partons are
in the initial state.  For concreteness, we study the Drell-Yan
process
\begin{equation}
  q(p_1) + \bar q(p_2) \to V(p_V). 
\end{equation}
Although it is well-known that the cross section of this process does
not receive linear power corrections~\cite{Beneke:1995pq}, we study it
using our formalism for completeness. We begin by considering a
suitable phase-space mapping for the process
\begin{equation}
  q(p_1) +\bar q(p_2) \to V(p_V) + g(k),
\end{equation}
where $k^2=\lambda^2$. We focus on local mappings.  We would
like to preserve the directions of both $p_1$ and $p_2$, so we look for
mappings of the form
\begin{equation}
  \begin{split}
    &p_1 = \big(1+(\kappa_1 k)\big)\,\tilde p_1,
    \\
    &p_2 = \big(1+(\kappa_2 k)\big)\,\tilde p_2,
    \\
    &p_V = \tilde p_V + (\kappa_1 k)\,\tilde p_1 +
    (\kappa_2 k)\,\tilde p_2 -k.
  \end{split}
  \label{eq:genmap_ii}
\end{equation}
We note that the above mappings automatically satisfy the momentum
conservation condition $p_1 + p_2 - (p_V+k) = \tilde p_1 + \tilde p_2
- \tilde p_V$ and also preserve the on-shell condition for the
incoming partons.  Requiring that the vector boson remains on the mass
shell $p_V^2 = \tp_V^2$, we obtain
\begin{equation}
  \big((\kappa_1 k)+(\kappa_2 k)\big)(\tp_1\tp_2) - (k\tp_1)-(k\tp_2) = 0.
  \label{eq3.55}
\end{equation}
Clearly, this equation
does not have a unique solution for the two vectors $\kappa_{1,2}$.
However, we can require that the
transformation does not change the rapidity of the vector boson $Y_V$
in the laboratory frame. Using $P_{1,2}$ to denote momenta of the colliding
hadrons, we find
\begin{equation}
  e^{-2Y_V} = \frac{(P_1 p_V)}{(P_2 p_V)} =
  \frac{ (P_1 \tilde p_V) + (\kappa_2
    k) (P_1 \tp_2) - (P_1 k)}
       { (P_2 \tilde p_V) + (\kappa_1 k) (P_2 \tp_1)
  -(P_2 k)}.
\end{equation}
Hence, if we choose
\begin{equation}
\kappa_1^\mu = \frac{P_2^\mu }{(P_2 \tp_1)} = \frac{\tp_2^\mu}{(\tp_1
  \tp_2)}, \;\;\;
\kappa_2^\mu = \frac{P_1^\mu }{(P_1 \tp_2)} =
\frac{\tp_1^\mu}{(\tp_1 \tp_2)},
\label{eq3.57}
\end{equation}
we find
\begin{equation}
e^{-2Y_V} = \frac{(P_1 p_V)}{(P_2 p_V)} = \frac{(P_1 \tp_V)}{(P_2 \tp_V)}.
\end{equation}
It is easy to check that the choice of $\kappa_i$-vectors 
in Eq.~(\ref{eq3.57}) satisfies the on-shell conditions Eq.~(\ref{eq3.55}).
Once again, Eq.~\eqref{eq3.57} leads to mappings which are
well-behaved in the soft and collinear limits. 

We now consider  the phase space. Since 
\begin{equation}
  \begin{split}
   \frac{\mathd^4 p_V}{(2\pi)^3} \delta_+(p_V^2-m_V^2) (2\pi)^4
   \delta^{(4)} (p_1+p_2-p_V-k) =
   \\
   \frac{\mathd^4 \tp_V}{(2\pi)^3}
   \delta_+( \tp_V^2 - m_V^2) (2\pi)^4 \delta^{(4)} (\tp_1+\tp_2-\tp_V),
  \end{split}
\end{equation}
there is no Jacobian factor in this case.  However, similar to the
initial-final case, we have to consider changes in the momenta of
the colliding partons. We interpret them as changes in Bjorken
$x_{1,2}$. The corresponding formulas read
\begin{equation}
  x_1 = \tilde x_1 + \frac{(P_2 k)}{(P_1 P_2)} + \mathcal O(k^2)
  \;\;\;\; x_2 = \tilde
  x_2 + \frac{(P_1 k)}{(P_1 P_2)} + \mathcal O(k^2).
\end{equation}
Then, similar to the initial-final case, we have to expand the parton
distribution functions in Taylor series to account for the difference
between 
$\tilde x_{1,2}$ and $x_{1,2}$.  The rest of the argument proceeds in
full analogy with final-final and initial-final cases. The
expansion of the amplitude squared leads to integrals over the gluon
momentum $k$ that are identical to the ones discussed in
Appendix~\ref{app:int}, where it is shown that they do not contain 
${\cal O}(\lambda)$ terms. This allows us to conclude that the cross section
and rapidity distribution  of a color singlet production at hadron colliders
is free of linear power correction.

\section{A first application to event-shape variables: the $C$-parameter}
\label{sec:shapevars-ex}
As we have mentioned, an interesting application of the framework
developed in the previous sections is the study of non-perturbative
corrections to $e^+e^-$ event-shape variables, for generic kinematic
configurations. In this section, we perform a semi-realistic analysis
of one of such variables, the so-called $C$-parameter. In the case of
a vector boson with momentum $q$ decaying into $N$ massless final
state particles with momenta $p_{1},...,p_{N}$, the $C$-parameter
reads
\begin{equation}
  C(\{p_1,...,p_N\};q) = 3 - 3\sum_{i>j}\frac{(p_i p_j)^2}{(p_i q)(p_j q)}.
  \label{eq:cdef}
\end{equation}
We will also use the same definition for the case of massive final-state
particles.

We are interested in computing power corrections to this observable in
a situation that approximates a three-jet configuration in $e^+ e^-$
annihilations. Since our formalism does not allow us to deal with
processes that contain gluons at leading order, we follow the same
approach as in the previous section and use photons as proxies for
hard gluons. We then consider the process
\begin{equation}
  V(q) \to q(p_1) + \bar q(p_2) + \gamma(p_3),
  \label{eq4.1}
\end{equation}
and study $\mathcal O(\Lambda_{\rm QCD}/Q)$ power corrections to the
$C$-parameter Eq.~(\ref{eq:cdef}) that may arise in this case. To this
end, we follow the approach described in the previous sections and study
${\cal O}(\alpha_s)$ corrections to the process in Eq.~(\ref{eq4.1})
in a theory where gluons are given a small mass $\lambda$.

We note that this way of computing ${\cal O}(\Lambda_{\rm QCD}/Q)$
corrections to the $C$-parameter is not fully justified since
the definition of this observable involves momenta of all particles in
the final state.  For this reason, if one pursues the standard
approach to power corrections that relates computations with large
number of massless fermions to calculations with massive gluons,
a computation of the $C$-parameter for the \emph{five-particle} final
state becomes necessary.
We discuss such a computation in Section~\ref{sec:shapevars}.  In this
section we consider a simplified setup where we neglect gluon
splitting $g^* \to q \bar q$ and consider the massive gluon as a
final state particle.  This allows us to directly apply ideas of the
previous sections to a relatively simple but non-trivial example and
to provide a connection between the general arguments of
Section~\ref{sec:gencase} and numerical calculations of event shapes in
Section~\ref{sec:shapevars}.

To compute ${\cal O}(\alpha_s)$ corrections to the process in
Eq.~(\ref{eq4.1}) we need to account for virtual and real-emission
contributions. We have argued in the previous section that virtual
corrections cannot produce linear terms in $\lambda$; for this reason
we discard them and focus only on the real-emission ones. Hence, we
consider the process
\begin{equation}
  V(q)\to q(p_1) + \bar q(p_2) + \gamma(p_3) + g(k),
  \label{eq4.3}
\end{equation}
with $k^2=\lambda^2$.
To define what needs to be computed, we consider the cumulative
distribution of the $C$-parameter
\begin{equation}
  \Sigma(c) = \int\limits_c^1 \mathd c'\frac{\mathd\sigma}{\mathd c'}.
  \label{eq:ccum}
\end{equation}
To determine the non-perturbative corrections to $\Sigma(c)$ we need
to calculate
\begin{equation}
\Sigma(c) = \int \mathd\sigma_R \,\theta\big[C(p_1,p_2,p_3,k;q)-c\big],
\end{equation}
where   ${\rm d} \sigma_R$ is the differential cross section of the process
in Eq.~(\ref{eq4.3})
\begin{equation}
  \mathd \sigma_R = N^{-1}{\rm dLips}(q;p_1,p_2,p_3,k)\,
  |\mathcal M(p_1,p_2,p_3,k)|^2.
\end{equation}

From our discussion in the previous section it follows that we only
need to study kinematic configurations where the gluon is
soft. We then  proceed as outlined in
Section~\ref{sec:gencase},  remap the momenta $p_i\to \tp_i$ and expand
the matrix element in the soft limit retaining next-to-leading
terms.
To discuss modifications of the observable, we split $C$ into two
contributions
\begin{equation}
  C(p_1,p_2,p_3,k;q) \equiv 3+C_3(p_1,p_2,p_3;q) +
    C_k(p_1,p_2,p_3,k;q),
  \label{eq:csplit}
\end{equation}
where 
\begin{equation}
  C_3(p_1,p_2,p_3;q) = -3\sum\limits_{i<j = 1}^{3} \frac{(p_i p_j)^2}{(p_i
    q)(p_j q)},
  ~~~
  C_k(p_1,p_2,p_3,k;q) = - 3\sum\limits_{i=1}^{3}\frac{(k p_i)^2}{(kq)(p_i q)}.
  \label{eq:c3ck}
\end{equation}
 Then,  it follows from Eq.~\eqref{eq:csplit}, that 
 \begin{equation}
   \begin{split} 
  & C_3(p_1,p_2,p_3;q) = C_3(\tp_1,\tp_2,\tp_3;q) + v^\mu k_\mu
  + \mathcal O(k^2),
\\
  & C_k(p_1,p_2,p_3,k;q) = C_k(\tp_1,\tp_2,\tp_3,k;q)  +   \mathcal O(k^2).
  \label{eq:cexp}
\end{split} 
   \end{equation}
where $v = v(\tp_1,\tp_2,\tp_3;q)$ is a vector whose specific
form  will not  be needed. 
Finally, to account  for changes in the $C$-parameter due to an emission of a soft gluon,  we expand the $\theta$-function
to first order in $k$ and write 
\begin{equation}
  \begin{split}
  \theta\big[C(p_1,p_2,p_3,k;q)-c\big] \approx &
  \theta\big[C(\tp_1,\tp_2,\tp_3;q)-c\big]
  \\
  &  + \delta\big(C(\tp_1,\tp_2,\tp_3;q)-c\big)\,
  \big[v^\mu k_\mu + C_k(p_1,p_2,p_3,k;q)\big].
  \end{split}
\end{equation}

We now combine the required changes in the phase space, the matrix
element and the observable and write
\begin{align}
  \Sigma(c) & \approx N^{-1}\int {\rm dLips}(q;\tp_1,\tp_2,\tp_3)
  \int [\mathd k] |\mathcal M(p_1,p_2,p_3,k)|^2\times
  \\
  &    \Big \{
  \theta\big[C(\tp_1,\tp_2,\tp_3;q)-c\big]
  J\, +
    \delta\big[C(\tp_1,\tp_2,\tp_3;q)-c\big]\left(v_\mu k^\mu 
  + C_k\right)   \Big\}.
  \nonumber
\end{align}
In the above formula
\begin{equation}
  [\mathd k] = \frac{\mathd^4 k}{(2\pi)^3}\delta_+(k^2-\lambda^2)
  \theta\big[(q-k)^2\big],
\end{equation}
and $J$ is the Jacobian of the transformation discussed in
Section~\ref{sec:gencase}.  Using arguments presented in
Section~\ref{sec:gencase} we conclude that the only potential source
of ${\cal O}(\lambda)$ corrections to $\Sigma(c)$ is the term
$C_k$. Since $C_k$ is proportional to the four-momentum of the soft
gluon, the amplitude $|\mathcal M(p_1,p_2,p_3,k)|^2$ in the relevant
terms can be taken in the leading soft approximation.  We find
\begin{equation}
  {\cal T}_\lambda \Sigma(c) = \frac{\alpha_s}{2\pi} C_F
  \int \frac{1}{N} {\rm dLips}(q;\{\tp_i\})
  \delta\big(C(\{\tp_i\};q)-c\big) |\mathcal M(\tp_1,\tp_2,\tp_3)|^2
            {\cal T}_\lambda   I_c(\{\tp_i\};q;\lambda),
            \label{eq4.13}
\end{equation}
where 
\begin{equation}
  I_c(\tp_1,\tp_2,\tp_3;q;\lambda) = 8\pi^2
  \int \frac{\mathd^4 k}{(2\pi)^3}
  \delta_+(k^2-\lambda^2)\theta\big[(q-k)^2\big]
  \frac{2(\tp_1\tp_2)}{(\tp_1 k)(\tp_2 k)}
  C_k(\tp_1,\tp_2,\tp_3,k),
  \label{eq4.14}
\end{equation}  
and the operator ${\cal T}_\lambda$ is defined to extract the ${\cal
  O}(\lambda)$ contribution from the function it acts upon.

We will now explain how the function $I_c$ can be computed.  To this
end, we use the definition of $C_k$ in Eq.~\eqref{eq:csplit} and write
\begin{equation}
  I_c(\tp_1,\tp_2,\tp_3;q;\lambda) = 3\sum\limits_{i=1}^{3}
  I_c^{(i)},
\end{equation}
with
\begin{equation}
  I_c^{(i)} = -8\pi^2\, \frac{2(\tp_1\tp_2)}{(q\tp_i)}
  \int\frac{\mathd^4 k}{(2\pi)^3}\delta_+(k^2-\lambda^2)
  \theta\big[(q-k)^2\big]
  \frac{(k\tp_i)^2}{(k\tp_1)(k\tp_2)(k q)}.
\end{equation}
It is convenient to compute  these integrals in the rest frame
of $q$. We find
\begin{equation}
  I_c^{(i)}=
  -\frac{4(\tp_1\tp_2)}{\sqrt{q^2}}
  \frac{(q\tp_i)}{(q\tp_1)(q\tp_2)}
  \int\limits_{\lambda}^{\omega_{\rm max}}
  \mathd \omega\, \beta\; W^{(i)},
  \label{eq:icint}
\end{equation}
where $\beta = \sqrt{1-\lambda^2/\omega^2}$ and $\omega_{\rm max}$ is
an  upper integration limit imposed by the condition $(q-k)^2>0$.
Since we are only interested in the linear dependence on $\lambda$, the
explicit form of $\omega_{\rm max}$ is irrelevant. Also, we have defined
\begin{equation}
  W^{(i)} = \int\frac{\mathd\Omega_k}{4\pi}
  \frac{\big(1-\beta\vec n\cdot\vec n_i\big)^2}
       {\big(1-\beta\vec n\cdot\vec n_1\big)
         \big(1-\beta\vec n\cdot\vec n_2\big)}
\end{equation}
where $\vec n_i$ and $\vec n$ are unit vectors that
define the directions of  the spatial components of the
momentum $\tp_i$ and of the gluon momentum $k$, respectively.  
The functions $W^{(i)}$  can be written as linear combinations of 
 three integrals. They are 
\begin{equation}
  \begin{split}
  & I_{12} = \int\frac{\mathd\Omega_k}{4\pi}
  \frac{1}{(1-\beta\vec n\cdot \vec n_1)(1-\beta\vec n\cdot \vec n_2)},
  \\
  & I_{1} = \int\frac{\mathd\Omega_k}{4\pi}
  \frac{1}{(1-\beta\vec n\cdot\vec n_1)} =
  \int\frac{\mathd\Omega_k}{4\pi}
  \frac{1}{(1-\beta\vec n\cdot\vec n_2)},
  \\
  & I_{0} = \int\frac{\mathd\Omega_k}{4\pi}.
  \label{eq:iint}
  \end{split}
\end{equation}
Indeed, using the momentum conservation condition
$\sum\limits_{i=1}^{3}E_i\,\vec n_i = 0$, 
one can write
\begin{equation}
  \begin{split}
    W^{(1)} & = W^{(2)} = \left[1-\frac{2(1-x_3)}{x_1 x_2}\right] I_0
    + \frac{2(1-x_3)}{x_1 x_2} I_1, \\
    W^{(3)} & = \frac{4
      I_{12}}{x_3^2} + \frac{2(x_1^2+x_2^2)(1-x_3)}{x_1 x_2
      x_3^2} (I_1-I_0)
    - \frac{4(x_1+x_2)}{x_3^2} I_1 +
      \left(\frac{x_1+x_2}{x_3}\right)^2 I_0,
  \end{split}
\end{equation}
where $x_i = 2 E_i/\sqrt{q^2} = 2(\tp_i q)/q^2$. 

To compute $I_c^{(i)}$ we require the following integrals, see
Eq.~\eqref{eq:icint}:
\begin{equation}
  \left\{ \mathcal I_0, \mathcal I_1, \mathcal I_{12}\right\} = 
  \int\limits_\lambda^{\omega_{\rm max}}\mathd\omega\,\beta\times
  \left\{
  I_0,I_1,I_{12}
  \right\}.
  \label{eq:genlin}
\end{equation}
Their calculation is described in
Appendix~\ref{app:cali}, where we show that
\begin{equation}
  \mathcal T_\lambda\; \mathcal I_0 = -\frac{\pi}{2}\lambda,
  ~~~
  \mathcal T_\lambda\; \mathcal I_1 =  0,~~~
  \mathcal T_\lambda\; \mathcal I_{12} = 0.
\end{equation}
Hence, the linear-$\lambda$ dependence of the functions $W^{(i)}$ 
reads
\begin{equation}
{\cal T}_\lambda W^{(i)} = \frac{\pi \lambda}{2} \overline W^{(i)},
\end{equation}
where
\begin{equation}
  \overline W^{(1)} = \overline W^{(2)} = \frac{2(1-x_3)}{x_1 x_2}-1,
  ~~~
  \overline W^{(3)} =
  \frac{2(x_1^2+x_2^2)(1-x_3)}{x_1 x_2
    x_3^2}
  -\left(\frac{x_1+x_2}{x_3}\right)^2.
\end{equation}
Putting everything together, we find
\begin{equation}
   {\cal T}_\lambda I_c =
           -6\pi\left(\frac{\lambda}{\sqrt{q^2}}\right)\left[
            \frac{(\tp_1\tp_2)}{(q\tp_2)}  \overline W^{(1)}
            +\frac{(\tp_1\tp_2)}{(q\tp_1)} \overline W^{(2)}
            +\frac{(\tp_1\tp_2)(q\tp_3)}{(q\tp_1)(q\tp_2)} \overline W^{(3)}
            \right].
           \label{eq:tlambdaIc}
\end{equation}
We can use this  result in Eq.(\ref{eq4.13}) to compute
the ${\cal O}(\lambda)$ correction to  $\Sigma(c)$. 
Finally, we note that it is
customary to present results for the non-perturbative corrections as a
shift with respect to the perturbative differential distribution. In
our case, this reads
\begin{equation}
  \begin{split}
    \delta_{\rm NP} &\equiv -
    \frac{\mathcal T_\lambda\Sigma(c)}{{\mathd\sigma}/{\mathd c}}
    \\& =-
    \frac{\alpha_s}{2\pi}C_F\times
    \frac{\int {\rm
        dLips}(q;\tp_1,\tp_2,\tp_3)\delta\big(C(\tp_1,\tp_2,\tp_3)-c\big)
      |\mathcal M(\tp_1,\tp_2,\tp_3)|^2 \,\mathcal T_\lambda I_c}
         {\int {\rm
             dLips}(q;\tp_1,\tp_2,\tp_3)\delta\big(C(\tp_1,\tp_2,\tp_3)-c\big)
           |\mathcal M(\tp_1,\tp_2,\tp_3)|^2}.
  \end{split}
  \label{eq:deltaC_nosplit}
\end{equation}
Eq.~\eqref{eq:deltaC_nosplit} allows one to immediately compute
$\mathcal O(\Lambda_{\rm QCD}/Q)$ corrections to the $C$-parameter for
a generic three-jet configuration.  However, as we have already said
the analysis of this section is only semi-realistic since we are
neglecting $g^*\to q\bar q$ splitting. We deal with the fully
realistic case in the next section.

\section{Event-shape variables: the general case}
\label{sec:shapevars}
In the previous section, we explained how to simplify a computation of
linear power corrections to the $C$-parameter in a generic three-jet
configuration using an improved understanding of potential sources of
${\cal O}(\lambda)$ terms.  However, the scope of the semi-analytic
computation described there is limited since we neglected the splitting of
a massive gluon into a $q \bar q$ pair.  The goal of this section is
to develop a general framework that will utilize findings reported
earlier in this paper and will allow us to compute linear power
corrections numerically for (almost) any shape variable in general
kinematic configurations in a straightforward way.

\subsection{Shape variables in the
  $\gamma^* \rightarrow d \bar{d} \gamma$ process in the large-$n_f$
  approximation}
\begin{figure}[t]
  \centering
  \begin{displaymath}
    \stackrel{\includegraphics[height=0.125\textheight]{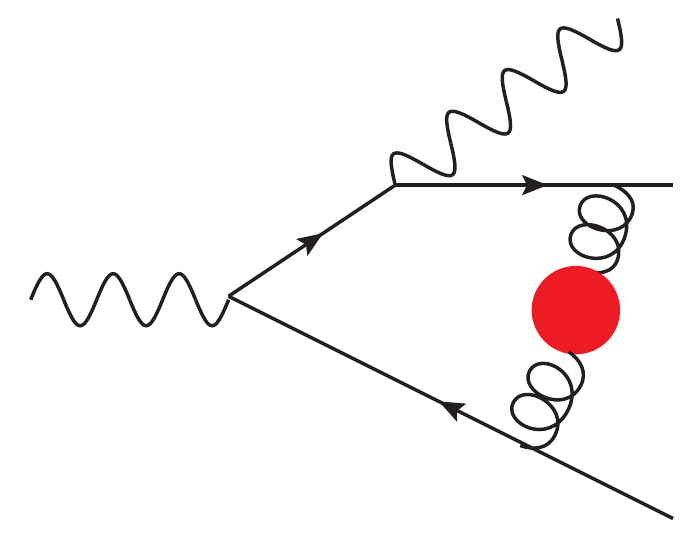} %
      \qquad\includegraphics[height=0.125\textheight]{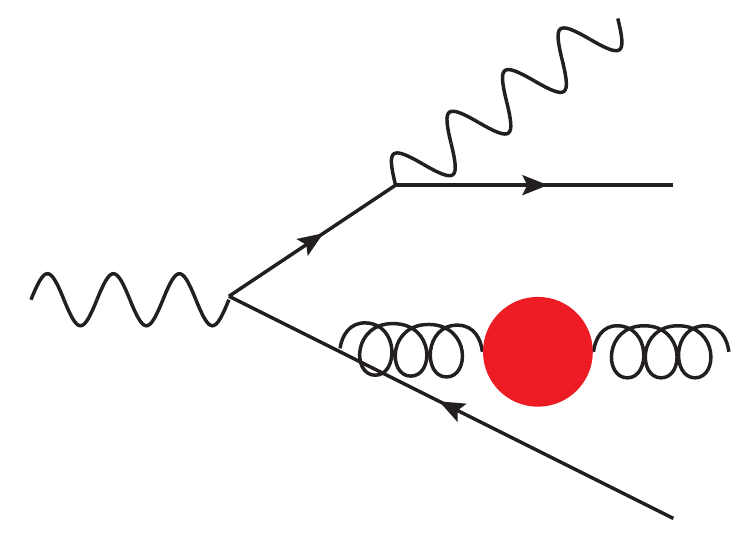} %
      \qquad\includegraphics[height=0.125\textheight]{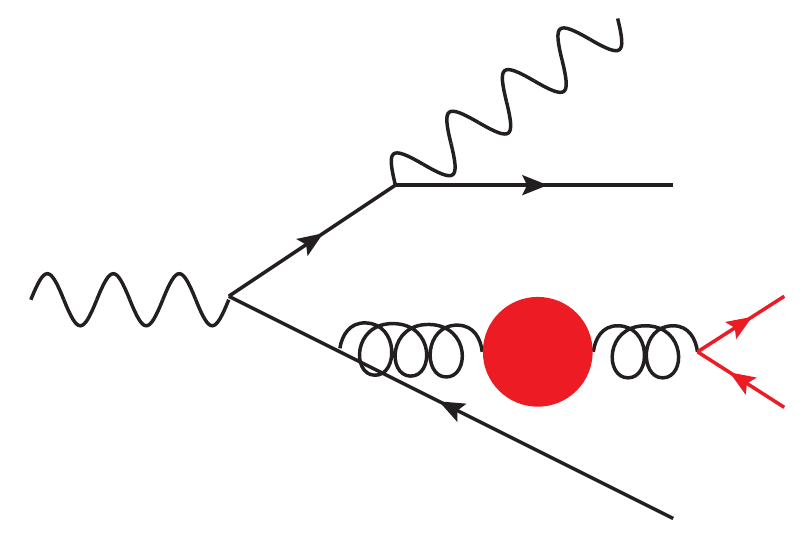}}{\phantom{zzzz}({\rm
        v}) \phantom{zzzzzzzzzzzzzzzzzzzzzz}({\rm
        g})\phantom{zzzzzzzzzzzzzzzzzzzzz}({\rm q\bar{q}})}
  \end{displaymath}
  \caption{A sample of the radiative corrections that need to be
    included in order to compute the all-order $\alpha_s(\alpha_s
    n_f)^n$ corrections to the process $\gamma^* \to d \bar{d}
    \gamma$.  }
  \label{fig:diagrams}
\end{figure}

To explain our approach, we consider the process $\gamma^* \rightarrow
d \bar{d} \gamma$.  We assume that only $d$-quarks couple to photons
and all other $n_f$ quarks only couple to gluons.  In the large-$n_f$
limit, the dominant corrections arise from the emission of virtual or
real gluons dressed with fermion bubbles~\cite{Beneke:1998ui}, see
Fig.~\ref{fig:diagrams}.
The solid blob in the gluon propagator in that figure implies that
fermion loops have been accounted for to all orders; its exact
definition follows from the recursion relation
\begin{equation}
\raisebox{-0.5cm}{\includegraphics[height=0.05\textheight]{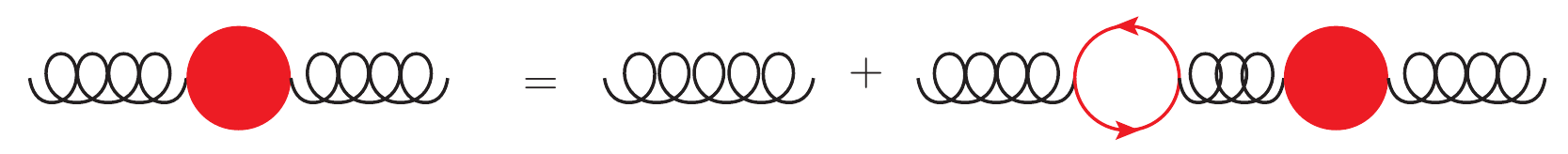}}.
\label{eq:dressed_gluon}
\end{equation}

As we already mentioned in the introduction, the large-$\nf$
prediction for an observable ${O}$ can be obtained by computing
the NLO QCD corrections to its expectation value; such computation,
however, should be performed with a massive gluon. Before discussing how
the findings of the previous sections allow us to easily obtain 
predictions for a wide class of observables, we introduce
some notation. 
We
denote the expectation value of $O$ with
$\langle O\rangle$ and indicate with a superscript the
perturbative order at which $\langle O\rangle$ is computed.
For example, 
$\langle O\rangle^{(0)}$ represents the Born-level result,
$\langle O\rangle^{(1)}$ represents the $\mathcal O(\as)$ correction
and so on. Finally, a subscript on $O$ indicates the number of
final-state particles that are used for the calculation of the observable. 
To give a concrete example, we consider the cumulative distribution
of the $C$-parameter, see Eq.~\eqref{eq:ccum}.
In this notation, the Born-level result is denoted by
\begin{equation}
  \begin{split}
    \langle O \rangle^{(0)} = \mathcal N
    \int{\rm dLips}(q;p_1,p_2,p_3) |\mathcal M (p_1,p_2,p_3)|^2
    O_3,
  \end{split}
  \label{eq:5.2}
\end{equation}
where $O_3 = \theta\big[C(p_1,p_2,p_3;q)-c\big]$ 
and $\mathcal N$ is a normalization factor that we will specify shortly.
The calculation of $\mathcal O(\as)$ corrections due to a gluon
with mass $\lambda$ is instead denoted by
\begin{equation}
  \begin{split}
    \langle O\rangle^{(1)}_{g^*} & =
    \mathcal N\int{\rm dLips}(q;p_1,p_2,p_3)\, 2\Re\left[
      \mathcal M (p_1,p_2,p_3) \mathcal M^*_{\rm1-loop}(p_1,p_2,p_3,\lambda)
      \right] O_3 
    \\
    &+
    \mathcal N\int{\rm dLips}(q;p_1,p_2,p_3,k) |
    \mathcal M (p_1,p_2,p_3,k)|^2 O_{3+1},
  \end{split}
\end{equation}
with $O_3$ defined after Eq.~\eqref{eq:5.2} and
$O_{3+1} = \theta\big[C(p_1,p_2,p_3,k;q)-c\big]$. In what follows, we
choose to normalize our results to the LO rate for the process
$\gamma^*\to d\bar d$, i.e.\footnote{As shown earlier, radiative corrections
  to the total cross section do not lead to linear terms.}
\begin{equation}
  \mathcal N^{-1} = \int {\rm dLips}(q;p_1,p_2) |\mathcal M(p_1,p_2)|^2
  \equiv \sigma_0.
\end{equation}
Finally, we introduce the short-hand notation
\begin{equation}
  \mathd\Phi_3 = {\rm dLips}(q;p_1,p_2,p_3),
  ~~~
  \mathd\Phi_{3+1} = {\rm dLips}(q;p_1,p_2,p_3,k),
\end{equation}
where $k$ denotes the massive gluon with $k^2=\lambda^2$ and
\begin{equation}
  \begin{split}
    & B(\Phi_3) = |\mathcal M(p_1,p_2,p_3)|^2,
    \\
    & V^{(\lambda)}(\Phi_3) = 
    2\Re\left[
      \mathcal M (p_1,p_2,p_3) \mathcal M^*_{\rm1-loop}(p_1,p_2,p_3,\lambda)
      \right],
    \\
    & R_{g^*}^{(\lambda)}(\Phi_{3+1})  = 
    |\mathcal M(p_1,p_2,p_3,k)|^2.
  \end{split}
  \label{eq:BRVdef}
\end{equation}

The connection between the massive-gluon calculation and non-perturbative
corrections is well-known. One can write~\cite{Beneke:1998ui}
\begin{equation}
\label{eq:xsec1}
\langle O \rangle=\langle O \rangle^{(0)}
+\frac{\kappa Q}{2\as\beta_0} \left[\frac{\mathd \langle
  O\rangle^{(1)}_{g^*}}{\mathd\lambda}\right]_{\lambda=0}
\left(\frac{\Lambda_{\rm QCD}}{Q}\right) + \dots
\end{equation}
where $\as = \as(\mu)$, $\beta_0$ is the first coefficient of the
$\beta$-function and $\kappa$ is an overall normalization which
depends on the resummation prescription.
Ellipses in Eq.~\eqref{eq:xsec1} stand for higher-order corrections,
both perturbative and non-perturbative. A linear power correction is
present in Eq.~(\ref{eq:xsec1}) provided that the derivative of
$\langle O \rangle^{(1)}_\lambda$ with respect to $\lambda$ does not
vanish for $\lambda\to0$. In what follows, we will discuss
non-perturbative corrections to various quantities. However,
unless stated otherwise we
will always present results for $\lambda\,\mathd\langle
O\rangle^{(1)}_{g^*}/\mathd\lambda$, without multiplying them by the
other coefficients in Eq.~\eqref{eq:xsec1}.

So far, we have only considered final states with a massive
gluon. However, if the definition of the observable ${O}$ is
sensitive to the presence of final-state quarks, then contributions
where a massive gluon splits into a $q \bar q$ pair have to be
accounted for. A detailed discussion of how to do this can be found
e.g. in Refs.~\cite{FerrarioRavasio:2018ubr,FerrarioRavasio:2020guj},
which use the same notation that we have just introduced. Here, we
limit ourselves to quote the final result. To account for $g^*\to
q\bar q$ splitting, one has to replace
\begin{equation}
  \langle O
  \rangle^{(1)}_{g^*} 
  \longrightarrow
  \langle O
  \rangle^{(1)}_\lambda = 
  T_{\rm{V}}(\lambda)+T_{\rm{R}}(\lambda)+T_{\rm{R}}^{\Delta}(\lambda),
  \label{eq:Tlambda}
\end{equation}
in Eq.~\eqref{eq:xsec1}, 
with~\cite{FerrarioRavasio:2018ubr,FerrarioRavasio:2020guj}
\begin{eqnarray}
T_{\rm{V}}(\lambda)&=& \mathcal N \int\mathd\Phi_3
V^{(\lambda)}(\Phi_3)O_3,
 \label{eq:TVlambda}
\\ T_{\rm{R}}(\lambda)&=& \mathcal N \int\mathd\Phi_{3+1}
R^{(\lambda)}_{g^*}(\Phi_{3+1})O_{3+1},
 \label{eq:TRlambda}
\\ T_{\rm{R}}^{\Delta}(\lambda)&=&\mathcal N \frac{3\pi}{\as
  \Tf}\lambda^2\int\mathd\Phi_{3+2}\delta\left(\lambda^2-(l_1+l_2)^2\right)R_{q\bar{q}}(\Phi_{3+2})[O_{3+2}-O_{3+(2)}].
\label{eq:Tqqlambda}
\end{eqnarray}
In Eq.~\eqref{eq:Tqqlambda}, $\mathd\Phi_{3+2} = {\rm
  dLips}(q;p_1,p_2,p_3,l_1,l_2)$ is the phase space for the process
\begin{equation}
  \gamma^*(q) \to 
  d(p_1)+\bar d(p_2)+\gamma(p_3) + q(l_1)+\bar q(l_2),
\end{equation}
and $R_{q\bar q}$ is the corresponding matrix element squared
\begin{equation}
R_{q\bar q}(\Phi_{3+2}) = |\mathcal M(p_1,p_2,p_3,l_1,l_2)|^2.
\end{equation}
Also, $O_{3+2}$ is the observable computed with the momenta of the
$d\bar d \gamma
q\bar q$ final state, while $O_{3+(2)}$ is the observable computed
using the momenta $p_1,p_2,p_3$ and $k=l_1+l_2$. In our example,
$O_{3+2} = \theta\big[C(p_1,p_2,p_3,l_1,l_2)-c\big]$ and
$O_{3+(2)} = \theta\big[C(p_1,p_2,p_3,l_1+l_2)-c\big]$.
As we have said, in general $O$ can be any (infrared safe) function of
the final-state kinematics. The only requirement that we impose at this
stage is that it should vanish in the two-jet limit where the three-jet
calculation diverges. 

We note that Eqs.~(\ref{eq:TVlambda}) and~(\ref{eq:TRlambda}) exhibit
logarithmic singularities in the $\lambda \to 0$ limit that, however,
cancel in the sum.  Since Eq.~(\ref{eq:Tlambda}) should be finite in
that limit, Eq.~(\ref{eq:Tqqlambda}) should be finite as well;
technically, this happens because the expression in the square bracket
in that equation vanishes in the soft limit.
Also, we point out that for observables {\it inclusive} with respect
to $g^* \to q\bar{q}$ splittings, the contribution shown in
Eq.~(\ref{eq:Tqqlambda}) would vanish and in this case the computation
with massive gluons and the large-$n_f$ calculation would be
identical.  However, since for observables that we are interested in
final states with a massive gluon never appear, the contribution shown
in Eq.~(\ref{eq:Tqqlambda}) can be seen as a required correction to
the calculation with a massive gluon, where the difference in the
observable computed with the $q{\bar q}$ pair and with the massive
gluon is added. In fact, as we will see in more detail below, the term
proportional to $O_{3+(2)}$ in Eq.~(\ref{eq:Tqqlambda}) and the one
proportional to $O_{3+1}$, Eq.~(\ref{eq:TRlambda}), cancel each other
exactly.

A careful reader could have noticed that there are unregulated soft
and collinear divergences in contributions that we account for even if
we make sure to stay in the 3-jet region by, e.g.  imposing a cut on
the $C$-parameter. These singularities arise when the final state
photon is collinear to one of the primary quarks or when it is soft
and, instead, the radiated gluon (eventually decaying into the
$q\bar{q}$ pair) is hard and at large angle. It should be clear from
the results of the previous sections that kinematic regions with a
hard gluon cannot produce linear power corrections and we ignore them
here. A more extensive discussion of how we treat these contributions
is given in the next subsection and in Appendix~\ref{app:largenfcalc},
where we also provide more details about the full large-$\nf$ calculation
including its connection to the all-orders perturbative expansion
shown in Fig.~\ref{fig:diagrams}.

\subsection{Simplified computation of the linear term}
The results presented in the previous subsection provide a simple
recipe for studying non-perturbative corrections to event shapes.
Indeed, one has to perform a calculation
with a massive gluon, which eventually splits into a $q\bar q$ pair,
and then extrapolate the result of
such computation to small values of $\lambda$. To do so, one requires
the full matrix element for the $\gamma^* \to d\bar d \gamma (q \bar
q)$ process as well as virtual corrections to the $\gamma^* \to d\bar d
\gamma$ process. For this reason, such a calculation is as complicated as any
computation with a multiparticle final state can be.

In this subsection we will explain how this procedure can be
dramatically simplified provided that the goal is to determine ${\cal
  O}(\lambda)$ terms only.  Indeed, we will show
that in order to determine linear power correction to a generic
infrared-safe observable all that needs to be known is the amplitude
 of the Born process $ \gamma^* \to d \bar d \gamma$, the
eikonal current for the emission of an off-shell soft gluon and
the matrix element for its splitting into a quark-antiquark pair.

To prove this statement, we proceed as follows. First we notice that
the 5-body phase space for the final state $d\bar{d}\gamma q{\bar q}$
can be factorized into a product of a 4-body phase space for the
production of a virtual gluon together with a $d \bar d \gamma$ final
state and a 2-body phase space that describes the decay of this gluon
into a $q \bar q$ pair. We write
\begin{equation}
  \begin{split}
    \mathd\Phi_{3+2}\delta(\lambda^2-(l_1+l_2)^2)&= {\rm
      dLips}(q;p_1,p_2,p_3,l_1,l_2)
    \delta\big(\lambda^2-(l_1+l_2)^2\big)  \\ &=
    \left.\frac{1}{2\pi} {\rm dLips}(q;p_1,p_2,p_3,k)\, \label{eq:dlips4}
    \mathd \Phi_{\tmop{split}}\right|_{k^2=\lambda^2},
  \end{split}
\end{equation}
with
\begin{equation}
  \label{eq:phiqqdef} \mathd
  \Phi_{\tmop{split}}={\rm dLips}(k;l_1,l_2)\,.
\end{equation}
For ease of notation we did not indicate that $\mathd
\Phi_{\tmop{split}}$ depends on $l_1$ and $l_2$. 
Furthermore, as we discussed in the preceding sections, the 4-body
phase space for the $d\bar{d}\gamma g^*$ final state  can be factorized
into a 3-body phase space for $d\bar{d}\gamma$ (the
underlying Born configuration) and a radiation phase space for the gluon
\begin{equation}
  {\rm dLips}(q;p_1,p_2,p_3,k) = {\rm
    dLips}(q;\tilde{p}_1,\tilde{p}_2,\tilde{p}_3) \,\mathd
  \Phi_{\tmop{rad}} \label{eq:phigstardef}.
\end{equation}
Again for ease of notation we do not show the dependence of $ \mathd
\Phi_{\tmop{rad}}$ on $\tilde{p}_{1\ldots 3}$ and $k$.  As explained
in the previous sections, this factorization is performed by
expressing the momenta of the 4-body phase space as a function of the
underlying Born four-momenta $\tilde{p}_{1\ldots 3}$ and the gluon
momentum $k$, see Eq.~\eqref{eq3.11}.
For convenience, we define $ \mathd \Phi_{\tmop{rad}}$
to include also the Jacobian of this momenta transformation. Finally,
we identify
\begin{equation}
  {\rm dLips}(q;\tilde{p}_1,\tilde{p}_2,\tilde{p}_3)= \mathd \Phi_3.
\end{equation}

Using the notation introduced above, we now show that
one can write
$\langle O\rangle_\lambda^{(1)}$ in 
Eq.~(\ref{eq:Tlambda}) as follows
\begin{equation}
  \langle O \rangle_{\lambda}^{(1)} = \frac{1}{\sigma_0} \int \mathd
  \Phi_3 \Bigg\{V_{\lambda} O_3 + \int \mathd \Phi_{\tmop{rad}} M_{\mu
    \nu} (k, \lambda)\int \mathd \Phi_{\tmop{split}} P^{\mu
    \nu}_{\tmop{split}} O_{3+2} \Bigg\}.  \label{eq:Shape1}
\end{equation}
In this equation, $M^{\mu\nu}$ is the amplitude squared for the
production of the $d\bar{d} \gamma g^*$ final state stripped of the
polarization vectors of the virtual gluon $g^*$.  Thus
\begin{equation}
  \sum_{\lambda} M^{\mu\nu} \epsilon^{*,\lambda}_{\mu}
  \epsilon^{\lambda}_{\nu} = -M^{\mu\nu}
  g_{\mu\nu}=R^{(\lambda)}_{g^*}(\Phi_{3+1}),
\end{equation}
where $R^{(\lambda)}_{g^*}$ is defined in Eq.~(\ref{eq:BRVdef}).  
The $P^{\mu \nu}_{\tmop{split}}$ factor in Eq.~(\ref{eq:Shape1}) 
is proportional to the matrix element squared for the decay of a virtual
gluon with mass $\lambda$ into a $q \bar q$ pair with momenta $l_1$ and $l_2$.
More precisely, we define it to be 
\begin{equation}
  P^{\mu\nu}_{\tmop{split}} = \frac{6\pi}{\lambda^2} {\rm
    tr}(\slashed{l}_1 \gamma^\mu \slashed{l}_{2} \gamma^\nu),
\end{equation}
so that the following equation holds
\begin{equation}\label{eq:psplitnorm}
  \int \mathd \Phi_{\tmop{split}} P^{\mu \nu}_{\tmop{split}} = -g^{\mu
    \nu}+\frac{k^\mu k^\nu}{\lambda^2}.
\end{equation}
Since, on the other hand, 
\begin{equation}
  \frac{ 4\pi \as\Tf}{\lambda^4} M_{\mu\nu} {\rm tr}(\slashed{l}_1 \gamma^\mu \slashed{l}_{2} \gamma^\nu)  = R^{(\lambda)}_{q{\bar q}},
\end{equation}
it also follows that 
\begin{equation}
\frac{3\pi \lambda^2}{\as \Tf} R^{(\lambda)}_{q{\bar q}}=  (2\pi) M_{\mu\nu} \, P^{\mu\nu}_{\rm split}\,.
\end{equation}
This equation, combined with the normalization condition for
$P^{\mu\nu}_{\tmop{split}}$ in Eq.~\eqref{eq:psplitnorm}, makes it clear
that the terms proportional to $O_{3+1}$ and $O_{3+(2)}$ in
Eqs.~(\ref{eq:TRlambda}) and (\ref{eq:Tqqlambda}) cancel out and
disappear from Eq.~(\ref{eq:Shape1}).

To proceed further, we rewrite Eq.~(\ref{eq:Shape1}) as
\begin{eqnarray}
  \langle O \rangle_{\lambda}^{(1)} & = & \frac{1}{\sigma_0} \int
  \mathd \Phi_3 \Bigg\{ \int \mathd \Phi_{\tmop{rad}} M_{\mu \nu} (k,
  \lambda) \bigg[\int \mathd \Phi_{\tmop{split}} P^{\mu
      \nu}_{\tmop{split}} O_{3+2} + O_3\, g^{\mu \nu} \bigg] \Bigg\}
  \nonumber\\ & + & \frac{1}{\sigma_0} \int \mathd \Phi_3 \left\{ \int
  \mathd \Phi_{\tmop{rad}} M_{\mu \nu} (k, \lambda)(- g^{\mu \nu}) +
  V_{\lambda} \right\} O_3 .
  \label{eq:shapevar2}
\end{eqnarray}
It is clear from the results of the previous sections that no linear
power corrections can arise from the second line of the above
equation, that includes virtual corrections and real emission
contribution integrated over the radiation phase space.  We therefore can
write
\begin{equation}
  {\cal T}_\lambda \langle O \rangle_{\lambda}^{(1)} = {\cal T}_\lambda \; 
 \sigma_0^{-1}  \int \mathd \Phi_3
  \Bigg\{ \int \mathd \Phi_{\tmop{rad}} M_{\mu \nu} (k, \lambda) \bigg[ \int
  \mathd \Phi_{\tmop{split}} P^{\mu \nu}_{\tmop{split}}  O_{3+2}
  + O_3\, g^{\mu \nu} \bigg]  \Bigg\},
  \label{eq:shapevar3}
\end{equation}
where the operator ${\cal T}_\lambda$, introduced in the previous section,
extracts ${\cal O}(\lambda)$ terms from the expression that it acts
upon.  We note that the second line in Eq.~(\ref{eq:shapevar2}) has a
finite $\lambda \to 0$ limit since virtual and integrated real
corrections are combined there. Hence, since the complete result is
infrared finite, also the first line in Eq.~(\ref{eq:shapevar2})
should have a finite $\lambda\to0$ limit. This implies that the quantity
upon which $\mathcal T_\lambda$ acts in Eq.~\eqref{eq:shapevar3} starts
at $\mathcal O(\lambda^0)$ and contains higher-order terms in the
$\lambda$-expansion. 

In principle, Eq.~(\ref{eq:shapevar3}) already provides a method for
computing linear power correction to shape variables that is much
simpler than the full large-$\nf$ calculation. However,
it can be simplified
even further.  Indeed, an important simplification arises if we observe
that in Eq.~(\ref{eq:shapevar3}) the term in square brackets vanishes
when $k$ becomes collinear to the primary quarks, as long as the shape
variable is infrared and collinear safe.  In fact, in this limit
$O_{3+2}$ becomes equal to $O_3$, and the integral of $P^{\mu\nu}_{\rm
  split}$ becomes equal to $-g^{\mu\nu}$.  It seems reasonable to
assume that in the hard collinear limit the left-over of the
collinear cancellation does not yield terms linear in $\lambda$. This
is easy to see in the thrust case, where a hard collinear splitting
changes the momentum of the splitting parton by an amount that is
proportional to the square of the splitting angle, and the sum of the
projections of the momenta of the pair onto the thrust axis is equal to
the projection of the total. We should, however, worry that this may
not be the case for all shape variables.  Indeed, a generic shape
variable may involve terms
\begin{equation}
  \int \frac{\mathd^2\vec k_\perp  }{\vec k_\perp^2+\lambda^2} |\vec k_\perp
  |f(\varphi),
\end{equation}
where $\vec k_\perp$ is the transverse momentum of the splitting, $\varphi$ is
its azimuthal angle, and $f(\varphi)$ is a function that does not vanish
upon azimuthal integration. In this case, hard collinear region may
produce ${\cal O}(\lambda)$ terms.  In what follows, we assume that
shape variables that we consider do not give rise to such terms, i.e.
that if any term linear in the absolute value of the transverse
momentum does arise, it vanishes upon azimuthal integration.

Given the above clarifications about admissible shape variables, we
conclude that for them linear power corrections can only arise from
the emission of soft massive gluons. However, as we pointed out
already, the expression in the square bracket in
Eq.~(\ref{eq:shapevar3}) vanishes in the soft limit so that the full
integral does not yield ${\cal O}(\ln \lambda )$ terms.  It
follows then that linear terms can only arise from the leading
soft-singular part of $M^{\mu\nu}$. Therefore we can substitute 
\begin{equation}
  M_{\mu\nu}(k,\lambda) \to B(\Phi_3) P_{\mu\nu}^{\rm soft}(\Phi_{3+1}),
\end{equation}
where $B$ is the Born matrix element, see Eq.~\eqref{eq:BRVdef}, and $P^{\mu
  \nu}_{\tmop{soft}} (\Phi_{3+1})$ is the soft factor that arises from
the product of eikonal currents that describe emission of a soft {\it
  massive} gluon in the above process.
Eq.~(\ref{eq:shapevar3}) can then be rewritten as
\begin{equation}
  \begin{split}
    {\cal T}_\lambda \langle O \rangle_{\lambda}^{(1)} =
    &
    {\cal T}_\lambda\,
    \sigma_0^{-1}
    \int \mathd
    \Phi_3\,B(\Phi_3)\times
    \\
    &
    \int \mathd \Phi_{\rm rad}\,  P_{\mu
      \nu}^{(\tmop{soft})} (\Phi_{3+1}) \Bigg[ \int \mathd
      \Phi_{\tmop{split}} P^{\mu \nu}_{\tmop{split}} O_{3+2} +g^{\mu
        \nu}\, O_3 \Bigg].
    \label{eq:shapeVarsFin}
  \end{split}
\end{equation}
Since the term in the square bracket vanishes in the soft limit, in
principle it is not necessary to use an exact phase space to compute
${\cal O}(\lambda)$ terms in Eq.~(\ref{eq:shapeVarsFin}).
However, for a numerical computation it is often convenient to
integrate over the exact phase space and in this case additional
issues arise since the soft factor $P^{\mu \nu}_{\tmop{soft}}
(\Phi_{3+1})$ may develop unwanted singularities as we now explain.
Indeed, $P^{\mu \nu}_{\tmop{soft}} (\Phi_{3+1})$ can be expressed in
terms of the original momenta
\begin{equation} \label{eq:softapprox}
  P^{\mu \nu}_{\tmop{soft}} (k)= 4 g_s^2 \Cf
  \left(\frac{p_1^\mu}{(p_1+k)^2}-\frac{p_2^\mu}{(p_2+ k)^2}\right)
  \left(\frac{p_1^\nu}{(p_1+k)^2}-\frac{p_2^\nu}{(p_2+ k)^2}\right)\,,
\end{equation}
or the momenta after the mapping
\begin{equation} \label{eq:approxeoik}
  \tilde{P}^{\mu \nu}_{\tmop{soft}} (k)=4g_s^2 \Cf\left\{
  \left(\frac{\tilde{p}_1^\mu}{(\tilde{p}_1
    k)}-\frac{\tilde{p}_2^\mu}{(\tilde{p}_2 k)}\right)
  \left(\frac{\tilde{p}_1^\nu}{(\tilde{p}_1
    k)}-\frac{\tilde{p}_2^\nu}{(\tilde{p}_2 k)}\right)\right\}.
\end{equation}
If the integration over $k$ is restricted to soft momenta, the two
equations are equivalent. However, if one uses the soft approximation
outside its range of validity, spurious divergences may appear. We now
discuss two examples of this, and how we deal with it.
Consider the situation where $p_1$ becomes soft
so that in the rest frame of $\tp_1 + \tp_2$ (i.e. of $p_1+p_2+k)$
the gluon recoils against $p_2$ and becomes collinear to
$\tilde{p}_1$. Although this is \emph{not} a singular configuration
of the full process, Eq.~(\ref{eq:approxeoik}) develops a collinear $\tp_1||k$
divergence, even if the original $p_1$ and $k$ are not collinear to each other.
This singularity in Eq.~\eqref{eq:approxeoik} is spurious, and it would
be removed by terms in the momentum mapping beyond the soft approximation
that we are neglecting.
To remedy this situation, it is sufficient to restrict the
integration over the radiation phase space, to exclude regions where $p_1$
or $p_2$ are soft. We do this by inserting a
$\theta$-function into $\mathd \Phi_{\rm rad}$ in
Eq.~(\ref{eq:shapeVarsFin})
\begin{equation}
  \mathd\Phi_{\rm rad} \to \mathd\Phi_{\rm rad} \;
  \theta\left(\eta-\frac{(\tilde{p}_1 k)+(\tilde{p}_2 k)}
             {(\tilde{p}_1 \tilde{p}_2)}\right),
\end{equation}
with $0<\eta<1$.\footnote{In the numerical implementation we use
  $\eta=1/2$.}
We will not show this $\theta$-function in what follows but it is
always assumed to be present in $\mathd\Phi_{\rm rad}$. 

Similarly, care is needed to deal with kinematic regions where emitted
photon is either soft or collinear to one of the primary quarks, but
the gluon is hard.  This region also contributes to shape
variables in the three-jet regions, and its contribution is divergent.
However, since
in this case the gluon must be hard, no linear terms in $\lambda$ can arise
in this case.\footnote{Including electromagnetic virtual corrections the
  divergence would cancel. But again these would involve a hard gluon,
  and thus would not lead to any linear term.} We thus suppress this
region multiplying the amplitude by the factor
\begin{equation}
  \frac{1}{(\tilde{p}_1+k)^2(\tilde{p}_2+k)^2} \times \left[
    \frac{1}{(\tilde{p}_1+\tilde{p}_3)^2(\tilde{p}_2+\tilde{p}_3)^2}+\frac{1}{(\tilde{p}_1+k)^2(\tilde{p}_2+k)^2} \right]^{-1},
\end{equation}
that dampens the photon-(anti)quark collinear singularity and
approaches one if the gluon is unresolved,
so that it does not affect ${\cal O}(\lambda)$ terms.

Finally, since the integration over $k$ in Eq.~(\ref{eq:shapeVarsFin})
is not restricted to the soft region, there are, in principle, terms
associated with hard gluons that contribute at ${\cal
  O}(\lambda^0)$. To remove them, we write 
\begin{eqnarray}
  && {\cal T}_\lambda \langle O \rangle^{(1)}_{\lambda} =
  {\cal T}_\lambda
  \sigma_0^{-1} \; \int \mathd \Phi_3 B(\Phi_3) \Bigg\{
  \Bigg[
  \int
  \mathd
  \Phi_{\rm rad} P_{\mu\nu}^{(\tmop{soft})}
  \left[
   \int \mathd \Phi_{\tmop{split}} P^{\mu\nu}_{\tmop{split}}
    O_{3+2}+ g^{\mu\nu}O_3 \right]\Bigg]\nonumber 
  \\
  &&\quad\quad
  -  \Bigg[
  \int
  \mathd
  \Phi_{\rm rad} P_{\mu\nu}^{(\tmop{soft})}
  \left[
   \int \mathd \Phi_{\tmop{split}} P^{\mu\nu}_{\tmop{split}}
   O_{3+2}+ g^{\mu\nu}O_3 \right]\Bigg]^{\lambda=0}\Bigg\}.
\end{eqnarray}
We now note that, since the observable $O$ is infrared safe,
then $O_{3+2}\to O_{3+1}$ if $\lambda\to 0$. This allows us to rewrite
this equation as
\begin{eqnarray}
    &&{\cal T}_\lambda \langle O \rangle^{(1)}_{\lambda} =
    {\cal T}_\lambda
    \sigma_0^{-1} \; \int \mathd \Phi_3 B(\Phi_3)\times
    \Bigg\{
    \Bigg[
      \int
      \mathd
      \Phi_{\rm rad} P_{\mu\nu}^{(\tmop{soft})}
      \left[
        \int \mathd \Phi_{\tmop{split}} P^{\mu\nu}_{\tmop{split}}
        O_{3+2}+ g^{\mu\nu}O_3 \right]\Bigg]
    \nonumber
    \\
    &&\quad\quad
    - \bigg[ \int \mathd \Phi_{\rm rad}
      P_{\mu\nu}^{(\tmop{soft})}(-g^{\mu\nu}) (O_{3+1} - O_3)
      \bigg]^{\lambda=0} \Bigg\} .\phantom{aaaaaaa}
    \label{eq:numerics}
\end{eqnarray}
Eq.~\eqref{eq:numerics} is our final result for the calculation of
linear power corrections to shape variables. Compared to a full
large-$\nf$ calculation, the formula in Eq.~\eqref{eq:numerics} is
remarkably simple. Indeed, it only requires the knowledge of the
matrix element of the Born process and eikonal factors that describe
the soft emission of a massive gluon and its splitting into a $q \bar
q$ pair.

We now proceed with the discussion of how to implement
Eq.~(\ref{eq:numerics}) in a numerical program.
First, we note that 
the integration over radiation variables can be suitably arranged so that
the cancellation among the two terms in the curly brackets occurs 
locally  enhancing the efficiency of the numerical integration.
To perform such an integration, we generate random phase-space points
in the $\mathd \Phi_3\,\mathd\Phi_{\rm rad}$ phase space. We then
compute the weight associated with the phase-space Jacobians and the
product of the Born amplitude squared and the contracted soft factor $P^{({\rm
    soft})}_{\mu\nu}(-g^{\mu\nu})$.  Given the $d\bar{d}\gamma g^*$
kinematic configuration, the $q\bar{q}$ splitting kinematics is
instead generated by a hit-and-miss technique, exploiting the
normalization of the $P^{\mu\nu}_{\rm split}$ factor. The integration weight
is then used to fill the histograms of the shape
variables, computed both for the $3+2$ and for the underlying Born
phase space. 

Although it is obvious that ${\cal O}(\lambda)$ power corrections
cannot depend on the phase-space mapping, such independence provides a
non-trivial check on the implementation of the numerical computation.
Hence, we have used the following mappings in our computer program:
\begin{enumerate}[label=(\roman*)]
\item a mapping that preserves the direction of $p_1$,
  i.e. such that $\tilde{p}_1\propto p_1$. This
    corresponds to the general mapping discussed in Section~\ref{sec:ffdip}
    with $\alpha=0$;
\item \label{it:diffMap} a mapping that preserves the direction of the
  difference $\vec{p}_2-\vec{p}_1 \propto
  \vec{\tilde{p}}_2-\vec{\tilde{p}}_1$ in the dipole rest frame.
  This corresponds to the general mapping discussed in Section~\ref{sec:ffdip}
  with $\alpha=1/2$;
\item a mapping that preserves the thrust direction of the dipole
  system in the dipole rest frame. In fact, this mapping is not linear
  in $k$ for small $k$.  However, the non-linear term cancels after
  the azimuthal integration over $k$, and thus also this mapping is
  acceptable.
\end{enumerate}
The above mappings are all of dipole-local type as defined in
Section~\ref{sec:gencase}.  Besides these mappings, we have also
considered the so-called global mapping of
Ref.~\cite{Dasgupta:2020fwr}. All these mappings can be expanded
linearly in $k$ for small $k$, and we have checked that all of them
give compatible results when used for the computation of ${\cal
  O}(\lambda)$ terms, as expected.

As a further check, we compare the two alternative formulae for the
soft eikonal factors Eqs.~(\ref{eq:softapprox},\ref{eq:approxeoik})
and found no significant differences. 

\subsection{Comparison with the result of Section~\ref{sec:shapevars-ex}}
As a first non-trivial check of
the numerical approach based on Eq.~\eqref{eq:numerics},
we compute the $C$-parameter distribution
neglecting the splitting of the virtual gluon into a $q\bar{q}$ pair
and compare it with the result of Section~\ref{sec:shapevars-ex}.
To this end, we need to remove the $g^*\to q\bar q$ splitting
from Eq.~\eqref{eq:numerics}.
This can be done by simply replacing $O_{3+2}$ with
$O_{3+(2)}$ there, which corresponds
to the computation of the $C$-parameter using momenta $p_1,p_2,p_3,k$,
where $k = l_1 + l_2$, instead of $p_1,p_2,p_3,l_1,l_2$.  As in
Section~\ref{sec:shapevars-ex}, we adopt Eq.~(\ref{eq:cdef}) for the
definition of the $C$-parameter in the massive case.  We then compute
$\delta_{\rm NP}$ defined in Eq.~\eqref{eq:deltaC_nosplit} both from a
direct numerical integration of Eq.~\eqref{eq:deltaC_nosplit} and from
a general-purpose numerical code based on Eq.~\eqref{eq:numerics}.
In what follows, we refer to the approach
based on Eq.~\eqref{eq:deltaC_nosplit}
as semi-analytic, and to the one based on Eq.~\eqref{eq:numerics}
as numerical.

\begin{figure}[t]
  \centering
  \includegraphics[width=0.5\textwidth]{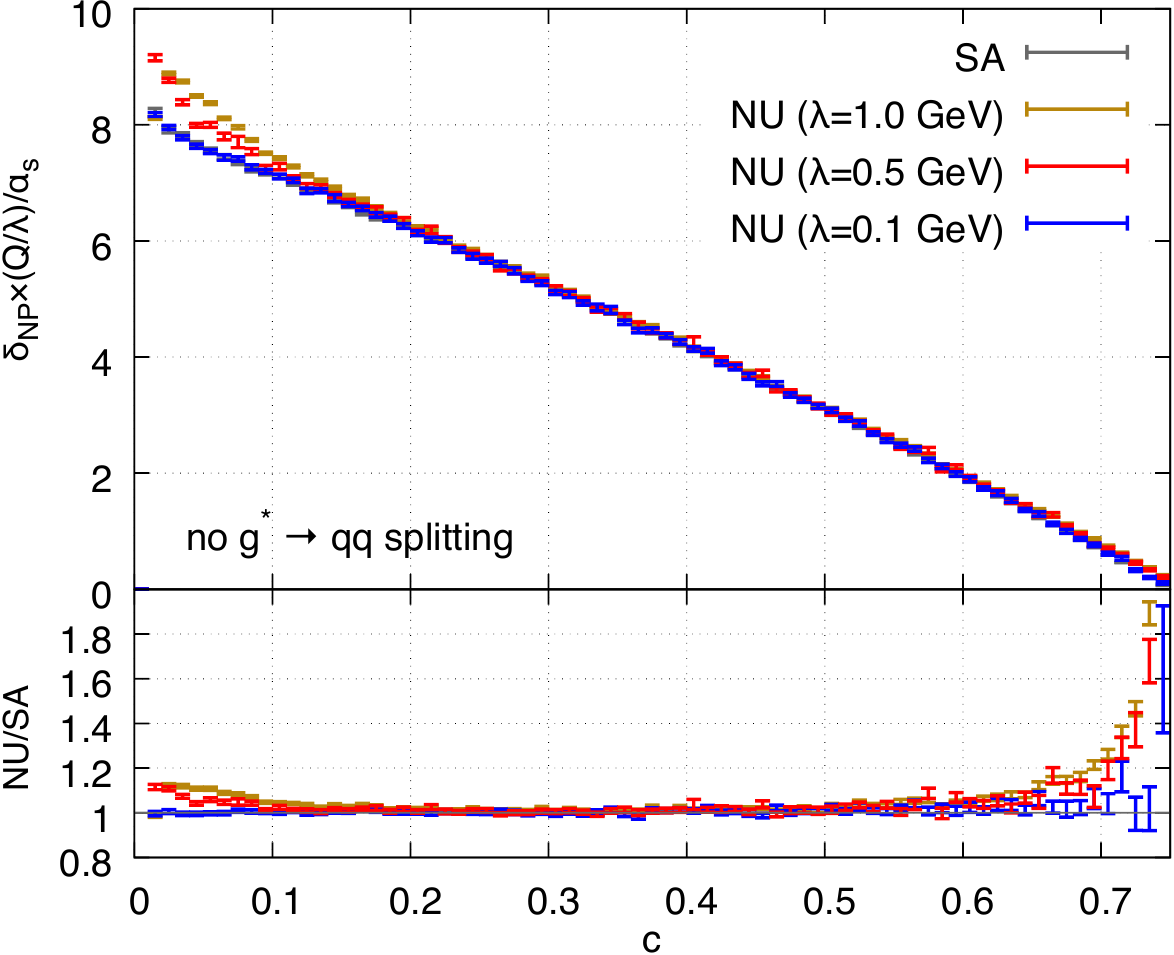} \hskip
  0.04\textwidth
  \caption{\label{fig:Cparamqq_nosplit} The non-perturbative shift
    $\delta_{\rm NP}$ for the $C$-parameter defined in
    Eq.~\eqref{eq:deltaC_nosplit}, stripped of the $\lambda/Q$ and
    $\as$ factor, computed with the semi-analytic result of
    Eq.~(\ref{eq:deltaC_nosplit}) (labeled SA), and using the
    numerical implementation of Eq.~(\ref{eq:numerics}) (labeled
    NU). The splitting $g^*\to q\bar{q}$ is \emph{not}
    included. For the numerical calculation, we use $Q=100$~GeV and
    three different values of $\lambda$.}
\end{figure}

Results for $\delta_{\rm NP}$ obtained with the two methods are
reported in Fig.~\ref{fig:Cparamqq_nosplit}.
While the semi-analytic result is linear in $\lambda$ by construction,
the numerical one also contains higher powers of $\lambda$ so that the
linear term only dominates in the $\lambda \to 0$ limit.  This
explains the differences between the numerical results obtained with
different values of $\lambda$, and also the residual differences
between the semi-analytic and numerical results.  We also notice that,
as we approach the endpoint regions $c=0$ and $c=3/4$,\footnote{The
  value of the $C$-parameter $c=3/4$ corresponding to the symmetric
  configuration of three thin jets of equal energy and equal angular
  separation.}  subleading powers of $\lambda$ become more important,
thus explaining the larger differences between the semi-analytic and
numerical results
there.
Overall, we observe good agreement between the results obtained
with the two methods. 

\subsection{Comparison with the full large-$\nf$ calculation}
\begin{figure}[t]
  \includegraphics[width=0.47\textwidth]{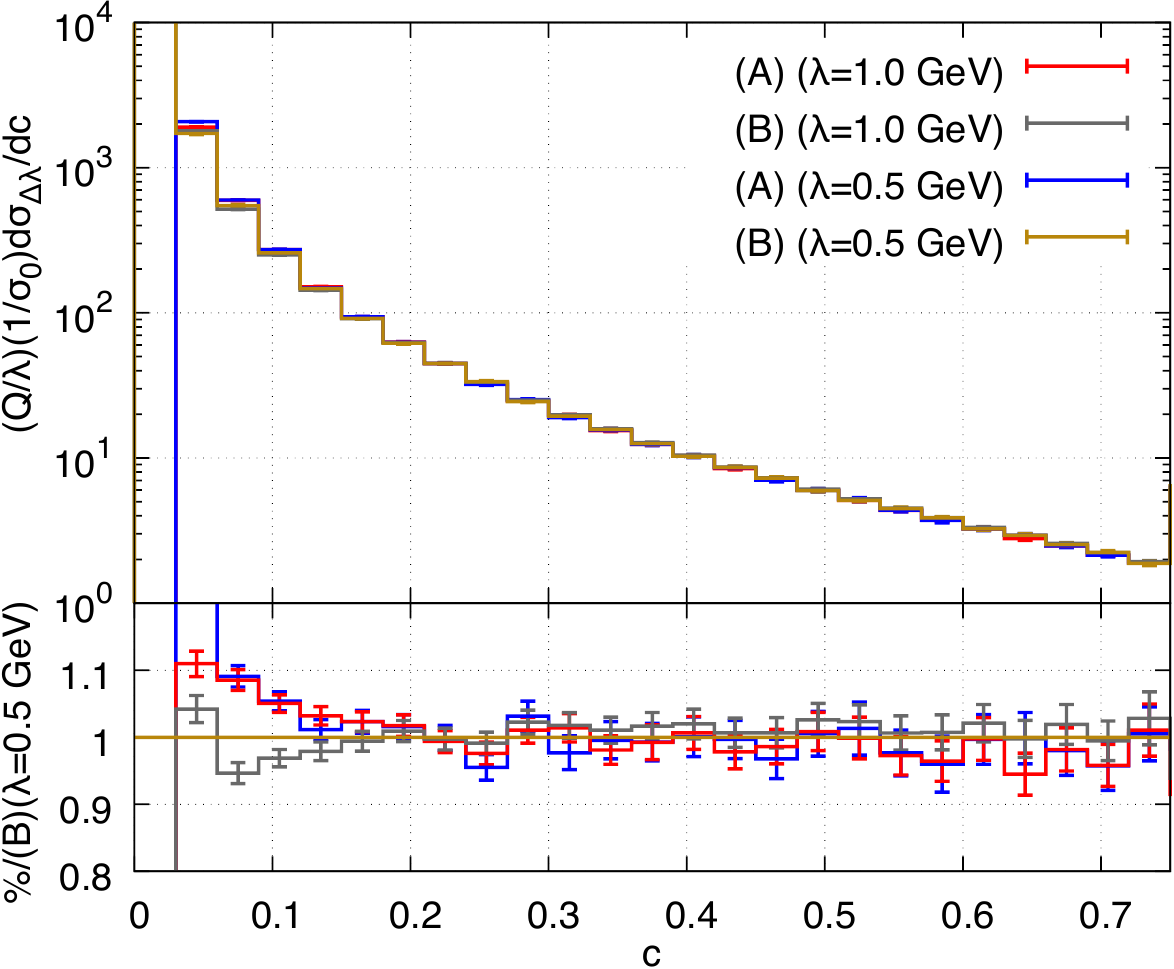} \hskip
  0.04\textwidth
  \includegraphics[width=0.47\textwidth]{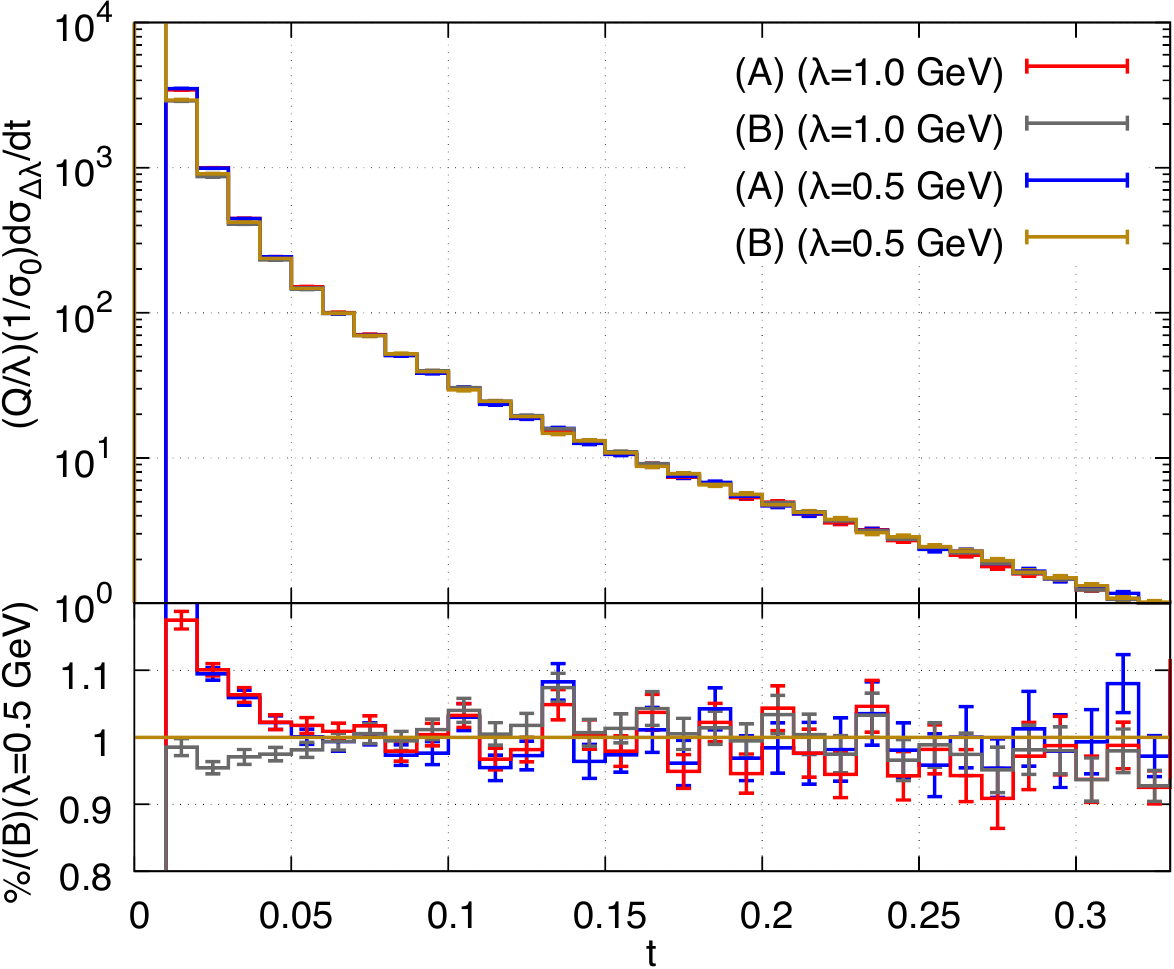}
  \caption{\label{fig:diff_shift}Non-perturbative shift in the
    differential distributions for the $C$-parameter (left) and the
    thrust (right), obtained from Eq.~\eqref{eq:numerics} (A), and
    from a full calculation in the large-$\nf$ limit (B). Results are
    shown for the process $\gamma^*\to d\bar{d} \gamma$, with
    $Q=100$~GeV and $\lambda=1$~GeV, $\lambda=0.5$~GeV.}
\end{figure}

As a second test of our approach, 
we compute coefficients of ${\cal O}(\lambda)$ terms for various
observables using
Eq.~(\ref{eq:numerics}) and compare them with the result of a full
numerical calculation performed in the large-$\nf$ approximation.
This comparison is shown in Fig.~\ref{fig:diff_shift} for the differential
distributions of the $C$-parameter and the thrust.
In all cases, we perform the computation for $\lambda=0.5~{\rm
  GeV}$ and $\lambda=1~{\rm GeV}$.  For the numerical approach based on
Eq.~\eqref{eq:numerics}, we use the mapping \ref{it:diffMap} and
Eq.~(\ref{eq:softapprox}) for the soft amplitude in the numerical
implementation of Eq.~(\ref{eq:numerics}).  Details of the full
calculation are reported in Appendix~\ref{app:largenfcalc}.

Since the results shown in Fig.~\ref{fig:diff_shift} are divided by
$\lambda$, the agreement between the $\lambda=0.5~{\rm GeV}$ and $\lambda =
1~{\rm GeV}$ cases indicates that the dependence of the observable on
$\lambda$ is indeed linear and that Eq.~(\ref{eq:numerics}) captures
the $\lambda$-dependence correctly.
For values of the $C$-parameter $c\lesssim 0.15$ and the thrust
$t\lesssim 0.07$, the results of the exact calculation performed for
two values of $\lambda$ deviate from each other and from the result
obtained with the help of Eq.~(\ref{eq:numerics}). This is an
indication of the fact that higher powers of $\lambda$ become important in
these regions so that smaller values of $\lambda$ need to be used for
these values of $c$ and $t$ to enable the extraction of
$\mathcal O(\lambda)$ terms from the large-$\nf$ computation. Apart from this
caveat, Fig.~\ref{fig:diff_shift} gives strong evidence that one can
use Eq.~\eqref{eq:numerics} to compute linear power corrections to generic
shape variables.

\subsection{Non-perturbative correction as a shift in the shape variable}
Having verified a simplified method for computing linear power
corrections to generic shape observables, we can now use it to derive
non-perturbative corrections to them.  We will start with a brief
overview of the history of such computations.

Non-perturbative corrections to shape variables in the two-jet limit
have been considered in
Refs.~\cite{Manohar:1994kq,Webber:1994cp,Dokshitzer:1995zt,
  Nason:1995np,Dokshitzer:1995qm,
  Dasgupta:1996ki,Nason:1996pk,Beneke:1997sr,Dokshitzer:1997ew,
  Dokshitzer:1997iz,Dokshitzer:1998pt,Campbell:1998qw,Luisoni:2020efy}
(for a review see Ref.~\cite{Beneke:1998ui}). These non-perturbative
corrections are usually employed together with the perturbative ones,
as well as with resummations, to extract the strong coupling constant
$\alpha_s$ from data on $e^+e^-$ annihilation into
hadrons~\cite{Abbate:2010xh,Hoang:2015hka,Catani:1998sf,Gehrmann:2012sc,Davison:2009wzs}.
Non-perturbative corrections are usually fitted in the two-jet
region and then extrapolated to the three-jet region, where the value
of the strong coupling constant is determined.  This approach
relies on the assumption that non-perturbative corrections in the
three- and two-jet regions are the same.

In a recent paper~\cite{Luisoni:2020efy}, an attempt
has been made to gain some insight into the behaviour of these
power corrections away from the two-jet limit. The authors of Ref.~\cite{Luisoni:2020efy}
studied  the $C$-parameter distribution, that, besides the Sudakov region at $c=0$, has a second Sudakov
region at $c=3/4$, corresponding to the symmetric
three-jets configuration. The presence of this
second region allows for a calculation of non-perturbative effects using 
techniques identical to the ones  used for the two-jet region.
It was found~\cite{Luisoni:2020efy} that there is
significant  difference between power corrections in two
Sudakov regions. Moreover, it was observed in Ref.~\cite{Luisoni:2020efy}
that  power corrections in the region where $\as$ is measured strongly 
depend on  the model used to interpolate between the two Sudakov regions.
Clearly these results call for a better understanding of the dependence of
non-perturbative corrections on the three-jet kinematics. 

In the previous sections, we have shown how to compute linear power
corrections in the three-jet region in a simplified model with
$d\bar d\gamma$ final state. Hence, we are in the
position to compare our findings in this simplified setup
with the approximate results of
Ref.~\cite{Luisoni:2020efy}.  Conversely, we should be able to
reproduce the ratio of non-perturbative corrections in the three-jet
symmetric point to the non-perturbative corrections in the two-jet
limit obtained in Ref.~\cite{Luisoni:2020efy}; such a comparison
should provide a further test of our numerical approach.

To set up the comparison, we follow the same approach as discussed in
Section~\ref{sec:shapevars-ex} where it was shown that the
non-perturbative corrections to a cumulant of the $C$-parameter can be
computed as follows
\begin{equation}
  \delta_{\rm NP}(c) = -\frac{ {\cal T}_\lambda
  \Sigma(c)}{ {\rm d} \sigma(c)/{\rm d} c }.
\label{eq5.35}
\end{equation}
Although this result was derived in Section~\ref{sec:shapevars-ex}
for a massive gluon in the final state, it is clear that it also holds
if the $g^*\to q\bar q$ splitting is accounted for.

The non-perturbative correction $\delta_{\rm NP}(c)$ defined in
Eq.~(\ref{eq5.35}) can be computed directly as a function of $c$ using
numerical approach described earlier in this section.  However, it is
customary to separate it into a normalization factor $h$ that
describes non-perturbative corrections in the two-jet region and a
$c$-dependent function $\zeta_{q\bar q}(c)$ that parametrizes the
dependence of non-perturbative corrections on three-jet kinematics.
Hence, we write
\begin{equation}
  \delta_{\rm NP}(c) = h \zeta_{q\bar q}(c).
\end{equation}

In principle, the non-perturbative correction in the two-jet region
can be computed in the same way as the one in the three-jet
region. However, since the two-jet cross section is proportional to
$\delta(c)$ in a fixed-order calculation, it is more convenient to
relate $h$ with the average value of the $C$-parameter computed in the
two-jet region. Indeed, since
\begin{equation}
  \frac{ {\rm d} \sigma}{{\rm d} c} =
  \sigma \delta(c+h),
\end{equation}
the average value of the $C$-parameter
computed for two-jet events is just $-h$.
Hence, we can write 
\begin{equation}
h = -{\cal T}_\lambda \langle C\rangle^{(1)}_{\lambda},
\end{equation}
where because of the two-jet constraints the expectation value has to be
computed starting from the Born process $\gamma^* \to d \bar d$.

We compute $h$ numerically for a multitude of different values of
$\lambda$; upon linear extrapolation to $\lambda = 0$, we obtain
\begin{equation}
  h= -9.21(1) \; \left ( \frac{ \alpha_s \lambda}{Q} \right ).
\end{equation}
This
result is consistent with the value $15\pi^2/16=9.253$ reported in
Ref.~\cite{Smye:2001gq}; we attribute the differences between
numerical and analytic results to higher powers of $\lambda$ that are
present in the numerical computation.\footnote{A similar computation
  for thrust yields $1.9457(8)$ as a linear slope in $\lambda$ at
  $\lambda=0$.  We can extract analytic result for this quantity,
  $5\pi/8=1.9635$, from
  Refs.~\cite{Webber:1994cp,Smye:2001gq}. Indeed, in
  Ref.~\cite{Webber:1994cp} the ratio of the non-perturbative shifts
  to $C$-parameter and thrust was computed. This, together with the
  result of Ref.~\cite{Smye:2001gq} for the $C$-parameter yields the
  value for the thrust slope quoted above.}

\begin{figure}[t]
  \includegraphics[width=0.48\textwidth]{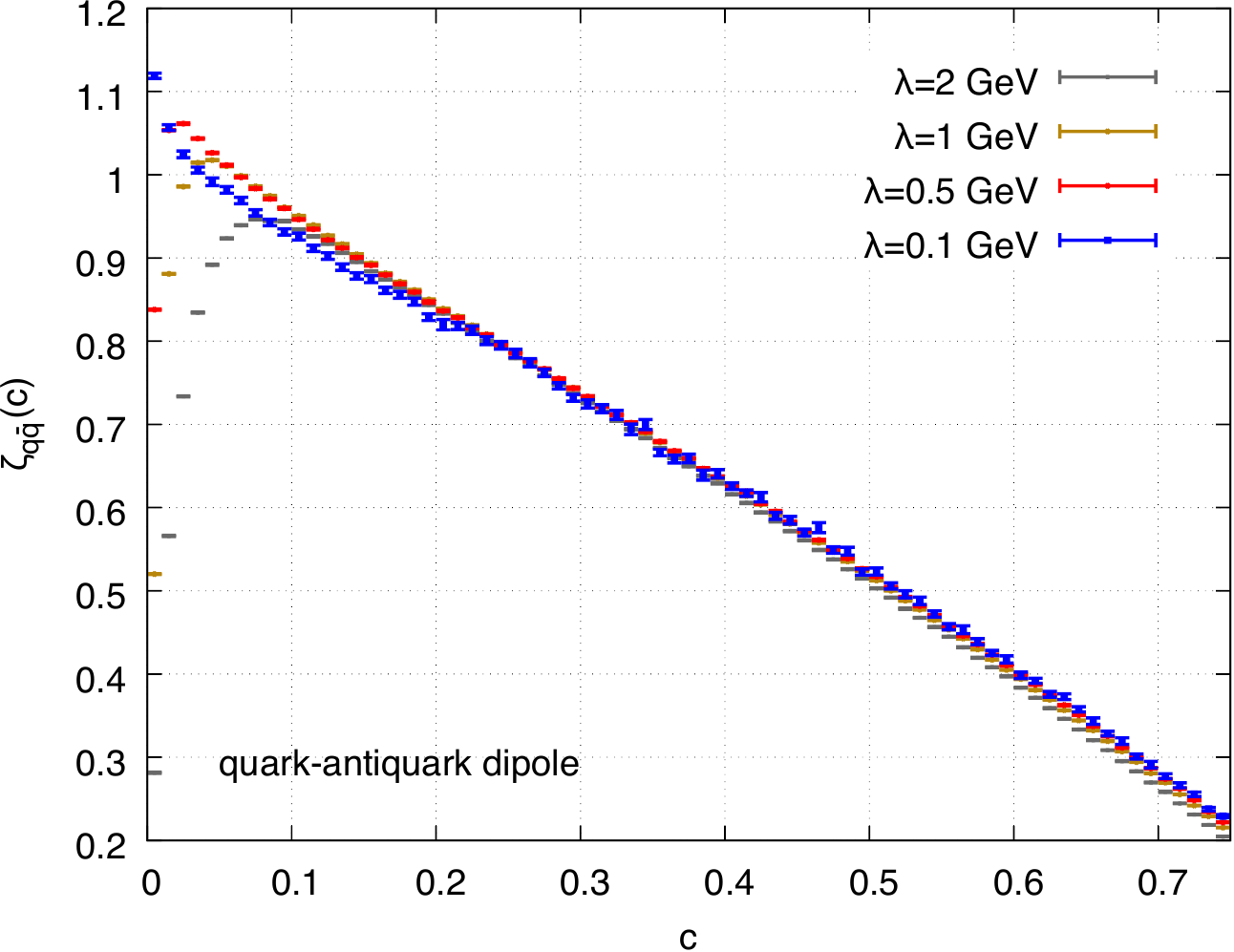}
  \includegraphics[width=0.48\textwidth]{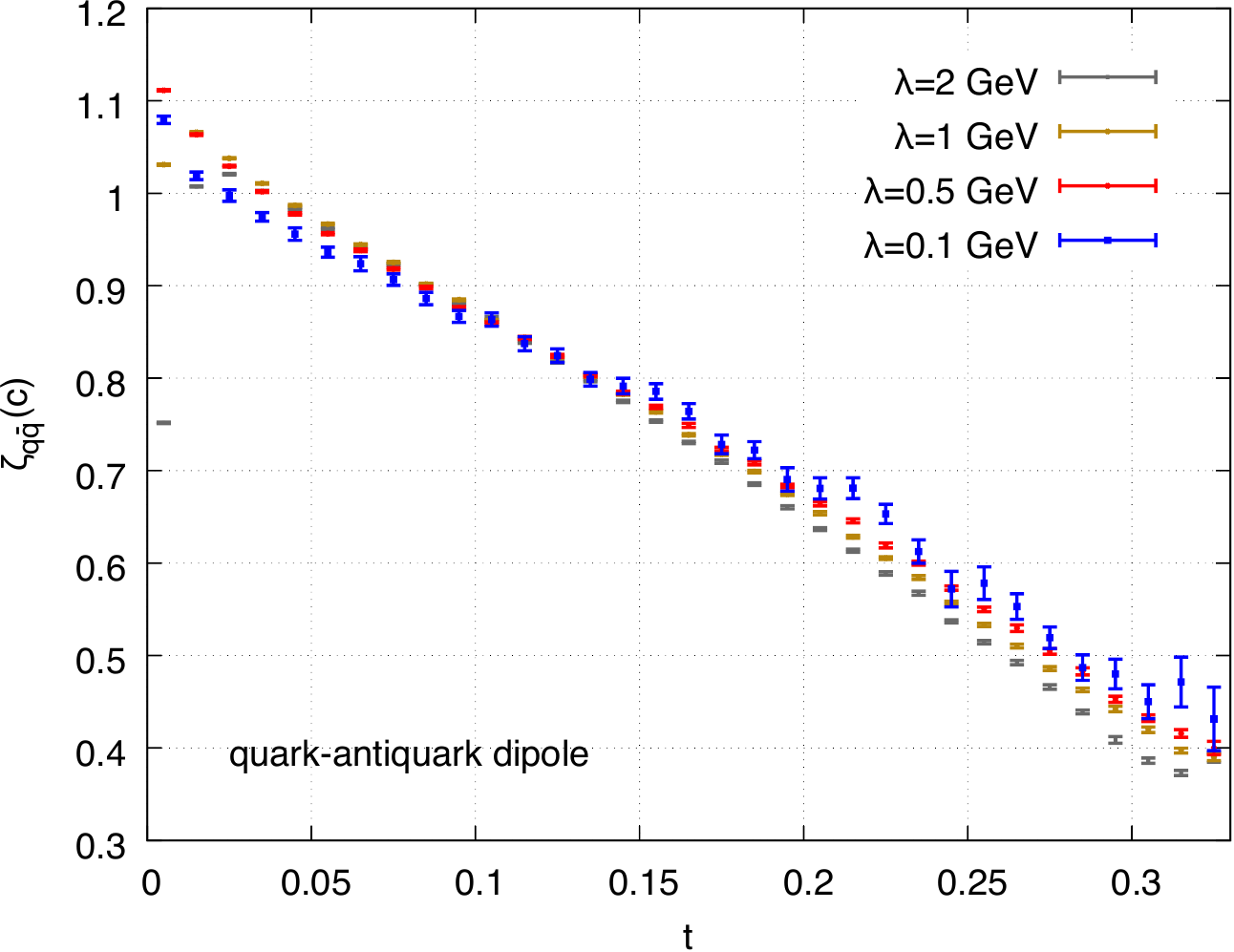}
  \caption{\label{fig:Zetaqq} The function $\zqq(c)$ for the
    $C$-parameter and for thrust $t=1-T$, obtained using $Q=100$~GeV
    and $\lambda=2$~GeV, $\lambda=1$~GeV, $\lambda=0.5$~GeV and
    $\lambda=0.1$~GeV.  The normalization factor $h$ was taken equal
    to $-(\lambda/Q)\alpha_s\times 15\pi^2/16$ for the $C$-parameter, and
    $-(\lambda/Q)\alpha_s\times 5\pi/8$ for thrust.  }
\end{figure}

Having determined the normalization coefficient, we can now turn to
the discussion of the function $\zqq$ that parametrizes the dependence
of non-perturbative corrections on the $C$-parameter.  We plot $\zqq$
in Fig.~\ref{fig:Zetaqq}; these results are obtained with
$\lambda=2~{\rm GeV}$ and $\lambda = 1$~GeV. We note that for small
values of $C$, $\zqq$ approaches unity.  This is explained by the fact
that soft emissions factorize independently. So, in the dominant
region where both the photon and the gluon are soft, the gluon behaves
as if it was radiated by a $q\bar{q}$ dipole (see
Appendix~\ref{app:twoJetLim}).

Near the three-jet symmetric point, that corresponds to $c=3/4$, we
find $\zqq(3/4) = 0.226(2)$ for $\lambda=0.1~{\rm GeV}$. This value is
consistent with the one found in Ref.~\cite{Luisoni:2020efy}, provided
that only the radiation of the quark dipole in the abelian limit is
considered. We note, however, that since the normalization used here
and in Ref.~\cite{Luisoni:2020efy} differ, we can only compare ratios
of $\zeta$-functions computed in~\cite{Luisoni:2020efy}; in what
follows we will always consider $\zeta_{\rm LMS}(c) =
\zeta(c)/\zeta(0)$ when we quote results of
Ref.~\cite{Luisoni:2020efy}.  With this clarification, and after
setting $C_A=0$ in Eq.~(18) of Ref.~\cite{Luisoni:2020efy}, we find
$\zeta_{\rm LMS}(3/4) =0.224$, consistent with our result.

\subsection{Including radiation from the quark-gluon dipoles}
The results described so far have been obtained for the $\gamma^* \to
d{\bar d} \gamma$ process and not for the much more interesting case
of $\gamma^* \to d \bar d g$.  As we explained in the introduction,
this is a well-known limitation of the large-$\nf$ approach to
computing non-perturbative corrections since processes with gluons at
the Born level cannot be dealt with in the theoretical framework
employed in this paper. 

Although we do not currently know how to overcome this limitation, the
structure of the results that we obtained allows us
to speculate that, perhaps, it is straightforward to do so.
Indeed, our final result shows that linear power corrections to 
shape variables are captured by the soft approximation to the
full matrix element of the $\gamma^* \to d \bar d \gamma + g$ process.
For the cases that we considered so far, the soft approximation
originates from a color dipole formed by the $d \bar d$ pair.
It is tempting to speculate that for the real three-jet
production process $\gamma^* \to d \bar d g$ we can compute linear
power corrections by simply considering the emission of an additional
soft massive gluon by \emph{all} the QCD dipoles $d \bar d$, $d g$ and $\bar d
g$ that are present in this case.
We emphasize that we cannot prove this statement at this point,
but we believe that it provides a reasonable conjecture.
 
Since the contributions of the three dipoles simply add  up, we can write 
\begin{equation}\label{eq:zetacomb}
  \zeta(x)=\zqq(x)\frac{\Cf-\Ca/2}{\Cf}+\zqg(x)\frac{\Ca}{\Cf}\,,
\end{equation}
where we have exploited the fact that the $d g$
and ${\bar d} g$ dipoles contribute equally. We have defined $\zqg(x)$
in the same way as
$\zqq(x)$ discussed in the previous sub-section,
except that we now assume that the radiating dipole is
$dg$ (or $\bar d g$). We keep, however, the same color factor and the same
normalization $h$ used for the $q\bar{q}$ case; hence the $1/\Cf$ factors
in Eq.~\eqref{eq:zetacomb}.

\begin{figure}[t]
  \includegraphics[width=0.48\textwidth]{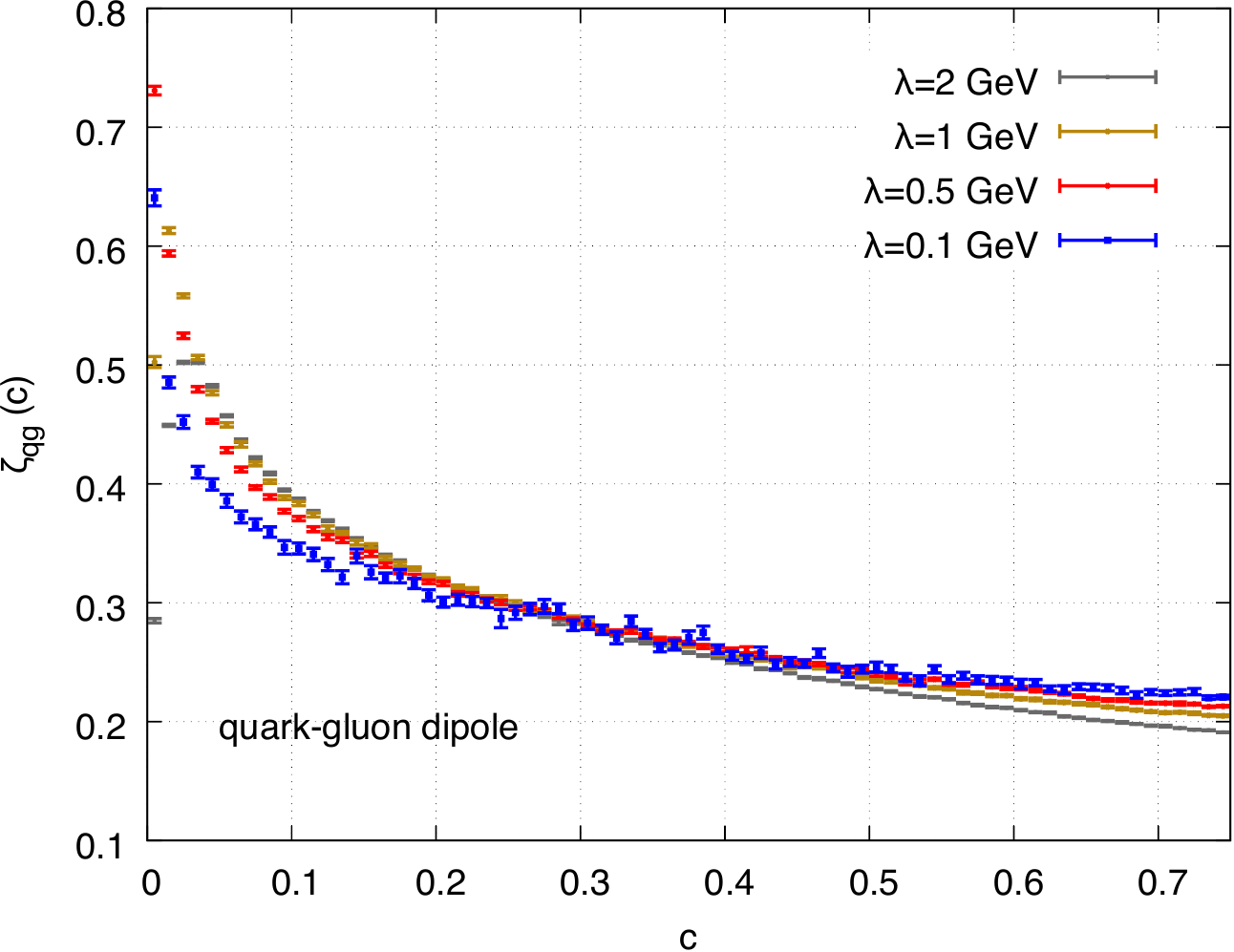}
  \includegraphics[width=0.48\textwidth]{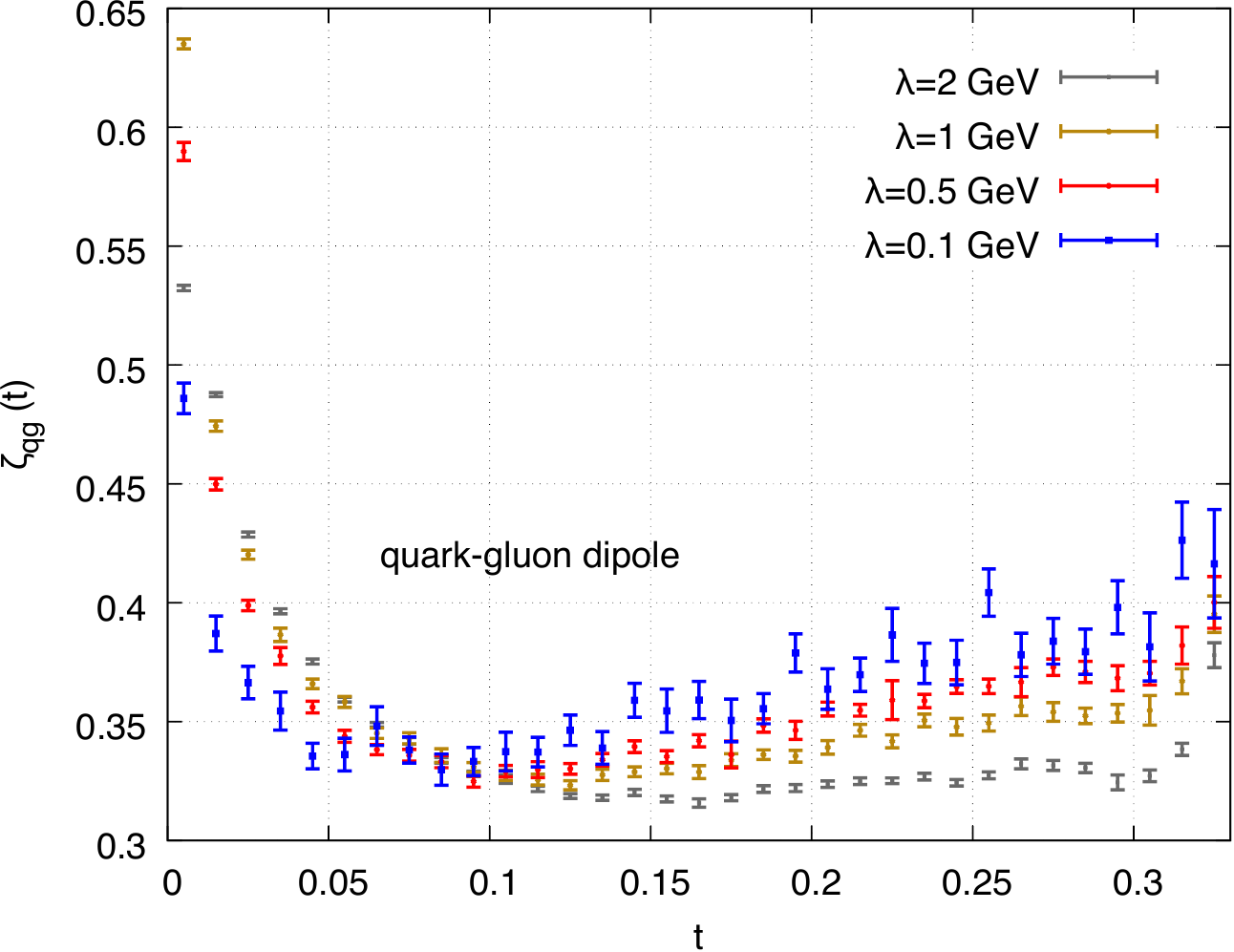}
  \caption{\label{fig:Zetaqg} The functions $\zqg(c)$ and $\zqg(t)$
    for the $C$-parameter and for thrust $t=1-T$, obtained using
    $Q=100$ GeV and $\lambda=2$~GeV, $\lambda=1$~GeV, $\lambda=0.5$~GeV 
    and $\lambda=0.1$~GeV. See text for details.}
\end{figure}

In Fig.~\ref{fig:Zetaqg} we display the function $\zqg$ for the
$C$-parameter and the thrust. We observe that in both cases $\zqg$
approaches $0.5$ for small $c$ and $t$. This is easily understood,
since in this limit $\zeta$ in Eq.~(\ref{eq:zetacomb}) should be one
by angular ordering arguments and, since $\zqq$ approaches one, it
follows that $\zqg$ approaches $0.5$ (see
Appendix~\ref{app:twoJetLim}). In the symmetric three-jet limit
$\zqg$ approaches the same value as $\zqq$. This is a consequence of
the fact that in the symmetric limit the $q{\bar q}$ and $q g$ dipoles
are geometrically equivalent and, once the color factors are removed,
they should give the same results.

We notice that the precision of the numerical result for the $q g$
dipole is inferior to the $q{\bar q}$ one and also that near the
symmetric point it is worse for thrust than for the $C$-parameter. The
first issue is probably related to the fact that the hard emitting
gluon is generally softer than the emitting quarks.  Thus the
effective $Q$ of the emission is smaller in the $q g$ case, leading to
larger non-perturbative effects, since they are proportional to
$\lambda/Q$. Regarding thrust, we recall that it vanishes in the
symmetric three-jet configuration at Born level. This is different for
the $C$-parameter, which approaches a constant there.

\begin{figure}[t]
  \includegraphics[width=0.48\textwidth]{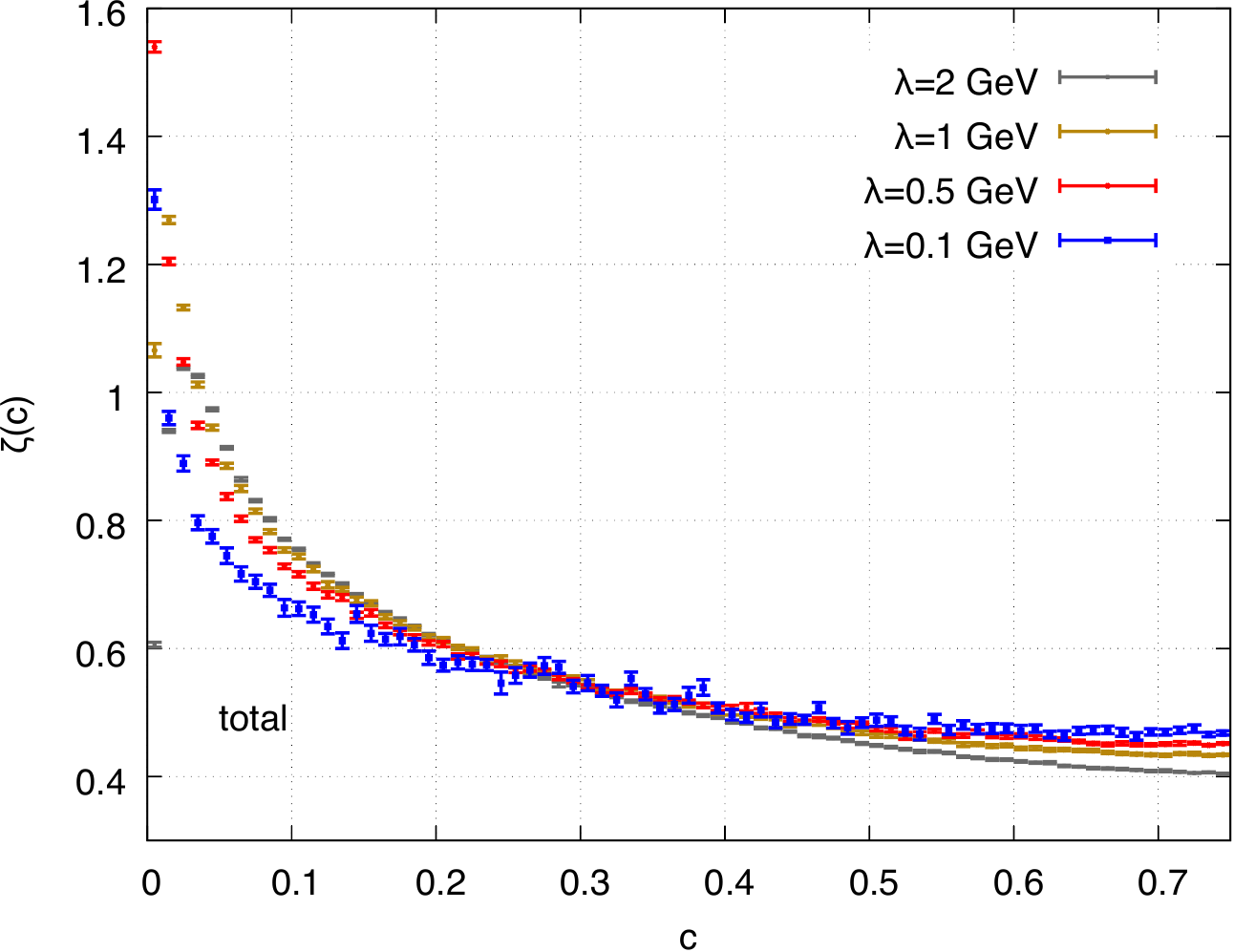}
  \includegraphics[width=0.48\textwidth]{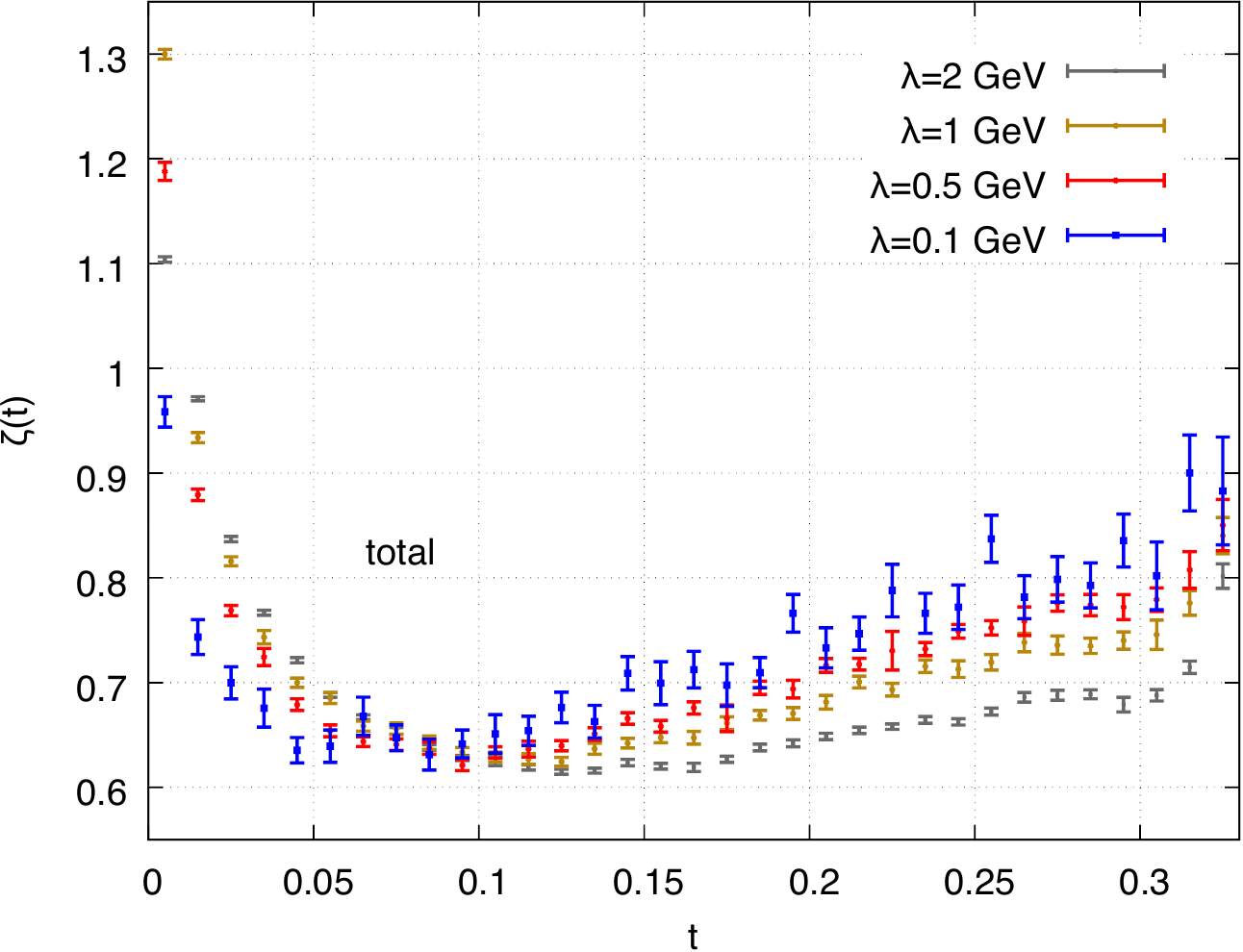}
  \caption{\label{fig:Zetacomb} The functions $\zeta(c)$ and
    $\zeta(t)$ for the $C$-parameter and for thrust $t=1-T$, obtained
    using $Q=100$ GeV and $\lambda=2$~GeV, $\lambda=1$~GeV,
    $\lambda=0.5$~GeV and $\lambda=0.1$~GeV.  See text for details.}
\end{figure}

In Fig.~\ref{fig:Zetacomb} we plot $\zeta$ defined in 
Eq.~(\ref{eq:zetacomb}) for the $C$-parameter and for thrust. The
results for the $C$-parameter can be compared to Figs.~1
and 3 of Ref.~\cite{Luisoni:2020efy}; we note again that predictions of
Ref.~\cite{Luisoni:2020efy} need to be rescaled so that they assume
the value $1$ at $c=0$. The normalized curves agree at the three-jet
symmetric point, $c=0.75$, where our result computed for $\lambda=0.1$
is $0.479(5)$, and the (re-scaled) result obtained in
Ref.~\cite{Luisoni:2020efy} is $0.476$; the difference can be
attributed to terms proportional to $\lambda^2$.  Among the various
extrapolations of the function $\zeta$ presented in
Ref.~\cite{Luisoni:2020efy}, their $\zeta_{{\rm b},3}$ curve seems to
be the closest to our result.

As a final comment, we notice that for both the $C$-parameter and the
thrust, the non-perturbative correction that we computed here is {\it
  smaller} than the one obtained by extrapolating it from the two-jet
region to a symmetric point, especially in the case of the
$C$-parameter.  In Ref.~\cite{Luisoni:2020efy} a fit to $\as$ using
the $C$-parameter was given under various assumptions about the shape
of the function $\zeta(c)$.  For the function $\zeta_{{\rm b},3}$
that, as we said, is closest to our results, the authors of
Ref.~\cite{Luisoni:2020efy} extract the value of the strong coupling
constant $\as = 0.117(3)$.  This result is in much better agreement
with the world average value $\as = 0.118(1)$ as compared to $\as =
0.112(2)$ obtained in Ref.~\cite{Hoang:2015hka} using a more
conventional treatment of non-perturbative effects.  It would be
interesting to see if also for the thrust a similar improvement can be
achieved.

\section{Conclusions}
\label{sec:conclusions}
Understanding non-perturbative corrections to collider processes is an
interesting problem in theoretical particle physics that received
surprisingly little attention in the recent past.  However, thanks to
the rapid development of the precision physics program at the LHC a
case for a better control of non-perturbative effects in hadron
collisions becomes stronger.

A possible way to investigate them is to make
use of the asymptotic nature of QCD perturbation theory and estimate
these effects by studying the ambiguities of a purely perturbative
treatment.
These ambiguities are related to the infrared pole in the running of the
coupling constant. For simple enough processes
they can be identified by computing ${\cal O}(\as)$ corrections
in an abelian version of QCD with a massive gluon, and
extracting terms that are non-analytic in $\lambda^2$, where
$\lambda$ is the gluon mass. From a phenomenological point of view,
of particular importance are linear
non-perturbative corrections $\mathcal O(\Lambda_{\rm QCD}/Q)$.
Their presence is exposed by the appearance of $\mathcal O(\lambda)$ terms
in the massive gluon calculation.

While many explicit computations within the massive gluon framework
have been performed in the past, we believe that there is a lack of
general understanding of how to approach computations of ${\cal
  O}(\lambda)$ corrections to a generic scattering process or
observable. In a certain sense, this is not surprising since
understanding of these $\mathcal O(\lambda)$ terms requires a theory
of soft effects at next-to-leading power which is more complicated
than the familiar soft limit of scattering amplitudes and cross
sections.

In this paper, we have shown that it is possible to demonstrate on
rather general grounds that many terms that arise at next-to-leading
power from the expansions of both phase spaces and matrix
elements for typical processes and observables
do not produce ${\cal O}(\lambda)$ terms.  This result 
allows us to argue that certain (simplified) collider processes and
observables cannot receive linear power corrections.  An interesting
example of this is the transverse momentum distribution of vector bosons in
proton-photon collisions that even if rapidity cuts are applied does not
contain $\mathcal O(\lambda)$ terms.

We have also shown that an improved understanding of how ${\cal
  O}(\lambda)$ terms may arise allows us to calculate non-perturbative
corrections to shape variables away from Sudakov regions both
analytically and numerically.  To this end, we have derived a formula
that allows us to compute linear power corrections to a generic shape
variable in a three-jet configuration. This remarkably simple formula
only involves the matrix element of the Born process and
soft-radiation eikonal functions of color dipoles.  As an application,
we have used our formalism to compute non-perturbative corrections to
the $C$-parameter and compared it with the result of
Ref.~\cite{Luisoni:2020efy} which is based on an interpolation between
the two-jet limit ($c=0$) and the three-jet symmetric point
($c=3/4$). As expected, we have found that we can reproduce the
results of Ref.~\cite{Luisoni:2020efy} for $c=0$ and $c=3/4$. Between
these two points, our results are close to one of the interpolations
presented in Ref.~\cite{Luisoni:2020efy} whereas they differ
significantly from a few other interpolations provided in that
reference. This, of course, is not unexpected since interpolations by
their very nature are subject to significant uncertainties.

Our analysis is based on the large-$\nf$ approach to the study of
non-perturbative corrections; for this reason currently
it cannot be applied
to processes with gluons at the Born level. It would be interesting
to understand how
to extend this formalism to deal with these cases as well. 
Such an extension is, of
course, very interesting for hadron collider processes. Also, as we
have shown, it may lead to improvements in the description of three-jet
events and to more reliable extractions of the strong coupling
constant from $e^+e^-$ data. We look forward to study this interesting
problem in the future.

\section*{Acknowledgements}
We are grateful to P.~F. Monni, G. Salam and G. Zanderighi for
many interesting discussions.  We would also like to thank P.~F. Monni and G. Salam
for their help in comparing against the results of
Ref.~\cite{Luisoni:2020efy}, and M. van Beekveld for useful comments on the manuscript.
The research of F.C. is supported by the ERC Starting Grant 804394
{\sc HipQCD} and the UK Science and Technology Facilities Council
(STFC) under grant ST/T000864/1.  S.F.R.’s work is supported by the
ERC Advanced Grant 788223 PanScales. P.N. acknowledges support from
Fondazione Cariplo and Regione Lombardia, grant 2017-2070, and from
INFN. K.M. is partially supported by the Deutsche
Forschungsgemeinschaft (DFG, German Research Foundation) under grant
396021762-TRR 257.

\appendix

\section{Soft integrals}
\label{app:int}
In this appendix, we discuss integrals of the form
\begin{equation}
    \vec I(v,\tilde p_{1,2}) = \int [\mathd k]
    \theta\left[(k-q)^2\right] \frac{(\tilde p_1\tilde
      p_2)}{\left(\tilde p_1 k\right)\left(\tilde p_2
      k\right)}\left\{1,\frac{(k\, v)}{q^2}, \frac{\lambda^2}{(k\,
      p_{1,2})}\right\},
    %
  \label{eq:softint}
\end{equation}
where
$[\mathd k] = \mathd^4 k\, \delta_+(k^2-\lambda^2)/(2\pi)^3$ and 
$v$ is a generic vector, $q=\tilde p_1+\tilde p_2$ and $\tilde
p_i^2 = 0$, $i=1,2$.
To compute these integrals, in a way that works for all types of
dipoles, it is convenient to use Sudakov decomposition.  We will
discuss the computation for final-final dipole but the calculation can
be repeated for initial-final and initial-initial ones with small
modifications.  We write
\begin{equation}
  k = \alpha\, \tp_1 + \beta\, \tp_2 + k_\perp.
\end{equation}
Since $2 (\tp_1 \tp_2) = q^2$, we find
\begin{equation}
  {\rm d}^4k\, \delta_+(k^2 - \lambda^2) \theta\big[ (q-k)^2\big] =
  \frac{q^2}{2} {\rm d} \alpha \, {\rm d} \beta \,{\rm d}^2 \vec
  k_\perp \delta (q^2 \alpha \beta - \vec k_\perp^2 - \lambda^2)
  \theta\big[q^2 -q^2(\alpha+\beta) + \lambda^2\big],
\end{equation}
and
\begin{equation}
  2 (\tp_1 k) = q^2 \beta,\;\;\; 2(\tp_2 k) = q^2 \alpha.
\end{equation}

When we apply the Sudakov decomposition to all integrals in
Eq.~(\ref{eq:softint}), we find
\begin{equation}
  (v\,k)  = (\tp_1 v ) \alpha + (\tp_2 v) \beta + (v\,k_\perp).
\end{equation}
Since this is the only dependence on the $k_\perp$-direction that
appears in the integrals, the last term in the above equation vanishes
after azimuthal integration. Therefore, for
the integrals in Eq.~(\ref{eq:softint}) the following replacement holds
true
\begin{equation}
  (v\,k) \to (\tp_1 v ) \alpha + (\tp_2 v) \beta.
\end{equation}
It follows that to compute Eq.~\eqref{eq:softint} we require the
following integrals
\begin{equation}
\frac{1}{16 \pi^2} \int {\rm d} \alpha \; {\rm d} \beta \;
\theta (q^2 \alpha \beta - \lambda^2) \theta\big[ q^2(1-\alpha - \beta) +
  \lambda^2\big] \frac{1}{\alpha \beta }
\left \{ 1, \alpha,
  \frac{\lambda^2}{\alpha},\frac{\lambda^2 \beta}{\alpha},
  \frac{\lambda^2 \beta^2}{\alpha}, \frac{\lambda^4 \beta}{\alpha^2}
  \right \}.
\end{equation}
To compute them, we need to know the integration boundaries.
They are found from the two $\theta$-functions in the above equation. 
Suppose that we first integrate over $\beta$. Then,
\begin{equation}
\frac{\lambda^2}{q^2 \alpha} < \beta < 1-\alpha + \frac{\lambda^2}{q^2}.
\end{equation}
Boundaries for the subsequent $\alpha$ integration follow from the condition
\begin{equation}
\frac{\lambda^2}{q^2 \alpha} < 1-\alpha + \frac{\lambda^2}{q^2}, 
\end{equation}
which can be re-written as
\begin{equation}
(\alpha - 1) \left ( \alpha - \frac{\lambda^2}{q^2} \right ) < 0. 
\end{equation}
It follows that the integration interval for $\alpha$ is 
\begin{equation}
\frac{\lambda^2}{q^2} <  \alpha < 1. 
\end{equation}
It is then straightforward to see that the various integrals shown in
Eq.~(\ref{eq:softint}) can be written in terms of the following ones
\begin{equation}
  \int
  \limits_{\lambda^2/q^2}^{1} \frac{ {\rm d} \alpha }{\alpha}\; \int
  \limits_{\lambda^2/(q^2\alpha)}^{1-\alpha + \lambda^2/q^2} \frac{{\rm
      d} \beta}{\beta} \left \{ 1, \alpha,
  \frac{\lambda^2}{\alpha},\frac{\lambda^2 \beta}{\alpha},
  \frac{\lambda^2 \beta^2}{\alpha}, \frac{\lambda^4 \beta}{\alpha^2}
  \right \}.
\end{equation}
This representation makes it completely obvious that
the integrals in Eq.~\eqref{eq:softint} are actually functions of $\lambda^2$.

\section{Computation of the $\mathcal I_{i}$ integrals}
\label{app:cali}
In this appendix, we compute the integrals
\begin{equation}
  \left\{ \mathcal I_0, \mathcal I_1, \mathcal I_{12}\right\} = 
  \int\limits_\lambda^{\omega_{\rm max}}\mathd\omega\,\beta\times
  \left\{
  I_0,I_1,I_{12}
  \right\}
  \label{eq:cali_app}
\end{equation}
introduced in Section~\ref{sec:shapevars-ex}, see
Eq.~\eqref{eq:genlin}. The $I_i$ integrals are defined in
Eq.~\eqref{eq:iint}, $\beta = \sqrt{1-\frac{\lambda^2}{\omega^2}}$ and
$\omega_{\rm max}$ is a kinematic bound whose precise form is not
important. 

We are interested in the small-$\lambda$ expansion of the 
integrals in Eq.~\eqref{eq:cali_app}.
The $\mathcal I_{0}$ integral does contain linear terms in $\lambda$. Indeed
\begin{equation}
  \mathcal I_0 =
  \int\limits_\lambda^{\omega_{\rm max}}\mathd\omega\,\beta\, I_0 =
  \int\limits_\lambda^{\omega_{\rm max}}\mathd\omega\,\beta =
  \omega_{\rm max} - \frac{\pi}{2}\lambda + \mathcal O(\lambda^2)
\end{equation}
We now study $\mathcal I_{1}$. A simple calculation shows that 
\begin{equation}
  I_1 = \int\frac{\mathd\Omega_k}{4\pi} \frac{1}{1-\beta\vec
    n\cdot\vec n_1} = \frac{1}{2\beta}\ln\frac{1+\beta}{1-\beta}.
\end{equation}
Integrating by parts, it is then straightforward to obtain
\begin{equation}
  \mathcal I_1 = \frac{1}{2}\int\limits_\lambda^{\omega_{\rm max}}
  \mathd\omega\ln\frac{1+\beta}{1-\beta} =
  \frac{1}{2} \left[
    \omega_{\rm max}
    \ln\frac{\omega_{\rm max}+\sqrt{\omega_{\rm
          max}^2-\lambda^2}}{\omega_{\rm max}-\sqrt{\omega_{\rm
          max}^2-\lambda^2}}
    -2\sqrt{w_{\rm max}^2-\lambda^2}
  \right],
\end{equation}
which implies that for small $\lambda$ the integral $\mathcal I_{1}$
can be expanded in powers of $\lambda^2$.

To understand whether $\mathcal I_{12}$ contains $\mathcal
O(\lambda)$ terms in the small-$\lambda$ expansion, we need to compute
$I_{12}$. To this end, we first combine the two 
propagators using Feynman parameters
\begin{equation}
  I_{12} = \int\limits_0^1\mathd x\int\frac{\mathd \Omega_k}{4\pi}
  \frac{1}{\big(1-\beta \vec n\cdot\vec\eta\big)^2},
\end{equation}
where
\begin{equation}
  \eta = x\vec n_1 + (1-x) \vec n_2.
\end{equation}
Choosing the $z$ axis along $\eta$ and integrating over the relative angle
we find
\begin{equation}
  I_{12} = \int\limits_0^1\frac{\mathd x}{1-\beta^2 \,\vec\eta\,^2}
  = \int\limits_0^1\frac{\mathd x}{1-\beta^2\big(1-4x(1-x) s_{12}^2\big)},
\end{equation}
where $s_{12} = \sin(\theta_{12}/2)$. Using the $x\to 1-x$ symmetry of the
integrand, it is then easy  to obtain
\begin{equation}
  I_{12} = \frac{1}{2\beta s_{12}\sqrt{1-\beta^2 c_{12}^2}}
  \ln\frac{\sqrt{1-\beta^2 c_{12}^2} + \beta s_{12}}
          {\sqrt{1-\beta^2 c_{12}^2} - \beta s_{12}},
          \label{eq:i12}
\end{equation}
where $c_{12}^2 = 1-s_{12}^2 = \cos^2(\theta_{12}/2)$. We can now investigate
the small-$\lambda$ behaviour of $\mathcal I_{12}$:
\begin{equation}
  \mathcal I_{12} =
  \frac{1}{2 s_{12}}\int\limits_\lambda^{\omega_{\rm max}}
  \frac{\mathd \omega}{\sqrt{1-\beta^2 c_{12}^2}}
      \ln\frac{\sqrt{1-\beta^2 c_{12}^2} + \beta s_{12}}
  {\sqrt{1-\beta^2 c_{12}^2} - \beta s_{12}}.
\end{equation}
To compute this integral, we change variables
$\omega\to\beta$, with
\begin{equation}
  \mathd\omega = \lambda\frac{\beta\mathd\beta}{(1-\beta^2)^{3/2}},
\end{equation}
and $\beta\to r$ with
\begin{equation}
  r = \frac{\beta s_{12}}{\sqrt{1-\beta^2 c_{12}^2}},
  ~~~\mathd\beta = \frac{s_{12}^2\mathd r}{(c_{12}^2 r^2+s_{12}^2)^{3/2}}.
\end{equation}
We obtain
\begin{equation}
  \mathcal I_{12} =
  \frac{\lambda}{2s_{12}^3}\int\limits_0^{r_{\rm max}}
  \frac{\mathd r\, r}{(1-r^2)^{3/2}} \ln\frac{1+r}{1-r} =
  \frac{\lambda}{2 s_{12}^3 \sqrt{1-r_{\rm max}^2}}
  \left(\ln\frac{1+r_{\rm max}}{1-r_{\rm max}} - 2 r_{\rm max}\right),
  \label{eq:i12a}
\end{equation}
where $r_{\rm max} = \beta_{\rm max}s_{12}/\sqrt{1-\beta_{\rm max}^2 c_{12}^2}$
and $\beta_{\rm max} = \sqrt{1-\lambda^2/w_{\rm max}^2}$. It is
straightforward to show that the expansion of Eq.~\eqref{eq:i12a} in
$\lambda$ does not contain $\mathcal O(\lambda)$ terms.

\section{Full calculation of the shape variables in the large-$n_f$ limit}
\label{app:largenfcalc}
In this appendix we describe computation of QCD corrections to the
process $\gamma^* \to d \bar d + \gamma$ in the large-$\nf$
approximation. The exact result of the large-$\nf$ calculation can
be expressed as~\cite{FerrarioRavasio:2018ubr}
\begin{equation}
\label{eq:xsec1_largenf}
\langle O \rangle=\langle O \rangle^{(0)}
-\frac{1}{b_0\as}\int_0^\infty\frac{\mathd \lambda}{
  \pi}\frac{\mathd \langle O
  \rangle^{(1)}_\lambda}{d\lambda}\arctan\frac{\pi b_0
  \as}{1+b_0\as\log\frac{\lambda^2}{\mu_c^2}},
\end{equation}
where $\langle O \rangle_\lambda^{(1)}$ is defined in Eq.~\eqref{eq:Tlambda}
and
\begin{equation}
  \mu_c = \mu e^{\frac{5}{6}},~~~ b_0 = -\frac{\nf
    \Tf}{3\pi}.
\end{equation}
By expanding Eq.~\eqref{eq:xsec1_largenf} in $\as$, one obtains a factorial
growth associated with a linear renormalon. This leads to an ambiguity
of the fully-resummed series corresponding to Eq.~\eqref{eq:xsec1}
(see Appendix A of Ref.~\cite{FerrarioRavasio:2020guj}).

For the full calculation, we start from Eq.~(\ref{eq:Tlambda}).  The
required amplitudes have been analytically calculated using
symbolic manipulation software MAXIMA~\cite{maxima}.  The 
scalar integrals needed for the computation of the virtual corrections
have been calculated with the \texttt{COLLIER}
library~\cite{Denner:2016kdg}.  The virtual contribution is infrared
finite, since the gluon mass regulates infrared singularities.  We
have used dimensional regularization to regularize and extract
ultraviolet divergences.

Integration over the phase space of final-state particles in the
process $\gamma^* \to d \bar d \gamma$ diverges in the two-jet limit.
To ensure that numerical computations are restricted to a three-jet
region, we introduced a suppression factor
\begin{equation}
  F_{\rm{supp}}=C^2,
  \label{eq:bornsupp}
\end{equation}
where $C$ is the $C$-parameter, that vanishes in the two-jet limit
regulating the integral.  This factor is then divided out when
computing distributions and cross sections with cuts.  As a result, we
are able to obtain correct results as long as we do not attempt to
compute observables sensitive to the two-jet region.

The real corrections are obtained by adding the emission of a massive
gluon in all possible ways to the Born diagram. Integrating
over its momentum, the real emission corrections are affected by
collinear divergences, that arise when the photon becomes collinear
to one of the primary quarks. These configurations can contribute
to the three-jet cross section if the radiated gluon is hard and not
collinear. Such singularities are dealt with routinely by the
\texttt{POWHEG-BOX} framework~\cite{Alioli:2010xd}, that we adopt
for our calculation.

Singularities associated with soft or collinear gluons are regulated
by the gluon mass $\lambda$ and manifest themselves as terms
proportional to $\log{\lambda}$ raised to second or first power.
Similar logarithmic contributions, but with the opposite sign, arise
from the virtual corrections, such that the sum of real and virtual
corrections is free of $\log \lambda$ terms.

A carefully-constructed importance sampling near the singular regions
is needed in order to reliably estimate the $\lambda\to 0$
behaviour. We divide the real contribution into three regions
\begin{align}\label{eq:real_sep}
  R=&R^{(1)}+R^{(2)}+R^{(3)},
\end{align}
where
\begin{align}
R^{(1)}=&\frac{f_{d \gamma}^2+f_{\bar{d}\gamma}^2}{f_{d\gamma}^2+f_{\bar{d}\gamma}^2+f_{dg}^2+f_{\bar{d}g}^2}R,\\
R^{(2)}=&\frac{f_{dg}^2}{f_{d\gamma}^2+f_{\bar{d}\gamma}^2+f_{dg}^2+f_{\bar{d}g}^2}R,\\
R^{(3)}=&\frac{f_{\bar{d}g}^2}{f_{d\gamma}^2+f_{\bar{d}\gamma}^2+f_{dg}^2+f_{\bar{d}g}^2}R,
\end{align}
and
\begin{equation}
f_{ij}=\frac{E_i+E_j}{(k_i+k_j)^2}.
\end{equation}
Here, $k_i$ and $E_i$ denote the four-momentum and the energy of the
particle $i$. $R$ is a short-hand notation for the
$R^{(\lambda)}_{g^*}(\Phi_{3+1})$ appearing in
Eq.~\eqref{eq:TRlambda}.  The different contributions $R^{(i)}$ in
Eq.~(\ref{eq:real_sep}) correspond to different kinematic
configurations of $d \bar d \gamma g$ final state.  For example
$R^{(1)}$ corresponds to the region where the final state photon
becomes collinear to either the $d$ or $\bar{d}$ quark, whereas
$R^{(2)}$ and $R^{(3)}$ project on regions where emitted gluon becomes
collinear to $d$ and $\bar{d}$, respectively.

The contribution of region (1) is handled in the {\tt
  POWHEG-BOX}~\cite{Alioli:2010xd}, which implements the required
subtractions of IR singularities associated with configurations
containing a soft or a collinear photon. The remaining two regions are
finite, but require dedicated importance sampling of the region that
becomes singular in the limit $\lambda \to 0$.

Finally, we compute the amplitude for the process $\gamma^* \to
d \bar{d} \gamma q \bar{q}$. This contribution
is IR finite in the $\lambda\to 0$ limit, but is affected
by QED singularity associated with the final state photon. We thus
proceed as for  region (1) case, by evaluating it
within the \texttt{POWHEG-BOX} framework. We also computed the
process $\gamma^*\to d\bar d\gamma$ at NLO with a massless gluon
and subtracted its result from the $\lambda$-dependent one in order
to isolate the linear term. 

The shape variable distributions are obtained in a standard way, by
computing each contribution to sufficient accuracy so that after the
cancellation of $\ln^2\lambda$, $\ln\lambda$ and $\lambda^0$ terms
one can extract the $\lambda$ dependence with enough precision. 

\section{On the two-jet limit of $C$}
\label{app:twoJetLim}
In this appendix we elaborate more on the two-jet limit of the
cumulant of shape variables within our framework.  For simplicity, we
focus upon the case of the $C$-parameter, following the calculation of
Section~\ref{sec:shapevars-ex}.

We consider first the process $\gamma^*\to q(p_1)+\bar
q(p_2)+\gamma(p_3)$.  Extending the notation of
Section~\ref{sec:shapevars-ex}, we define the ``double underlying''
Born momenta $\dtp_1$ and $\dtp_2$ as follows:
\begin{itemize}
\item If $\tp_3$ becomes collinear to $\tp_1$, then
  $\dtp_1\approx\tp_1+\tp_3$ and $\dtp_2\approx \tp_2$,
\item If $\tp_3$ becomes collinear to $\tp_2$, then
  $\dtp_2\approx\tp_2+\tp_3$ and $\dtp_1\approx \tp_1$,
\item If $\tp_3$ becomes soft, then $\dtp_2\approx\tp_2$ and
  $\dtp_1\approx \tp_1$.
\end{itemize}
We want to show that, in the two-jet limit, the non-perturbative
correction of the cumulant of $C$ becomes proportional to the
non-perturbative correction to the average value of $C$ in the
$\gamma^*\to q\bar{q}$ process.  Three observations are needed to
prove this:
\begin{itemize}
\item
  As $c$ approaches zero, the Born cross section has two
  collinear-singular regions, when the photon is collinear to either
  primary quark; a soft singular region, when the photon is soft; and
  two soft-collinear regions, when the photon is both collinear and
  soft.
\item
  The correction to the $C$-parameter due to the emission of a soft
  gluon, i.e. the $C_k$ function of Eq.~\eqref{eq:c3ck}, has a
  smooth limit if any pair of the 1, 2 and 3 particles become
  collinear, as well as if one of them becomes soft. In particular, if
  $\tp_3$ becomes soft or collinear to either $\tp_1$ or $\tp_2$ we
  have
  \begin{equation}
    C_k(\Phi_3,k)=-3\sum_{i=1}^3 \frac{(k \tp_i)^2}{(k q)(\tp_i q)}
    \rightarrow -3\sum_{i=1}^2 \frac{(k \dtp_i)^2}{(k q)(\dtp_i
      q)}=C_k(\Phi_2,k)
  \end{equation}
  where we have written collectively $\{\tp_1,\tp_2,\tp_3\}=\Phi_3$
  and $\{\dtp_1,\dtp_2\}=\Phi_2$.
\item
  The eikonal factor in Eq.~\eqref{eq4.14} only depends upon the direction of
  the radiating partons, and not upon the absolute value of their
  momenta.
\end{itemize}
For the $q\bar q\gamma$ final state, we only have to consider the
the case of emission from the quark-antiquark dipole. Our result
for the non-perturbative correction, Eq.~(\ref{eq:deltaC_nosplit}) can
be written concisely as
\begin{equation}
    \delta_{\rm NP} \equiv - \frac{\mathcal
      T_\lambda\Sigma(c)}{{\mathd\sigma}/{\mathd C}} =-
    \frac{\alpha_s}{2\pi}C_F\times \frac{\int \mathd \Phi_3
      \delta\big(C(\Phi_3)-c\big) |\mathcal M(\Phi_3)|^2 \,\mathcal
      T_\lambda I_c(\Phi_3)} {\int \mathd \Phi_3
      \delta\big(C(\Phi_3)-c\big) |\mathcal M(\Phi_3)|^2},
  \label{eqApp:deltaC_nosplit}
\end{equation}
where
\begin{equation}
  I_c(\Phi_3;q;\lambda) = 8\pi^2 \int \frac{\mathd^4 k}{(2\pi)^3}
  \delta_+(k^2-\lambda^2)\theta\big[(q-k)^2\big]
  \frac{2(\tp_1\tp_2)}{(\tp_1 k)(\tp_2 k)}
  C_k(\Phi_3,k), \label{eqapp:Ic}
\end{equation}
and $\mathcal T_\lambda$ is an operator that takes out the term
linear in $\lambda$ from the expression it is applied to.  For both the
collinear and the soft limits of the Born configuration, the integrand
in Eq.~(\ref{eqapp:Ic}) becomes
\begin{equation}
  \frac{(\tp_1\tp_2)}{(\tp_1 k)(\tp_2 k)} C_k(\Phi_3,k)
  \rightarrow
  \frac{(\dtp_1\dtp_2)}{(\dtp_1k)(\dtp_2k)} C_k(\Phi_2,k).
  \label{eqapp:eiktimesC}
\end{equation}
Thus, in these limits
\begin{equation}
 I_c(\Phi_3;q;\lambda) \rightarrow  I_c(\Phi_2;q;\lambda),
\end{equation}
that can be taken out of the integrand in Eq.~(\ref{eqApp:deltaC_nosplit}),
yielding
\begin{equation}
    \delta_{\rm NP} \approxczero - \frac{\alpha_s}{2\pi}C_F\times
    \mathcal T_\lambda I_c(\Phi_2;q;\lambda).
  \label{eqApp:deltaC_nosplit2}
\end{equation}
Eq.~\eqref{eqApp:deltaC_nosplit2} can be immediately interpreted as the
non perturbative correction to the average value of $C$ in the two jet
case. Note that $c\to 0$ does not imply that $\tp_3$ is either soft
or collinear.  We could also have $\tp_1$ or $\tp_2$ soft, or
collinear to each other.  What is important is that for all dominant
singular contributions in the amplitude $I_c$ can be taken out of the
integral.

We now consider the $\gamma^*\to q\bar q g$ process, and use the 
conjecture for non-perturbative corrections that we have described in the
text. In comparison to the $\gamma^*\to q\bar q \gamma$ case, we now 
have to consider $qg$ dipoles as well. For the $q(p_1)g(p_3)$ dipole, 
the eikonal factor in Eq.~(\ref{eqapp:eiktimesC}) becomes
\begin{equation}
  \frac{(\tp_1\tp_3)}{(\tp_1 k)(\tp_3 k)} C_k(\Phi_3,k).
\end{equation}
When  $\tp_2$ is collinear to $\tp_3$, it reduces to
\begin{equation}
  \frac{(\tp_1\tp_3)}{(\tp_1 k)(\tp_3 k)} C_k(\Phi_3,k) \underset{p_2\parallel p_3}{\approx}  \frac{(\dtp_1\dtp_2)}{(\dtp_1k)(\dtp_2k)} C_k(\Phi_2,k),
\end{equation}
as before. However, when  $\tp_1$ is collinear to $\tp_3$ it becomes zero.
In this case then
\begin{equation}
    \delta_{\rm NP} \approxczero - \frac{1}{2} \times
    \frac{\alpha_s}{2\pi}C_F\times \mathcal T_\lambda I_c(\Phi_2;q;\lambda).
  \label{eqApp:deltaC_nosplit3}
\end{equation}
This result follows because the most enhanced regions when $c\to 0$ are
the soft-collinear ones, but only one of the two contributes, hence the
factor of one half.

The above reasoning is corroborated by an explicit evaluation of the
final result of Section~\ref{sec:shapevars-ex}, Eq.~(\ref{eq:tlambdaIc}).
Setting\footnote{We recall that $x_i = 2(\tp_i q)/q^2$, $x_1+x_2+x_3=2$.}
\begin{equation}
  x_1=1-z x_3,\quad x_2=1-(1-z)x_3
\end{equation}
in that equation, we find that
\begin{equation}
  {\cal T}_\lambda I_c \approx -6\pi \left(\frac{\lambda}{\sqrt{q^2}}\right)
  \times 2,
\end{equation}
for both the soft $x_3\to 0$ and the two collinear $z\to 0$,  $z\to 1$
limits.
To extend this result to the quark-gluon dipole, 
we still start from Eq.~\eqref{eq:tlambdaIc} but assume 
that 1 and 2 are the quark and the gluon, and 3 is the antiquark.
We now use the parametrization
\begin{equation}
  x_1=1-z x_2,\quad x_3=1-(1-z)x_2.
\end{equation}
The singular regions of the Born amplitude are then given by
 $x_2\to 0$ and $z\to 0$,  $z\to 1$. In this case we find \begin{equation}
  {\cal T}_\lambda I_c \approx -6\pi \left(\frac{\lambda}{\sqrt{q^2}}\right)
  \times \xi,
\end{equation}
with $\xi \to 2$ for $z\to 0$, $\xi \to 0$ for  $z\to 1$, and
$\xi \to 4z^2-6z+2$ for $x_2 \to 0$. Restricting ourself to the most
singular regions, i.e. the soft-collinear ones, we see that indeed
only one of the two regions contributes, as we have said before. 

We further remark that the soft-collinear approximation is enough to
get these results, that can thus be considered consequences of angular
ordering.  In the quark-antiquark dipole case, the result follows also from the
full soft factorization that applies in abelian theories. In this case
we expect that the limit is reached earlier.  This is not the case for
the $q g$ dipole, since the Born level gluon and gluon emitted by the
$q g$ dipole do not factorize simultaneously in the soft limit.


\bibliographystyle{JHEP}
\bibliography{renorm}

\providecommand{\href}[2]{#2}\begingroup\raggedright\begin{thebibliography}{10}

\bibitem{Heinrich:2020ybq}
G.~Heinrich, \emph{{Collider Physics at the Precision Frontier}},
  \href{http://dx.doi.org/10.1016/j.physrep.2021.03.006}{\emph{Phys. Rept.}
  {\bf 922} (2021) 1--69}, [\href{http://arxiv.org/abs/2009.00516}{{\tt
  2009.00516}}].

\bibitem{Bigi:1993fe}
I.~I.~Y. Bigi, M.~A. Shifman, N.~G. Uraltsev and A.~I. Vainshtein, \emph{{QCD
  predictions for lepton spectra in inclusive heavy flavor decays}},
  \href{http://dx.doi.org/10.1103/PhysRevLett.71.496}{\emph{Phys. Rev. Lett.}
  {\bf 71} (1993) 496--499}, [\href{http://arxiv.org/abs/hep-ph/9304225}{{\tt
  hep-ph/9304225}}].

\bibitem{Beneke:1998ui}
M.~Beneke, \emph{{Renormalons}},
  \href{http://dx.doi.org/10.1016/S0370-1573(98)00130-6}{\emph{Phys. Rept.}
  {\bf 317} (1999) 1--142}, [\href{http://arxiv.org/abs/hep-ph/9807443}{{\tt
  hep-ph/9807443}}].

\bibitem{Manohar:1994kq}
A.~V. Manohar and M.~B. Wise, \emph{{Power suppressed corrections to hadronic
  event shapes}},
  \href{http://dx.doi.org/10.1016/0370-2693(94)01504-6}{\emph{Phys. Lett. B}
  {\bf 344} (1995) 407--412}, [\href{http://arxiv.org/abs/hep-ph/9406392}{{\tt
  hep-ph/9406392}}].

\bibitem{Webber:1994cp}
B.~R. Webber, \emph{{Estimation of power corrections to hadronic event
  shapes}}, \href{http://dx.doi.org/10.1016/0370-2693(94)91147-9}{\emph{Phys.
  Lett. B} {\bf 339} (1994) 148--150},
  [\href{http://arxiv.org/abs/hep-ph/9408222}{{\tt hep-ph/9408222}}].

\bibitem{Akhoury:1995sp}
R.~Akhoury and V.~I. Zakharov, \emph{{On the universality of the leading, 1/Q
  power corrections in QCD}},
  \href{http://dx.doi.org/10.1016/0370-2693(95)00866-J}{\emph{Phys. Lett. B}
  {\bf 357} (1995) 646--652}, [\href{http://arxiv.org/abs/hep-ph/9504248}{{\tt
  hep-ph/9504248}}].

\bibitem{Dokshitzer:1995zt}
Y.~L. Dokshitzer and B.~R. Webber, \emph{{Calculation of power corrections to
  hadronic event shapes}},
  \href{http://dx.doi.org/10.1016/0370-2693(95)00548-Y}{\emph{Phys. Lett. B}
  {\bf 352} (1995) 451--455}, [\href{http://arxiv.org/abs/hep-ph/9504219}{{\tt
  hep-ph/9504219}}].

\bibitem{Nason:1995np}
P.~Nason and M.~H. Seymour, \emph{{Infrared renormalons and power suppressed
  effects in e+ e- jet events}},
  \href{http://dx.doi.org/10.1016/0550-3213(95)00461-Z}{\emph{Nucl. Phys. B}
  {\bf 454} (1995) 291--312}, [\href{http://arxiv.org/abs/hep-ph/9506317}{{\tt
  hep-ph/9506317}}].

\bibitem{Dokshitzer:1995qm}
Y.~L. Dokshitzer, G.~Marchesini and B.~R. Webber, \emph{{Dispersive approach to
  power behaved contributions in QCD hard processes}},
  \href{http://dx.doi.org/10.1016/0550-3213(96)00155-1}{\emph{Nucl. Phys. B}
  {\bf 469} (1996) 93--142}, [\href{http://arxiv.org/abs/hep-ph/9512336}{{\tt
  hep-ph/9512336}}].

\bibitem{Dasgupta:1996ki}
M.~Dasgupta and B.~R. Webber, \emph{{Power corrections and renormalons in
  $e^{+} e^{-}$ fragmentation functions}},
  \href{http://dx.doi.org/10.1016/S0550-3213(96)00622-0}{\emph{Nucl. Phys. B}
  {\bf 484} (1997) 247--264}, [\href{http://arxiv.org/abs/hep-ph/9608394}{{\tt
  hep-ph/9608394}}].

\bibitem{Beneke:1997sr}
M.~Beneke, V.~M. Braun and L.~Magnea, \emph{{Phenomenology of power corrections
  in fragmentation processes in e+ e- annihilation}},
  \href{http://dx.doi.org/10.1016/S0550-3213(97)00251-4}{\emph{Nucl. Phys. B}
  {\bf 497} (1997) 297--333}, [\href{http://arxiv.org/abs/hep-ph/9701309}{{\tt
  hep-ph/9701309}}].

\bibitem{Dokshitzer:1997ew}
Y.~L. Dokshitzer and B.~R. Webber, \emph{{Power corrections to event shape
  distributions}},
  \href{http://dx.doi.org/10.1016/S0370-2693(97)00573-X}{\emph{Phys. Lett. B}
  {\bf 404} (1997) 321--327}, [\href{http://arxiv.org/abs/hep-ph/9704298}{{\tt
  hep-ph/9704298}}].

\bibitem{Dokshitzer:1997iz}
Y.~L. Dokshitzer, A.~Lucenti, G.~Marchesini and G.~P. Salam,
  \emph{{Universality of 1/Q corrections to jet-shape observables rescued}},
  \href{http://dx.doi.org/10.1016/S0550-3213(97)00650-0}{\emph{Nucl. Phys. B}
  {\bf 511} (1998) 396--418}, [\href{http://arxiv.org/abs/hep-ph/9707532}{{\tt
  hep-ph/9707532}}].

\bibitem{Dokshitzer:1998pt}
Y.~L. Dokshitzer, A.~Lucenti, G.~Marchesini and G.~P. Salam, \emph{{On the
  universality of the Milan factor for 1 / Q power corrections to jet shapes}},
  \href{http://dx.doi.org/10.1088/1126-6708/1998/05/003}{\emph{JHEP} {\bf 05}
  (1998) 003}, [\href{http://arxiv.org/abs/hep-ph/9802381}{{\tt
  hep-ph/9802381}}].

\bibitem{Agarwal:2020uxi}
N.~Agarwal, A.~Mukhopadhyay, S.~Pal and A.~Tripathi, \emph{{Power corrections
  to event shapes using eikonal dressed gluon exponentiation}},
  \href{http://dx.doi.org/10.1007/JHEP03(2021)155}{\emph{JHEP} {\bf 03} (2021)
  155}, [\href{http://arxiv.org/abs/2012.06842}{{\tt 2012.06842}}].

\bibitem{Beneke:1994sw}
M.~Beneke and V.~M. Braun, \emph{{Heavy quark effective theory beyond
  perturbation theory: Renormalons, the pole mass and the residual mass term}},
  \href{http://dx.doi.org/10.1016/0550-3213(94)90314-X}{\emph{Nucl. Phys. B}
  {\bf 426} (1994) 301--343}, [\href{http://arxiv.org/abs/hep-ph/9402364}{{\tt
  hep-ph/9402364}}].

\bibitem{Bigi:1994em}
I.~I.~Y. Bigi, M.~A. Shifman, N.~G. Uraltsev and A.~I. Vainshtein, \emph{{The
  Pole mass of the heavy quark. Perturbation theory and beyond}},
  \href{http://dx.doi.org/10.1103/PhysRevD.50.2234}{\emph{Phys. Rev. D} {\bf
  50} (1994) 2234--2246}, [\href{http://arxiv.org/abs/hep-ph/9402360}{{\tt
  hep-ph/9402360}}].

\bibitem{Beneke:1995pq}
M.~Beneke and V.~M. Braun, \emph{{Power corrections and renormalons in
  Drell-Yan production}},
  \href{http://dx.doi.org/10.1016/0550-3213(95)00439-Y}{\emph{Nucl. Phys. B}
  {\bf 454} (1995) 253--290}, [\href{http://arxiv.org/abs/hep-ph/9506452}{{\tt
  hep-ph/9506452}}].

\bibitem{Dasgupta:1999zm}
M.~Dasgupta, \emph{{Power corrections to the differential Drell-Yan
  cross-section}},
  \href{http://dx.doi.org/10.1088/1126-6708/1999/12/008}{\emph{JHEP} {\bf 12}
  (1999) 008}, [\href{http://arxiv.org/abs/hep-ph/9911391}{{\tt
  hep-ph/9911391}}].

\bibitem{Korchemsky:1996iq}
G.~P. Korchemsky, \emph{{Power corrections in Drell-Yan production beyond the
  leading order}},  in \emph{{28th International Conference on High-energy
  Physics}}, 10, 1996.
\newblock \href{http://arxiv.org/abs/hep-ph/9610207}{{\tt hep-ph/9610207}}.

\bibitem{Korchemsky:1994is}
G.~P. Korchemsky and G.~F. Sterman, \emph{{Nonperturbative corrections in
  resummed cross-sections}},
  \href{http://dx.doi.org/10.1016/0550-3213(94)00006-Z}{\emph{Nucl. Phys. B}
  {\bf 437} (1995) 415--432}, [\href{http://arxiv.org/abs/hep-ph/9411211}{{\tt
  hep-ph/9411211}}].

\bibitem{Dasgupta:2007wa}
M.~Dasgupta, L.~Magnea and G.~P. Salam, \emph{{Non-perturbative QCD effects in
  jets at hadron colliders}},
  \href{http://dx.doi.org/10.1088/1126-6708/2008/02/055}{\emph{JHEP} {\bf 02}
  (2008) 055}, [\href{http://arxiv.org/abs/0712.3014}{{\tt 0712.3014}}].

\bibitem{FerrarioRavasio:2018ubr}
S.~Ferrario~Ravasio, P.~Nason and C.~Oleari, \emph{{All-orders behaviour and
  renormalons in top-mass observables}},
  \href{http://dx.doi.org/10.1007/JHEP01(2019)203}{\emph{JHEP} {\bf 01} (2019)
  203}, [\href{http://arxiv.org/abs/1810.10931}{{\tt 1810.10931}}].

\bibitem{FerrarioRavasio:2020guj}
S.~Ferrario~Ravasio, G.~Limatola and P.~Nason, \emph{{Infrared renormalons in
  kinematic distributions for hadron collider processes}},
  \href{http://dx.doi.org/10.1007/JHEP06(2021)018}{\emph{JHEP} {\bf 06} (2021)
  018}, [\href{http://arxiv.org/abs/2011.14114}{{\tt 2011.14114}}].

\bibitem{Luisoni:2020efy}
G.~Luisoni, P.~F. Monni and G.~P. Salam, \emph{{$C$-parameter hadronisation in
  the symmetric 3-jet limit and impact on $\alpha_s$ fits}},
  \href{http://dx.doi.org/10.1140/epjc/s10052-021-08941-z}{\emph{Eur. Phys. J.
  C} {\bf 81} (2021) 158}, [\href{http://arxiv.org/abs/2012.00622}{{\tt
  2012.00622}}].

\bibitem{vanBeekveld:2019prq}
M.~van Beekveld, W.~Beenakker, E.~Laenen and C.~D. White,
  \emph{{Next-to-leading power threshold effects for inclusive and exclusive
  processes with final state jets}},
  \href{http://dx.doi.org/10.1007/JHEP03(2020)106}{\emph{JHEP} {\bf 03} (2020)
  106}, [\href{http://arxiv.org/abs/1905.08741}{{\tt 1905.08741}}].

\bibitem{Ebert:2018gsn}
M.~A. Ebert, I.~Moult, I.~W. Stewart, F.~J. Tackmann, G.~Vita and H.~X. Zhu,
  \emph{{Subleading power rapidity divergences and power corrections for
  q$_{T}$}}, \href{http://dx.doi.org/10.1007/JHEP04(2019)123}{\emph{JHEP} {\bf
  04} (2019) 123}, [\href{http://arxiv.org/abs/1812.08189}{{\tt 1812.08189}}].

\bibitem{Passarino:1978jh}
G.~Passarino and M.~J.~G. Veltman, \emph{{One Loop Corrections for e+ e-
  Annihilation Into mu+ mu- in the Weinberg Model}},
  \href{http://dx.doi.org/10.1016/0550-3213(79)90234-7}{\emph{Nucl. Phys. B}
  {\bf 160} (1979) 151--207}.

\bibitem{Low:1958sn}
F.~E. Low, \emph{{Bremsstrahlung of very low-energy quanta in elementary
  particle collisions}},
  \href{http://dx.doi.org/10.1103/PhysRev.110.974}{\emph{Phys. Rev.} {\bf 110}
  (1958) 974--977}.

\bibitem{landau}
V.~B. Berestetskiĭ, E.~M. Lifshits, L.~P. Pitaevskiĭ, J.~B. Sykes and J.~S.
  Bell, \emph{Quantum electrodynamics}.
\newblock Landau, L. D., 1908-1968. Teoreticheskaia fizika (Izd. 2-e). English;
  v. 4. Butterworth-Heinemann, Oxford, 2nd ed.~ed., 1982.

\bibitem{DelDuca:2019ctm}
V.~Del~Duca, N.~Deutschmann and S.~Lionetti, \emph{{Momentum mappings for
  subtractions at higher orders in QCD}},
  \href{http://dx.doi.org/10.1007/JHEP12(2019)129}{\emph{JHEP} {\bf 12} (2019)
  129}, [\href{http://arxiv.org/abs/1910.01024}{{\tt 1910.01024}}].

\bibitem{Smirnov:1997gx}
V.~A. Smirnov, \emph{{Asymptotic expansions of two loop Feynman diagrams in the
  Sudakov limit}},
  \href{http://dx.doi.org/10.1016/S0370-2693(97)00545-5}{\emph{Phys. Lett. B}
  {\bf 404} (1997) 101--107}, [\href{http://arxiv.org/abs/hep-ph/9703357}{{\tt
  hep-ph/9703357}}].

\bibitem{Dasgupta:2020fwr}
M.~Dasgupta, F.~A. Dreyer, K.~Hamilton, P.~F. Monni, G.~P. Salam and G.~Soyez,
  \emph{{Parton showers beyond leading logarithmic accuracy}},
  \href{http://dx.doi.org/10.1103/PhysRevLett.125.052002}{\emph{Phys. Rev.
  Lett.} {\bf 125} (2020) 052002}, [\href{http://arxiv.org/abs/2002.11114}{{\tt
  2002.11114}}].

\bibitem{Nason:1996pk}
P.~Nason and B.~R. Webber, \emph{{Nonperturbative corrections to heavy quark
  fragmentation in e+ e- annihilation}},
  \href{http://dx.doi.org/10.1016/S0370-2693(97)00129-9}{\emph{Phys. Lett. B}
  {\bf 395} (1997) 355--363}, [\href{http://arxiv.org/abs/hep-ph/9612353}{{\tt
  hep-ph/9612353}}].

\bibitem{Campbell:1998qw}
J.~M. Campbell, E.~W.~N. Glover and C.~J. Maxwell, \emph{{Determination of
  Lambda(QCD) from the measured energy dependence of \ensuremath{<}1 -
  thrust\ensuremath{>}}},
  \href{http://dx.doi.org/10.1103/PhysRevLett.81.1568}{\emph{Phys. Rev. Lett.}
  {\bf 81} (1998) 1568--1571}, [\href{http://arxiv.org/abs/hep-ph/9803254}{{\tt
  hep-ph/9803254}}].

\bibitem{Abbate:2010xh}
R.~Abbate, M.~Fickinger, A.~H. Hoang, V.~Mateu and I.~W. Stewart, \emph{{Thrust
  at $N^{3}LL$ with Power Corrections and a Precision Global Fit for
  $\alpha_{s}(mZ)$}},
  \href{http://dx.doi.org/10.1103/PhysRevD.83.074021}{\emph{Phys. Rev. D} {\bf
  83} (2011) 074021}, [\href{http://arxiv.org/abs/1006.3080}{{\tt 1006.3080}}].

\bibitem{Hoang:2015hka}
A.~H. Hoang, D.~W. Kolodrubetz, V.~Mateu and I.~W. Stewart, \emph{{Precise
  determination of $\alpha_s$ from the $C$-parameter distribution}},
  \href{http://dx.doi.org/10.1103/PhysRevD.91.094018}{\emph{Phys. Rev. D} {\bf
  91} (2015) 094018}, [\href{http://arxiv.org/abs/1501.04111}{{\tt
  1501.04111}}].

\bibitem{Catani:1998sf}
S.~Catani and B.~R. Webber, \emph{{Resummed C parameter distribution in e+ e-
  annihilation}},
  \href{http://dx.doi.org/10.1016/S0370-2693(98)00359-1}{\emph{Phys. Lett. B}
  {\bf 427} (1998) 377--384}, [\href{http://arxiv.org/abs/hep-ph/9801350}{{\tt
  hep-ph/9801350}}].

\bibitem{Gehrmann:2012sc}
T.~Gehrmann, G.~Luisoni and P.~F. Monni, \emph{{Power corrections in the
  dispersive model for a determination of the strong coupling constant from the
  thrust distribution}},
  \href{http://dx.doi.org/10.1140/epjc/s10052-012-2265-x}{\emph{Eur. Phys. J.
  C} {\bf 73} (2013) 2265}, [\href{http://arxiv.org/abs/1210.6945}{{\tt
  1210.6945}}].

\bibitem{Davison:2009wzs}
R.~A. Davison and B.~R. Webber, \emph{{Non-Perturbative Contribution to the
  Thrust Distribution in e+ e- Annihilation}},
  \href{http://dx.doi.org/10.1140/epjc/s10052-008-0836-7}{\emph{Eur. Phys. J.
  C} {\bf 59} (2009) 13--25}, [\href{http://arxiv.org/abs/0809.3326}{{\tt
  0809.3326}}].

\bibitem{Smye:2001gq}
G.~E. Smye, \emph{{On the 1/Q correction to the C - parameter at two loops}},
  \href{http://dx.doi.org/10.1088/1126-6708/2001/05/005}{\emph{JHEP} {\bf 05}
  (2001) 005}, [\href{http://arxiv.org/abs/hep-ph/0101323}{{\tt
  hep-ph/0101323}}].

\bibitem{maxima}
Maxima, \emph{Maxima, a Computer Algebra System. Version 5.43.2}.
\newblock http://maxima.sourceforge.net/, 2020.

\bibitem{Denner:2016kdg}
A.~Denner, S.~Dittmaier and L.~Hofer, \emph{{Collier: a fortran-based Complex
  One-Loop LIbrary in Extended Regularizations}},
  \href{http://dx.doi.org/10.1016/j.cpc.2016.10.013}{\emph{Comput. Phys.
  Commun.} {\bf 212} (2017) 220--238},
  [\href{http://arxiv.org/abs/1604.06792}{{\tt 1604.06792}}].

\bibitem{Alioli:2010xd}
S.~Alioli, P.~Nason, C.~Oleari and E.~Re, \emph{{A general framework for
  implementing NLO calculations in shower Monte Carlo programs: the POWHEG
  BOX}}, \href{http://dx.doi.org/10.1007/JHEP06(2010)043}{\emph{JHEP} {\bf 06}
  (2010) 043}, [\href{http://arxiv.org/abs/1002.2581}{{\tt 1002.2581}}].

\end{thebibliography}\endgroup

\end{document}